\documentclass[aps,prb,twocolumn,floatfix,superscriptaddress,longbibliography]{revtex4-2}
\usepackage{graphicx,txfonts,bm}
\usepackage[colorlinks,citecolor=magenta,linkcolor=blue]{hyperref}
\usepackage{float}
\usepackage[section]{placeins}
\begin{document}

\title{\emph{Ab initio} dynamical mean-field theory with natural orbitals renormalization group impurity solver: Formalism and applications}

\author{Jia-Ming Wang}
\thanks{These authors contributed equally to this work.}
\author{Jing-Xuan Wang}
\thanks{These authors contributed equally to this work.}
\author{Rong-Qiang He}
\email{rqhe@ruc.edu.cn}
\affiliation{Department of Physics, Renmin University of China, Beijing 100872, China}
\affiliation{Key Laboratory of Quantum State Construction and Manipulation (Ministry of Education), Renmin University of China, Beijing 100872, China}

\author{Li Huang}
\email{huangli@caep.cn}
\affiliation{Science and Technology on Surface Physics and Chemistry Laboratory, P.O. Box 9-35, Jiangyou 621908, China}

\author{Zhong-Yi Lu}
\email{zlu@ruc.edu.cn}
\affiliation{Department of Physics, Renmin University of China, Beijing 100872, China}
\affiliation{Key Laboratory of Quantum State Construction and Manipulation (Ministry of Education), Renmin University of China, Beijing 100872, China}
\affiliation{Hefei National Laboratory, Hefei 230088, China}

\date{\today}

\begin{abstract}
In this study, we introduce a novel implementation of density functional theory integrated with single-site dynamical mean-field theory to investigate the complex properties of strongly correlated materials. This comprehensive first-principles many-body computational toolkit, termed \texttt{Zen}, utilizes the Vienna \emph{ab initio} simulation package and the \texttt{Quantum ESPRESSO} code to perform density functional theory calculations and generate band structures for realistic materials. The challenges associated with correlated electron systems are addressed through two distinct yet complementary quantum impurity solvers: the natural orbitals renormalization group solver for zero temperature and the hybridization expansion continuous-time quantum Monte Carlo solver for finite temperature. Additionally, this newly developed toolkit incorporates several valuable post-processing tools, such as \texttt{ACFlow}, which employs the maximum entropy method and the stochastic pole expansion method for the analytic continuation of Matsubara Green's functions and self-energy functions. To validate the performance of this toolkit, we examine three representative cases: the correlated metal SrVO$_{3}$, the nickel-based unconventional superconductor La$_{3}$Ni$_{2}$O$_{7}$, and the wide-gap Mott insulator MnO. The results obtained demonstrate strong agreement with experimental findings and previously available theoretical results. Notably, we successfully elucidate the quasiparticle peak and band renormalization in SrVO$_{3}$, the dominance of Hund correlation in La$_{3}$Ni$_{2}$O$_{7}$, and the pressure-driven insulator-metal transition as well as the high-spin to low-spin transition in MnO. These findings suggest that \texttt{Zen} is proficient in accurately describing the electronic structures of $d$-electron correlated materials.   
\end{abstract}

\maketitle

\section{introduction\label{sec:intro}}

The physical and chemical properties of solid-state materials are remarkably varied and extensive. Nevertheless, upon tracking their origins, it becomes evident that the quantum behaviors of valence electrons are crucial. Consequently, a precise characterization of the microscopic behavior of valence electrons in solid-state materials is a fundamental requirement in the field of condensed matter physics.

Electromagnetic theory posits the existence of long-range Coulomb interactions among electrons. In scenarios where the interaction between electrons is significantly less than their kinetic energy, a single-particle approximation can be employed to analyze the motion of electrons within a solid. This foundational approach has led to the development of classical band theory, which has effectively elucidated the fundamental characteristics of conventional metals, semiconductors, and insulators. These materials are typically categorized as weakly correlated or non-correlated. In contrast, there exist strongly correlated materials~\cite{adma.201202018}, such as certain transition metal oxides~\cite{RevModPhys.70.1039}, copper-based~\cite{RevModPhys.78.17} and iron-based~\cite{RevModPhys.87.855} unconventional superconductors, as well as cerium-based~\cite{Weng_2016} and plutonium-based~\cite{annurev-conmatphys-031214-014508} heavy fermion systems. The electronic structures of these materials frequently exhibit narrow and partially filled $d$- or $f$-electron energy bands. In these cases, the Coulomb interaction among electrons is considerably greater than their kinetic energy, rendering the single-particle approximation inadequate~\cite{PhysRevB.44.943}. The collective motion of numerous electrons in strongly correlated materials leads to a variety of exotic phenomena, including colossal magnetoresistance, Mott metal-insulator transitions, unconventional superconductivity, non-Fermi liquid behavior, heavy fermion behavior, and the Kondo effect~\cite{RevModPhys.70.1039,RevModPhys.87.855,Weng_2016,annurev-conmatphys-031214-014508,RevModPhys.73.797,RevModPhys.78.17}. Classical band theory falls short in providing a coherent explanation for these phenomena. Consequently, the advancement of a theory for strongly correlated electrons that transcends classical band theory represents a prominent area of research within condensed matter physics~\cite{science.aat5975}.

The development of a theoretical framework for strongly correlated electrons typically commences with the examination of simplified models that encapsulate strong correlation effects. While this approach significantly alleviates the complexity of the problem, it remains a challenging task, as most models characterized by strong correlations lack exact analytical solutions. Compounding this challenge is the fact that in strongly correlated systems, the bare Coulomb interaction is substantial, rendering conventional perturbation theory inapplicable. The advent of dynamical mean-field theory (dubbed DMFT) has provided valuable insights into this pressing issue~\cite{andp.201100250}. The fundamental premise of DMFT is that, in the limit of infinite dimensions, the electron self-energy $\Sigma$ is local and momentum-independent. Consequently, a general interacting lattice model can be self-consistently transformed into a single quantum impurity model, from which the properties of the original lattice model can be inferred by solving the quantum impurity model~\cite{RevModPhys.68.13}. It is important to note that DMFT is inherently a local theory. Although it adopts a mean-field approximation to address spatial fluctuations of quantum states, which is equivalent to ignoring the non-local aspects of electron correlations, it rigorously accounts for temporal fluctuations and comprehensively captures the local components of electron correlations~\cite{andp.201100250}. In principle, for interacting models, local correlations are of primary significance, while non-local correlations are often negligible. Consequently, over the past two decades, DMFT has emerged as a pivotal tool in the investigation of strongly correlated models. Through the application of DMFT, the understanding of various strongly correlated systems, including the Hubbard model, the $t-J$ model, and the periodic Anderson model, has reached unprecedented levels, leading to the resolution of numerous longstanding physical challenges~\cite{RevModPhys.68.13}. Currently, DMFT has evolved into numerous extensions, enabling its application to the study of disordered, inhomogeneous, and non-equilibrium systems~\cite{PhysRevB.77.054202,acprof:oso,RevModPhys.86.779}.

Merely addressing strongly correlated models is insufficient, it is imperative to investigate the electronic structures and physical properties of realistic materials. The DMFT method is adept at managing correlation effects among electrons, yet it is primarily applicable to the analysis of model Hamiltonians~\cite{andp.201100250,RevModPhys.68.13}. On the contrary, classical band theory, exemplified by the widely utilized density functional theory (dubbed DFT), while inadequate in accurately addressing electron-electron interactions in strongly correlated materials, excels in elucidating crystal field splitting and the chemical environment of realistic materials without reliance on empirical parameters. Given that both methodologies possess distinct advantages and limitations that complement one another, it is natural to integrate them to formulate the DFT+DMFT approach. The standard procedure for DFT+DMFT calculations is delineated as follows~\cite{held2007,RevModPhys.78.865}. Initially, a conventional DFT calculation is performed to derive the band structure under the single-particle approximation, followed by the construction of a model Hamiltonian for the correlated orbitals of interest, devoid of empirical parameters. Subsequently, interaction terms, including Coulomb interactions and spin-orbit coupling, are incorporated into this model Hamiltonian. Finally, the DMFT method is employed to solve the effective model Hamiltonian, thereby extracting the ground state and spectroscopic properties of realistic materials. The pioneering application of the DFT+DMFT method to investigate the photoelectron spectrum of La-doped SrTiO$_{3}$ was conducted by V. I. Anisimov \emph{et al.} in 1997~\cite{Anisimov_1997}. Since that time, the DFT+DMFT method has become a dominant approach in the realm of first-principles calculations for strongly correlated materials. Particularly in recent years, during the surge of research into various unconventional superconductors~\cite{PhysRevB.108.125105,PhysRevB.109.165140,PhysRevB.82.064504,PhysRevB.100.121101}, strongly correlated kagome materials~\cite{PhysRevB.102.125130,Liu2020,PhysRevB.105.155131}, and cerium-based heavy fermion materials~\cite{PhysRevB.102.155140,science.1149064,PhysRevB.94.075132}, the DFT+DMFT method has shown its unique brilliance.

In the past decade, significant efforts have been dedicated to the development of efficient first-principles software packages for DFT+DMFT calculations, thereby facilitating the broader application of this method. Currently, several open-source DFT+DMFT packages have been released, including \texttt{w2dynamics}~\cite{WALLERBERGER2019388}, \texttt{TRIQS}~\cite{AICHHORN2016200,PARCOLLET2015398,SETH2016274}, \texttt{ALPSCore}~\cite{GAENKO2017235,SHINAOKA2017128}, \texttt{eDMFT}~\cite{PhysRevB.81.195107,PhysRevB.75.155113,PhysRevLett.115.196403}, \texttt{DCore}~\cite{10.21468/SciPostPhys.10.5.117}, \texttt{Questaal}~\cite{PASHOV2020107065}, \texttt{Abinit}~\cite{PhysRevB.77.205112,Amadon_2012,GONZE2020107042}, \texttt{ABACUS}~\cite{acs.jctc.2c00472}, \texttt{ComDMFT}~\cite{CHOI2019277}, \texttt{DMFTwDFT}~\cite{SINGH2021107778}, and \texttt{solid\_dmft}~\cite{Merkel2022}, among others. Notably, the \texttt{w2dynamics}, \texttt{TRIQS}, and \texttt{ALPSCore} packages primarily incorporate high-performance quantum impurity solvers, utilizing the hybridization expansion continuous-time quantum Monte Carlo algorithm (dubbed CT-HYB)~\cite{RevModPhys.83.349,PhysRevB.74.155107,PhysRevLett.97.076405}. Other software packages, with the exceptions of \texttt{eDMFT}~\cite{PhysRevB.75.155113} and \texttt{ComDMFT}~\cite{MELNICK2021108075}, do not incorporate individual quantum impurity solvers; instead, they typically employ publicly available CT-HYB impurity solvers. These packages offer flexible interfaces to connect DFT codes with quantum impurity solvers, supporting both fully self-consistent and one-shot DFT+DMFT calculations aiming at investigating the electronic structures and lattice dynamics of strongly correlated materials~\cite{PhysRevB.102.245104,PhysRevMaterials.7.093801,PhysRevB.102.241108}. They are capable of producing various physical observables, such as spectral functions, Fermi surfaces, total energies, forces, and phonon band structures, which can be readily compared with experimental data. Notably, the \texttt{Questaal}~\cite{PASHOV2020107065} and \texttt{ComDMFT}~\cite{CHOI2019277} packages can incorporate the quasiparticle approximation (dubbed GW)~\cite{faryase1998} within the DFT part and support the GW+DMFT calculation mode~\cite{PhysRevB.94.201106,PhysRevX.11.021006}. Additionally, the \texttt{w2dynamics} package can interface with the \texttt{AbinitioD$\Gamma$A} code~\cite{GALLER2019106847} to facilitate \emph{ab initio} dynamical vertex approximation (D$\Gamma$A) calculations~\cite{RevModPhys.90.025003,PhysRevB.75.045118,PhysRevB.95.115107}. Overall, while these packages exhibit similarities, their distinctions primarily arise from the selection of DFT codes and the definitions of local basis sets employed in constructing low-energy effective Hamiltonians.

Though many DFT+DMFT software packages have been published, another open-source implementation is always beneficial for the community. In this paper, we are pleased to introduce \texttt{Zen}, a new DFT+DMFT toolkit. This toolkit offers the following features: (i) It supports two types of local orbitals: projected local orbitals (interfaced with \texttt{VASP}~\cite{PhysRevB.59.1758,PhysRevB.54.11169}) and maximally localized Wannier functions (interfaced with \texttt{Quantum ESPRESSO}~\cite{Giannozzi_2009,Giannozzi_2017} and \texttt{Wannier90}~\cite{Pizzi2020}). (ii) It includes two quantum impurity solvers: the natural orbitals renormalization group solver (dubbed NORG)~\cite{PhysRevB.89.085108,PhysRevB.91.155140} and the CT-HYB solver (interfaced with \texttt{iQIST}~\cite{HUANG2015140,HUANG2017423}). The NORG impurity solver operates at zero temperature, while the CT-HYB impurity solver works for finite temperature. These two quantum impurity solvers are complementary. (iii) It features a powerful analytic continuation backend, namely \texttt{ACFlow}~\cite{HUANG2023108863}, which implements several state-of-the-art analytic continuation methods, including the maximum entropy method, stochastic analytic continuation, and stochastic pole expansion, among others. \texttt{ACFlow} can convert single-particle or two-particle correlation functions from the imaginary time or imaginary frequency axis to the real-frequency axis. (iv) The \texttt{Zen} toolkit is developed using the Julia programming language and supports large-scale parallel calculations. Furthermore, it can be executed in interactive mode, allowing users to monitor calculations and dynamically adjust computational parameters. After extensive testing, we suggest that this toolkit is well-suited for first-principles calculations of correlated $d$-electron materials.

The remainder of this paper is organized as follows. In Section~\ref{sec:method}, we provide a brief overview of the core components of the \texttt{Zen} toolkit, including the flowchart, DFT codes, DMFT engine, quantum impurity solvers, and utilities for analytic continuation. In Section~\ref{sec:benchmarks}, we first summarize the computational parameters. We then present the benchmark results for the correlated metal SrVO$_{3}$, the unconventional superconductor La$_{3}$Ni$_{2}$O$_{7}$, and the Mott insulator MnO, comparing them with previously published theoretical and experimental results. Finally, a brief conclusion is provided in Section~\ref{sec:summary}.

\section{DFT+DMFT method\label{sec:method}}

\begin{figure*}[ht] 
\centering 
\includegraphics[width=\textwidth]{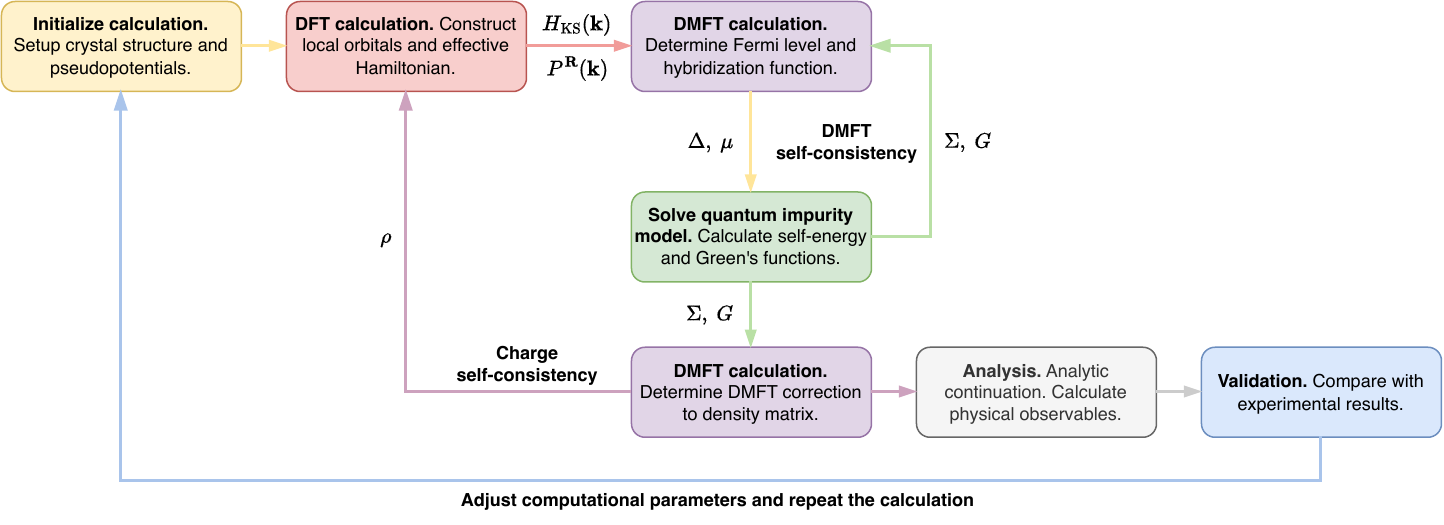}
\caption{Schematic picture of the standard DFT+DMFT calculation flowchart as implemented in the \texttt{Zen} toolkit. $H_{\text{KS}}(\mathbf{k})$: Low-energy effective Hamiltonian for correlated orbitals and ligand orbitals. $P^{\mathbf{R}}(\mathbf{k})$: Projection matrix between local basis and Kohn-Sham basis. $\Delta$: Hybridization function. $\Sigma$: Self-energy function. $G$: Green's function. $\mu$: Fermi level. $\rho$: Density matrix. The DFT interface (pink box) supports the \texttt{VASP} and \texttt{Quantum ESPRESSO} codes. The quantum impurity solvers (green box) can be NORG or CT-HYB. The analytic continuation calculations are done by the \texttt{ACFlow} toolkit (grey box). See main text for more details.}
\label{fig:dmft}
\end{figure*}

The theoretical foundation and computational procedure of the DFT+DMFT method are now well established~\cite{held2007,RevModPhys.78.865}. This section will focus on the overall framework of \texttt{Zen}, briefly describing the features and underlying formulas for the major computational components, with particular emphasis on the differences between \texttt{Zen} and the other DFT+DMFT calculation softwares. The technical details of \texttt{Zen} will be presented in a successive paper.

\subsection{Flowchart}

Here's a brief overview of the basic steps about how to perform charge fully self-consistent DFT+DMFT calculations using \texttt{Zen} (see Fig.~\ref{fig:dmft})~\cite{held2007}. (i) \emph{DFT calculation}. At first, we need to perform standard DFT calculations to obtain electronic structures of the materials. By choosing a reasonable local basis, the low-energy effective Hamiltonian for correlated orbitals and important ligand orbitals is constructed. This step provides a starting point for the DFT+DMFT calculation. (ii) \emph{DMFT self-consistency}. The low-energy effective Hamiltonian is supplemented with the local interaction term. Then it is solved self-consistently within the DMFT framework~\cite{RevModPhys.68.13}. To be more specific, it is mapped onto an effective impurity model, where the correlated electrons are treated as an impurity embedded in a non-interacting bath. The local impurity problem is solved using various quantum impurity solvers. The impurity solver generates new Green's function $G$ and self-energy function $\Sigma$ for the impurity model. $G$ and $\Sigma$ are then used to calculate new bath Green's function $G_0$ and hybridization function $\Delta$. They will be fed back into the impurity solver again. The process is repeated until the hybridization function and self-energy function converge to a consistent solution. (iii) \emph{Charge self-consistency}. In addition to the DMFT self-consistency loop, a charge self-consistency step is superimposed to ensure that the total charge density of the material is converged. This involves adjusting the charge density until the total charge matches the expected value. (iv) \emph{Analysis}. Once the self-consistent solution is converged, we should analyze the results to extract physical properties such as the density of states, spectral functions, and other relevant quantities. (v) \emph{Validation}. We have to validate the results by comparing them with experimental data or other theoretical results. If necessary, the computational model or parameters should be refined, and the calculation could be repeated.

In the \texttt{Zen} toolkit, the above computational steps are encapsulated in four components. They are the DFT, DMFT, quantum impurity solver, and post-processing components. The four components exchange parameters and data through files. They are manipulated by the \texttt{Zen} core library, which is written in Julia language. By using the \texttt{Zen} core library, the users can perform DFT+DMFT calculations step by step.  

\subsection{DFT codes}

Now the \texttt{Zen} toolkit is interfaced with two popular DFT codes, namely \texttt{VASP}~\cite{PhysRevB.59.1758,PhysRevB.54.11169} and \texttt{Quantum ESPRESSO}~\cite{Giannozzi_2009,Giannozzi_2017}. Both codes implement the pseudopotential plane-wave method. In the present work, the projector augmented wave (dubbed PAW) method is adopted~\cite{PhysRevB.50.17953}.

The \texttt{VASP} interface builds on the projected local orbitals (dubbed PLO) scheme~\cite{PhysRevB.77.205112}, where the resulting Kohn-Sham states $| \Psi \rangle$ from DFT calculations are projected on localized orbitals $| \chi \rangle$, which defines a basis for setting up a Hubbard-like model Hamiltonian. As is described in detail in Ref.[\onlinecite{PhysRevB.77.205112}], the projection matrix $P$ between $|\chi \rangle$ and $| \Psi \rangle$ in the PAW framework can be written as
\begin{equation}
P^{\mathbf{R}}(\mathbf{k}) = \sum_i 
\langle \chi^{\mathbf{R}} | \phi_i \rangle
\langle \tilde{p}_i | \tilde{\Psi}_{\mathbf{k}} \rangle.
\end{equation}
Here, the index $i$ denotes the PAW channel $n$, the angular momentum quantum number $l$, and its magnetic quantum number $m$. $|\chi^{\mathbf{R}}\rangle$ are localized basis functions associated with the correlated site $\mathbf{R}$. $|\tilde{\Psi}_{\mathbf{k}}\rangle$ are the pseudo-Kohn-Sham states. $|\phi_i\rangle$ and $|\tilde{p}_i\rangle$ are the all-electron partial waves and the standard PAW projectors, respectively. They can be extracted from the PAW dataset directly. The projection matrix $P$ will be calculated and written into the file ``LOCPROJ'' by the \texttt{VASP} code. The \texttt{Zen} toolkit will parse this file, read in the projection matrix, and filter out the Kohn-Sham states that don't belong to the given energy window. Then the projection matrix is further orthogonalized to make sure that the local basis in the restricted energy window is orthonormal.   

The \texttt{Quantum ESPRESSO} interface~\cite{Giannozzi_2017,Giannozzi_2009} should be used in conjunction with the \texttt{Wannier90} code~\cite{Pizzi2020}. It supports two different schemes. One is the traditional PLO scheme as introduced above~\cite{PhysRevB.77.205112}. When this scheme is used, the \texttt{Zen} toolkit will extract the projection matrix from the ``seedname.amn'' file. Another scheme builds on the maximally-localised Wannier functions (MLWF)~\cite{PhysRevB.56.12847}. The symmetry-adapted Wannier functions (SAWF) are also supported~\cite{PhysRevB.87.235109}. The $U^{\text{dis}(\mathbf{k})}$ and $U^{(\mathbf{k})}$ matrices are responsible for disentangling the correlated orbitals, and transforming Bloch bands $|u_\mathbf{k}\rangle$ into Wannier orbitals $|w_{\mathbf{R}}\rangle$. The \texttt{Zen} toolkit will read the two matrices from files ``seedname\_u\_dis.mat'' and ``seedname\_u.mat'', respectively. Finally, the two matrices are combined to build the projection matrix $P$.

We note that both the \texttt{VASP} and \texttt{Quantum ESPRESSO} interfaces support DFT+DMFT charge self-consistent calculations. That is to say the two codes can read the DMFT correction to the density matrix $\rho$, and then restart the DFT calculation to generate a new projection matrix $P$.   

\subsection{DMFT engine}

In the \texttt{Zen} toolkit, the DMFT engine is developed with the Fortran 90 language. The following tasks should be accomplished in the DMFT engine.

\emph{Lattice Green's function}. The expression for lattice Green's function is as follows~\cite{RevModPhys.78.865}:
\begin{equation}
\label{eq:glatt}
G^{\mathbf{R}}(i\omega_n) = \frac{1}{\Omega_{\text{bz}}}
\int d^3\mathbf{k}~
P^{\mathbf{R}}(\mathbf{k})
G(\mathbf{k},i\omega_n)
P^{\mathbf{R}*}(\mathbf{k}),
\end{equation}
\begin{equation}
G(\mathbf{k},i\omega_n) = 
\frac{1}{(i\omega_n + \mu)\mathbb{I} - H_{\text{KS}}(\mathbf{k}) - \Sigma(i\omega_n) + \Sigma_{\text{dc}}}.
\end{equation}
Here, $\Omega_{\text{bz}}$ is the volume of the first Brillouin zone, $P^{\mathbf{R}}(\mathbf{k})$ is the projection matrix, $\mathbb{I}$ is the identity matrix, $\mu$ is the Fermi level, $H_{\text{KS}}(\mathbf{k})$ is the Kohn-Sham Hamiltonian, and $\Sigma_{\text{dc}}$ is the double counting term for self-energy function. Note that the right-hand side of Eq.~(\ref{eq:glatt}) is a typical Brillouin zone integration in the complex-energy plane. The Lambin-Vigneron analytical tetrahedron method~\cite{PhysRevB.29.3430} is employed to calculate this integral. In order to accelerate the calculation, the integration algorithm is fully parallelized over the $k$-points.

\emph{Hybridization function}. The hybridization function $\Delta(i\omega_n)$ describes the hybridization effect between impurity electrons and conduction electrons. Its expression is as follows~\cite{RevModPhys.68.13}:
\begin{equation}
\Delta(i\omega_n) = i\omega_n - E_{\text{imp}} - G^{-1}(i\omega_n) - \Sigma(i\omega_n),
\end{equation}
where the impurity level $E_{\text{imp}}$ reads
\begin{equation}
E_{\text{imp}} = \frac{1}{\Omega_{\text{bz}}}
\int d^3\mathbf{k}~[H_{\text{KS}}(\mathbf{k}) - \Sigma_{\text{dc}} - \mu \mathbb{I}].
\end{equation}
Both the hybridization function and impurity level are essential inputs for the quantum impurity solvers.    

\emph{Orbital occupancy}. Given the Fermi level $\mu$, to calculate the orbital-resolved impurity occupancy $N_{\alpha}$ is not a trivial problem. In principle, the orbital occupancy is defined by
\begin{equation}
\label{eq:nalpha}
N_{\alpha} = T \sum_{\mathbf{k}} \sum_n 
\frac{1}{i\omega_n + \mu - \epsilon_{\alpha\mathbf{k}}(i\omega_n)},
\end{equation}
where $\alpha$ is the orbital index, $\epsilon_{\alpha\mathbf{k}}(i\omega_n)$ are the eigenvalues of the Hamiltonian $H_{\text{KS}}(\mathbf{k}) + \Sigma(i\omega_n) - \Sigma_{\text{dc}}$. But the above equation is seldom used in practical calculations because there is a long ``tail'' $\propto \frac{1}{i\omega_n}$. If we want to use Eq.~(\ref{eq:nalpha}), we have to consider a large number of Matsubara frequency points to obtain accurate $N_{\alpha}$. This is rather inefficient. In order to overcome this problem, we adopt the following equation to evaluate $N_{\alpha}$:
\begin{equation}
\label{eq:nalpha2}
N_{\alpha} = 
\sum_{\mathbf{k}} f(\epsilon^{\infty}_{\alpha\mathbf{k}} - \mu) 
+ 2T\sum_{\mathbf{k},n}
\left[
\frac{1}{i\omega_n + \mu - \epsilon_{\alpha\mathbf{k}}} - 
\frac{1}{i\omega_n + \mu - \epsilon^{\infty}_{\alpha\mathbf{k}}}
\right].
\end{equation}
Here, $f(\epsilon)$ is the Fermi-Dirac distribution function, and $\epsilon^{\infty}_{\alpha\mathbf{k}}$ are actually the eigenvalues of the Hamiltonian $H_{\text{KS}}(\mathbf{k}) + \Sigma(i\infty) - \Sigma_{\text{dc}}$. The first term in the right hand side of Eq.~(\ref{eq:nalpha2}) is the contribution of the ``tail'' part, while the second term is from the contributions of the residual part~\cite{PhysRevB.81.195107}. 

\emph{Double counting term}. When combining DFT and DMFT, the electron-electron interactions are included in both the DFT part (through the exchange-correlation functional) and the DMFT part (through the local quantum impurity solver). To avoid double counting these interactions, a correction term must be subtracted from the self-energy function $\Sigma$. This correction term is what we call the double counting term $\Sigma_{\text{dc}}$. Now the exact expression for $\Sigma_{\text{dc}}$ is not known. In the DMFT engine, the following empirical formulas are supported to subtract the double counting term: (i) Fully localized limit (FLL) scheme~\cite{PhysRevB.77.155104}.
\begin{equation}
\label{eq:fll}
\Sigma_{\text{dc}} = U\left(N-\frac{1}{2}\right) - \frac{J}{2} (N - 1).
\end{equation}
Here $N$ means the total impurity occupancy. The FLL scheme is usually used in the calculations for Mott insulators and band insulators. (ii) Around mean-field (AMF) scheme~\cite{PhysRevB.77.155104}.
\begin{equation}
\Sigma_{\text{dc}} = \frac{UN}{2} + (U-J) \frac{Nl}{2l+1}. 
\end{equation}
Here $l$ denotes the quantum number of angular momentum of correlated orbitals. For $d$-electron systems, $l = 2$. The AMF scheme is suitable for strongly correlated metals. (iii) Held's scheme~\cite{held2007}. 
\begin{equation}
\Sigma_{\text{dc}} = \bar{U} \left(N - \frac{1}{2}\right),
\end{equation}
where the averaged interaction $\bar{U}$ reads
\begin{equation}
\bar{U} = \frac{U + (M - 1)(2U - 5J)}{2M - 1}.
\end{equation}
Here $M$ is the number of correlated orbitals.

\emph{Fermi level}. During the DFT+DMFT iterations, the Fermi level $\mu$ should be adjusted dynamically, such that the impurity occupancy is equal to the nominal one. For given $\Sigma(i\omega_n)$, the eigenvalues $\epsilon_{\alpha\mathbf{k}}$ and $\epsilon^{\infty}_{\alpha\mathbf{k}}$ are at first calculated. Then Eq.~(\ref{eq:nalpha2}) is used to evaluate the impurity occupancy for the current Fermi level. By using the classic bisection algorithm, it is easy to determine the desired Fermi level. 

\subsection{NORG impurity solver}

The NORG method is an innovative approach within the field of quantum many-body physics, designed to address the complex problem of interacting electron systems. Traditional methods, such as exact diagonalization (ED) and quantum Monte Carlo (QMC), face severe limitations due to exponential scaling in computational complexity and the notorious sign problem, respectively. The NORG method offers a non-perturbative alternative that is particularly adept at handling strong electron correlations across various coupling regimes.

The NORG method is grounded in the concept of natural orbitals, which are the eigenvectors of the single particle density matrix. These natural orbitals provide a basis for representing many-body wave functions in an optimized manner. The NORG method involves an iterative renormalization group procedure that constructs a structured subspace by projecting onto active natural orbitals. This allows for the accurate solution of quantum impurity problems with multiple impurities, which is beyond the scope of traditional numerical renormalization group methods that are limited to at most two impurities~\cite{PhysRevB.91.155140}. The core of the NORG algorithm involves several steps: Starting with an arbitrary but complete set of natural orbitals, selecting a subset to form a subspace, constructing an effective Hamiltonian, diagonalizing it to obtain the ground state, forming the single particle density matrix, and diagonalizing it to obtain a new set of natural orbitals. This process is iterated until convergence~\cite{PhysRevB.89.085108}.

The \texttt{Zen} toolkit includes a high-performance NORG impurity solver written in C++. The impurity solver introduces a ``shortcut'' trick that significantly improves the efficiency of the NORG algorithm by imposing restrictions on the orbital occupancy distribution, thereby reducing the Hilbert space dimension. This ``shortcut'' NORG method has been demonstrated to be dramatically faster than the general NORG method, offering a powerful tool for studying ground state and low-energy properties of quantum cluster-impurity models. Quite recently, this NORG impurity solver has been embedded into the \texttt{eDMFT} package~\cite{PhysRevB.81.195107} to study unconventional superconductivity and electron correlations in La$_{3}$Ni$_{2}$O$_{7}$~\cite{ouyangzhenfeng2024}. The technical details of \texttt{NORG} will be presented in a successive paper.

\subsection{CT-HYB impurity solver}

The impurity solvers based on the quantum Monte Carlo (QMC) algorithms exhibit several advantages. First of all, they are built on top of the imaginary time action, in which the infinite bath has been integrated out. Second, they can treat arbitrary couplings, and can thus be applied to all kinds of phases, including the metallic phase, insulating state, and phases with spontaneous symmetry breaking. Third, the QMC impurity solvers are numerically exact with a controllable numerical error. These are the reasons why the QMC impurity solvers are considered as the method of choice in the DMFT and DFT+DMFT calculations~\cite{RevModPhys.68.13,RevModPhys.78.865}. The CT-HYB impurity solver is an important variation of the continuous-time quantum Monte Carlo (CT-QMC) method~\cite{RevModPhys.83.349}. In this impurity solver, the partition function of the quantum impurity problem is diagrammatically expanded in the impurity-bath hybridization term. Then the diagrammatic expansion series is evaluated by the Metropolis Monte Carlo algorithm. The continuous-time nature of the algorithm means that the impurity operators can be placed at any arbitrary position on the imaginary time interval, so that time discretization errors can be completely avoided. Perhaps the CT-HYB is the most popular and powerful QMC impurity solver so far, since it can be used to solve multi-orbital impurity models with general interactions at low temperature~\cite{PhysRevLett.97.076405,PhysRevB.74.155107,PhysRevB.75.155113}.

In the \texttt{Zen} toolkit, the well established \texttt{iQIST} package is imported to contribute highly optimized CT-HYB impurity solvers~\cite{HUANG2015140,HUANG2017423}, which support both the segment representation~\cite{PhysRevLett.97.076405} and general matrix representation~\cite{PhysRevB.74.155107} algorithms. The former is suitable for the impurity models with density-density type interaction term. While the latter suits the impurity models with general interaction terms (such as rotationally invariant interaction with spin-orbit coupling term). The segment representation algorithm is extremely efficient. But the general matrix representation algorithm needs much more effort. In the \texttt{iQIST} package, the following strategies are adopted to accelerate the calculations. (i) The local Hamiltonian is partitioned into sub-blocks, which are labeled by using good quantum numbers~\cite{PhysRevB.75.155113}. A smart auto-partition algorithm suggested by P. Seth \emph{et al.} is implemented~\cite{SETH2016274}. Of course, the users can specify the partition scheme manually. (ii) The Hilbert space of the impurity problem can be truncated dynamically or by the nominal impurity occupancy. (iii) The lazy trace evaluation trick~\cite{PhysRevB.90.075149}, sparse matrix multiplication, and red-black tree algorithm~\cite{SETH2016274} are implemented to speed the computation of trace term in the Monte Carlo transition probability~\cite{PhysRevB.74.155107,PhysRevB.75.155113}. (iv) The CT-HYB impurity solvers are fully parallelized by using the message passing interface (MPI). In addition, the \texttt{iQIST} package can measure the single-particle Green's functions, two-particle Green's functions, and vertex functions. It supports the Legendre orthogonal polynomial representation~\cite{PhysRevB.84.075145} and intermediate representation~\cite{PhysRevB.96.035147} for Green's functions and improved estimator for self-energy functions~\cite{PhysRevB.85.205106}. These tricks can suppress the random noise effectively.  

Just like the other QMC impurity solvers, the CT-HYB impurity solver suffers from the fermionic sign problem. This problem becomes severe when the hybridization function is non-diagonal, or spin-orbit coupling is present, or the system's temperature is relatively low. Thus, the calculated results by the CT-HYB impurity solver become unreliable. At this time, we could turn to the NORG impurity solver. 

\subsection{Analytic continuation}

Often quantum impurity solvers working on the Matsubara axis are used within the DFT+DMFT framework. Their outputs are usually Matsubara Green's functions $G(i\omega_n)$ and self-energy functions $\Sigma(i\omega_n)$. In order to compare with the experimental results, they must be converted into real-frequency axis. Especially, to calculate the momentum-resolved spectral function $A(\mathbf{k},\omega)$, Fermi surface, and optical conductivity $\sigma(\omega)$, real-frequency self-energy function $\Sigma(\omega)$ is an essential input. Notice that $G(i\omega_n)$ and the spectral function $A(\omega)$ are related by the following Laplace transformation:
\begin{equation}
\label{eq:laplace}
G(i\omega_n) = \int^{\infty}_{-\infty} \frac{A(\omega)}{i\omega_n - \omega} d\omega.
\end{equation}
Given $G(i\omega_n)$, solving Eq.~(\ref{eq:laplace}) to extract $A(\omega)$ is the so-called analytic continuation problem~\cite{JARRELL1996133}. Once $A(\omega)$ is obtained, the retarded Green's function $G(\omega)$ can be easily evaluated via the Kramers-Kronig transformation. To extract $\Sigma(\omega)$ from $\Sigma(i\omega_n)$, we should at first subtract the Hartree-Fock term $\Sigma_{\text{HF}}$ from $\Sigma(i\omega_n)$:
\begin{equation}
\label{eq:hf}
\tilde{\Sigma}(i\omega_n) = \Sigma(i\omega_n) - \Sigma_{\text{HF}}.
\end{equation}
And then an auxiliary Green's function $G_{\text{aux}}(i\omega_n)$ is constructed:
\begin{equation}
\label{eq:gaux}
G_{\text{aux}}(i\omega_n) = \frac{1}{i\omega_n - \tilde{\Sigma}(i\omega_n)}.
\end{equation}
Next, we perform analytic continuation for $G_{\text{aux}}(i\omega_n)$ to get $G_{\text{aux}}(\omega)$. By inverting Eq.~(\ref{eq:hf}) and (\ref{eq:gaux}), we finally obtain $\Sigma(\omega)$. 

Until now, analytic continuation is still a challenging, yet to be solved problem. In the \texttt{Zen} toolkit, a full-fledged analytic continuation package, namely \texttt{ACFlow}~\cite{HUANG2023108863}, is included. It supports various analytic continuation methods, including the maximum entropy method (MaxEnt)~\cite{JARRELL1996133}, Nevanlinna analytical continuation (NAC)~\cite{PhysRevLett.126.056402}, barycentric rational function approximation (BarRat), stochastic analytic continuation (SAC)~\cite{PhysRevB.57.10287,SHAO20231}, stochastic optimization method (SOM)~\cite{PhysRevB.62.6317}, and stochastic pole expansion (SPX)~\cite{PhysRevB.108.235143,PhysRevD.109.054508}, etc. These methods have their own pros and cons. For examples, the MaxEnt method is quite efficient, but it tends to generate smooth and broad spectral function~\cite{JARRELL1996133}. The BarRat method is even more efficient than the MaxEnt method. It can resolve both sharp and broad spectral functions. But sometimes the sum-rules about the spectral functions might be violated. The SPX method employs the pole representation to approximate Matsubara Green's function and relies on a simulated annealing algorithm to figure out the optimal pole representation. It is somewhat time-consuming, but it can resolve fine features in the spectra~\cite{PhysRevB.108.235143,PhysRevD.109.054508}. The \texttt{ACFlow} package provides some diagnostic tools. With them, we can easily compare and crosscheck the analytic continuation results obtained by various methods. The \texttt{ACFlow} package supports parallel computing. Furthermore, it is developed with the Julia language, which makes it more interactive with the other components of the \texttt{Zen} toolkit.

\section{benchmarks\label{sec:benchmarks}}

In this section, we would like to benchmark the \texttt{Zen} toolkit. Here, we consider three typical examples, including the correlated metal SrVO$_{3}$, unconventional superconductor La$_{3}$Ni$_{2}$O$_{7}$, and Mott insulator MnO. The calculated results are compared with the experimental data and previous DFT+DMFT results, if available. 

\begin{table*}[ht]
    \caption{The DFT+DMFT computational parameters used in the present work. Here, $N_{\text{site}}$ means the number of inequivalent impurity atoms in the unit cell, $N_{\text{imp}}$ means the nominal impurity occupancy, and $E_{\text{cut}}$ is the cutoff energy for plane-wave expansion. \label{tab:params}}
    \begin{ruledtabular}
    \begin{tabular}{lllllllll}
    Case & Correlated orbitals & $N_{\text{site}}$ & $N_{\text{imp}}$ & $E_{\text{cut}}$ & $k$-mesh & $U$ & $J$ & Projection window \\
    \hline
    SrVO$_{3}$              & V-$3d$  & 1 & 1.0 & 400 eV& 9 $\times$ 9 $\times$ 9 & 4.0~eV & 0.7~eV & [-1.4~eV, 6.0~eV]\\
    La$_{3}$Ni$_{2}$O$_{7}$ & Ni-$e_g$ ($d_{z^2}$ + $d_{x^2-y^2}$) & 1 & 7.5 & 400 eV& 9 $\times$ 9 $\times$ 9 & 5.0~eV & 1.0~eV & [-8.0~eV, 4.0~eV]\\
    MnO                     & Mn-$3d$ & 1 & 5.0 & 400 eV& 13 $\times$ 13 $\times$ 13 & 8.0~eV & 0.5~eV & [-8.0~eV, 3.0~eV]\\
\end{tabular}
\end{ruledtabular}
\end{table*}

\begin{figure*}[ht]
\centering
\includegraphics[width=\textwidth]{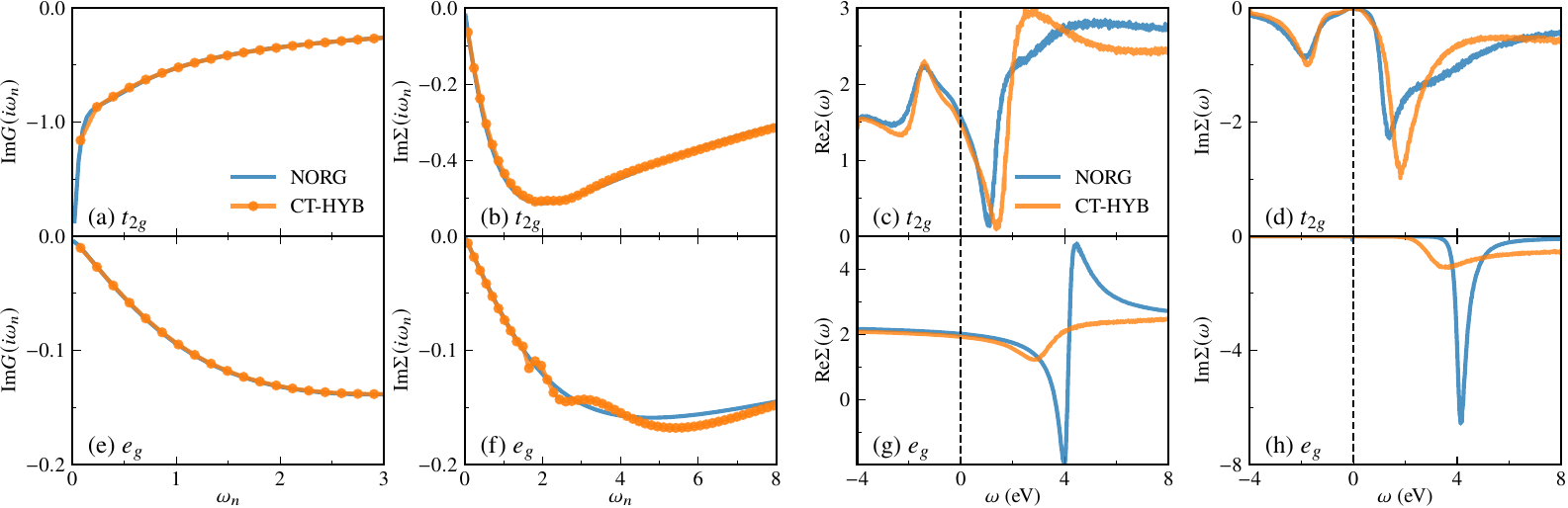}
\caption{Calculated Green's functions and self-energy functions for V's $t_{2g}$ and $e_{g}$ orbitals in SrVO$_{3}$. (a) and (e) $\text{Im}G(i\omega_n)$. (b) and (f) $\text{Im}\Sigma(i\omega_n)$. (c) and (g) $\text{Re}\Sigma(\omega)$. (d) and (h) $\text{Im}\Sigma(\omega)$. The upper panels are for the $t_{2g}$ orbitals, and the lower panels are for the $e_g$ orbitals. The vertical dashed lines denote the Fermi levels. \label{fig:svo_SE}}
\end{figure*}

\begin{figure}[ht]
\centering 
\includegraphics[width=\columnwidth]{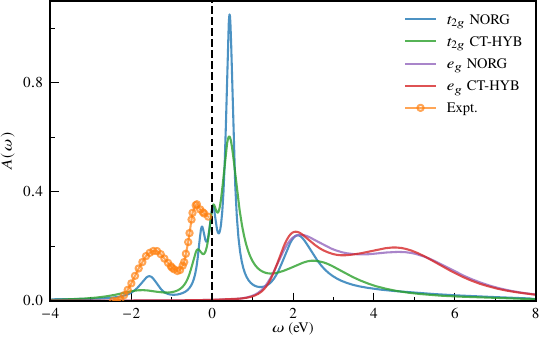}
\caption{Calculated spectral functions for V's $t_{2g}$ and $e_{g}$ orbitals in SrVO$_{3}$. They were extracted from Matsubara Green's functions $G(i\omega_n)$ by analytic continuation calculations. The experimental spectrum (empty circles) was taken from Ref.~[\onlinecite{YOSHIDA201611}]. The vertical dashed line denotes the Fermi level. \label{fig:svo_dos}}
\end{figure}

\begin{figure*}[ht]
\centering
\includegraphics[width=\textwidth]{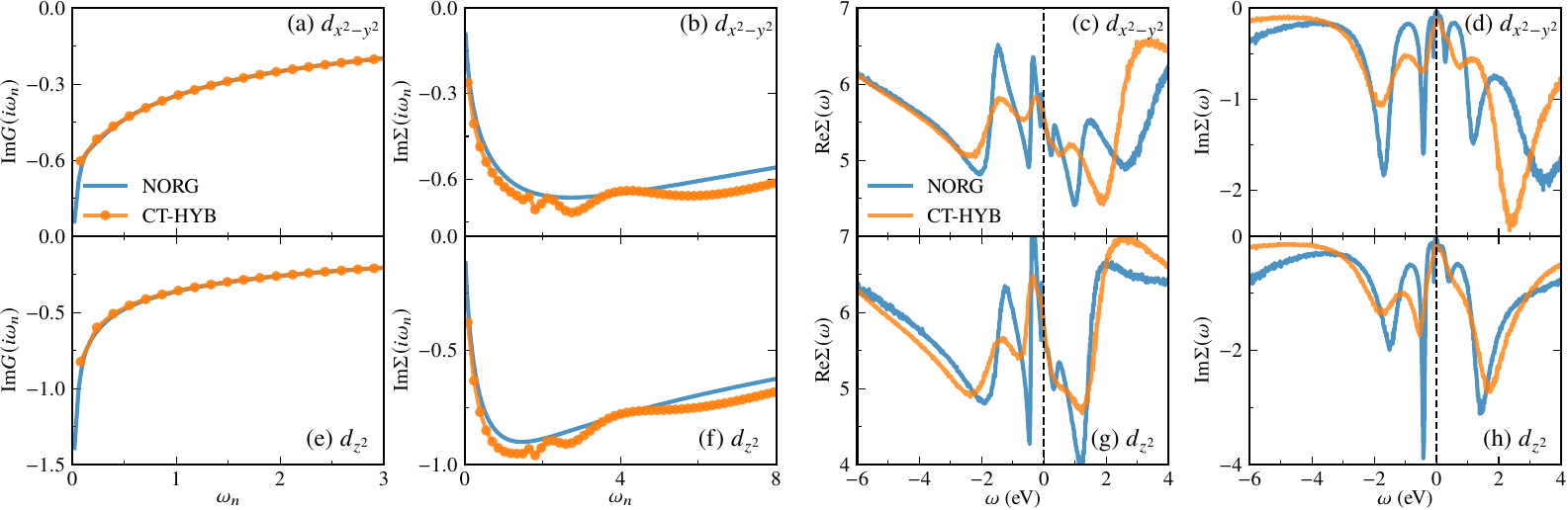}
\caption{Calculated Green's functions and self-energy functions for Ni's 3$d_{x^2-y^2}$ and 3$d_{z^2}$ orbitals in La$_{3}$Ni$_{2}$O$_{7}$. (a) and (e) $\text{Im}G(i\omega_n)$. (b) and (f) $\text{Im}\Sigma(i\omega_n)$. (c) and (g) $\text{Re}\Sigma(\omega)$. (d) and (h) $\text{Im}\Sigma(\omega)$. Panels (a)-(d) correspond to the Ni's 3$d_{x^2-y^2}$ orbitals, and panels (e)-(h) correspond to the Ni's 3$d_{z^2}$ orbitals. The vertical dashed lines denote the Fermi levels. \label{fig:lno_SE}}
\end{figure*}

\begin{figure}[ht]
\centering
\includegraphics[width=\columnwidth]{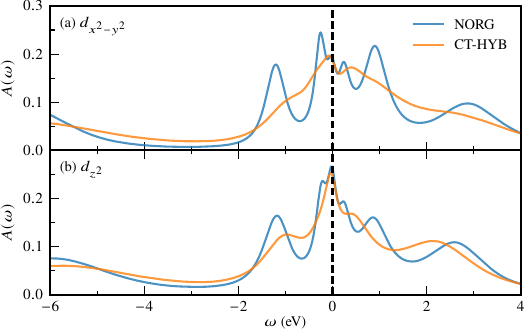}
\caption{Calculated spectral functions for Ni's $3d_{x^2-y^2}$ and $3d_{z^2}$ orbitals in La$_{3}$Ni$_{2}$O$_{7}$. The vertical dashed lines denote the Fermi levels.}
\label{fig:lno_dos}
\end{figure}

\begin{figure*}[ht]
\centering
\includegraphics[width=\textwidth]{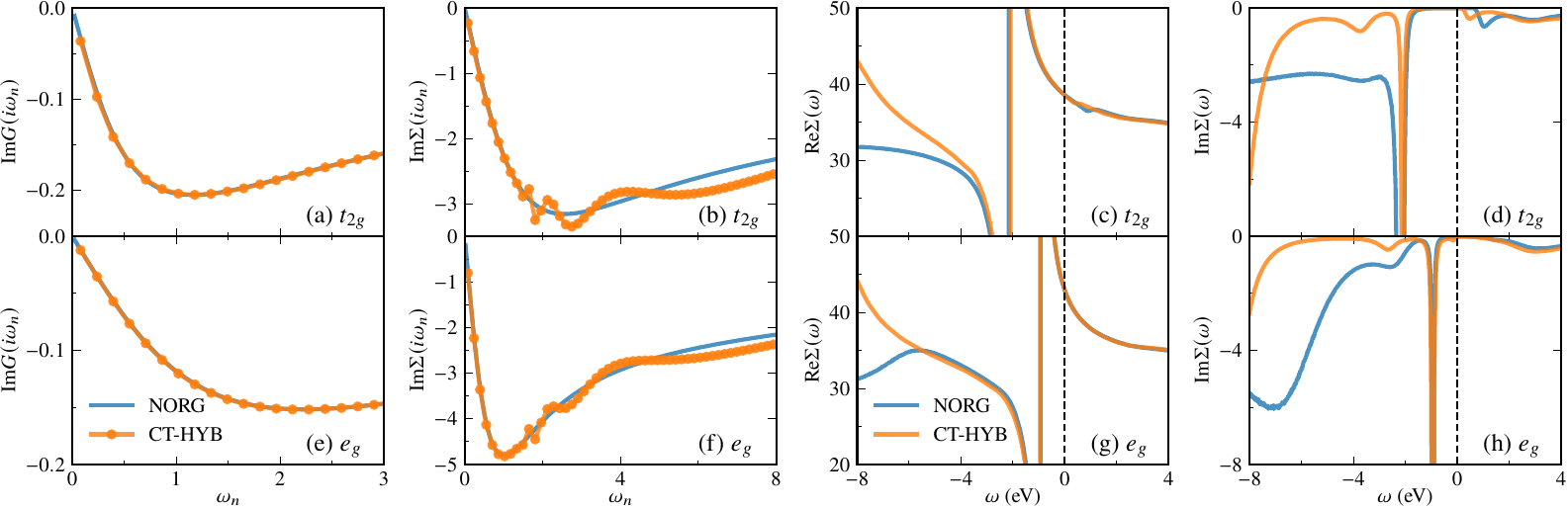}
\caption{Calculated Green's functions and self-energy functions for Mn-$3d$ orbitals in MnO at ambient pressure. (a) and (e) $\text{Im}G(i\omega_n)$. (b) and (f) $\text{Im}\Sigma(i\omega_n)$. (c) and (g) $\text{Re}\Sigma(\omega)$. (d) and (h) $\text{Im}\Sigma(\omega)$. Panels (a)-(d) correspond to the $t_{2g}$ orbitals, and panels (e)-(h) correspond to the $e_{g}$ orbitals. The vertical dashed lines denote the Fermi levels. \label{fig:mno_SE}}
\end{figure*}

\begin{figure}[ht]
\centering
\includegraphics[width=\columnwidth]{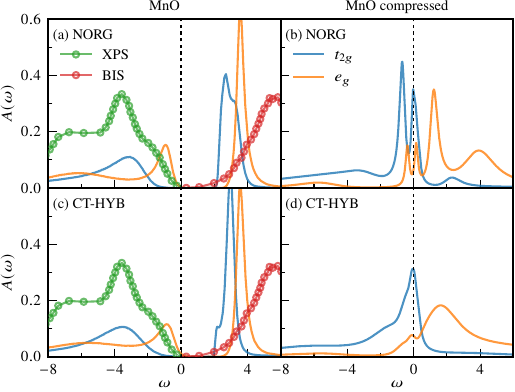}
\caption{Calculated spectral functions for Mn's $t_{2g}$ and $e_{g}$ orbitals in MnO at $V/V_0$ = 1.0 (left panels) and 0.53 (right panels). The experimental data was extracted from Ref.~[\onlinecite{PhysRevB.44.1530}]. The vertical dashed lines denote the Fermi level. \label{fig:mno_dos}}
\end{figure}

\subsection{General setup}

All the DFT+DMFT calculations were done by using the \texttt{Zen} toolkit. The relevant parameters are collected in Table~\ref{tab:params}.

For the DFT part, only the \texttt{VASP} interface was tested~\cite{PhysRevB.59.1758,PhysRevB.54.11169}. We chose the experimental crystal structures for the three compounds. Core and valence electrons were treated within the PAW formalism~\cite{PhysRevB.50.17953}. The generalized-gradient approximation (GGA) was used for the exchange-correlation functional within the Perdew-Burke-Ernzerhof (PBE) approach~\cite{PhysRevLett.77.3865}. The Brillouin zone was sampled using a $\Gamma$-centered Monkhorst-Pack $k$-point grid. For the sake of simplicity, the systems were restricted to being nonmagnetic. Once the DFT calculations were converged, the PLO scheme was applied to construct the projection matrices for V-$3d$, Ni-$3d$ and Mn-$3d$ orbitals. For the DMFT part, the double counting terms were built with the FLL scheme~\cite{PhysRevB.77.155104}. In Eq.~(\ref{eq:fll}), the impurity occupancy $N$ was fixed to the nominal one. The interaction parameters, including Coulomb repulsion interaction $U$ and Hund's exchange interaction $J$, were obtained from the references~\cite{PhysRevB.73.155112,Kunes2008,PhysRevB.109.115114}. For the quantum impurity solvers part, both the NORG and CT-HYB impurity solvers were used. Here, only the density-density type interactions were considered. In other words, the pair-hopping and spin-flip terms were ignored. The NORG impurity solver works at zero temperature, corresponding to $\beta=\infty$. We set an effective inverse temperature $\beta^{\rm eff} = 50 \pi$ eV$^{-1}$ for the effective Matsubara frequencies $\omega_n = (2n+1)\pi/\beta^{\rm eff} = 0.02(2n+1)$ eV in the calculation of Matsubara Green's functions. We found that four bath orbitals per impurity orbital are adequate to faithfully describe the non-interacting electron bath in the quantum impurity model derived from the DMFT ~\cite{Chen2407.13737}. So the number of bath orbitals per impurity orbital $n_{\rm b}$ was set to at least 4. A parallel Lanczos algorithm was employed to achieve high-performance computation in finding the ground state and Green's function. 64 CPU cores were utilized in the calculations. For the CT-HYB impurity solver, $\beta = 40.0$ $\rm{eV^{-1}}$, which corresponds to approximately 290~K. The Legendre orthogonal polynomial representation for the Green's function~\cite{PhysRevB.84.075145} and the improved estimator for self-energy function~\cite{PhysRevB.85.205106} were adopted to suppress the stochastic noises. The number of Monte Carlo sweeps was 10$^8$ per process, and 64~CPU cores were utilized in the calculations.

For SrVO$_{3}$ and La$_{3}$Ni$_{2}$O$_{7}$, we performed charge fully self-consistent DFT+DMFT calculations. For each DFT + DMFT iteration, 10 DFT internal cycles and a one-shot DMFT calculation were executed. In most cases, 60 DFT + DMFT iterations were adequate to obtain well converged charge density $\rho$ and total energy $E_{\text{tot}}$. The converged criteria for total energy was set to $10^{-6}$ eV, respectively. For MnO, we performed one-shot DFT+DMFT calculations. That is to say, the self-energy function, instead of total energy and charge density, was converged. The converged criterion for self-energy function was set to $10^{-4}$ eV. Usually 40 DMFT iterations were enough. Once the DFT+DMFT calculations were converged, we just used the \texttt{ACFlow} package~\cite{HUANG2023108863} to perform analytic continuation to extract the spectral functions $A(\omega)$ and real-frequency self-energy functions $\Sigma(\omega)$. The analytic continuation calculations were at first done by the SPX method~\cite{PhysRevB.108.235143} and then crosschecked by the MaxEnt method~\cite{JARRELL1996133}.
  
\subsection{Correlated metal SrVO$_{3}$}

SrVO$_{3}$ is a typical correlated metal. It crystallizes in a cubic perovskite structure. Its space group is ${Pm\overline{3}m}$. The experimental lattice parameter is $a = 3.8410\, \text{\AA}$~\cite{REY1990101}. The electronic structure of SrVO$_{3}$ is quite simple. Under cubic crystal fields, the five V-$3d$ orbitals should be split into triply degenerate $t_{2g}$ orbitals and double degenerate $e_{g}$ orbitals. The three $t_{2g}$ orbitals cross the Fermi level. They are well separated from the empty $e_{g}$ orbitals and the fully occupied O-$2p$ orbitals. Thus, a minimal model for SrVO$_{3}$ can include only the three $t_{2g}$ orbitals. This makes SrVO$_{3}$ an ideal system to examine various beyond-DFT methods~\cite{PhysRevB.77.205112,PhysRevB.72.155106,Huang_2012}. In the present work, both the $t_{2g}$ and $e_g$ orbitals were treated as correlated. The corresponding interaction parameters and projection window are summarized in Table~\ref{tab:params}. 

Figure~\ref{fig:svo_SE} shows the calculated Green's functions and self-energy functions. Let's focus on the Matsubara data at first. Though the calculations were done at different temperatures, overall the results obtained by the NORG and CT-HYB impurity solvers agree quite well with each other. The only exception lies in $\text{Im}\Sigma(i\omega_n)$ for $e_g$ orbitals as obtained by using the CT-HYB impurity solver. It exhibits obvious oscillations in the range of 1.5 $\sim$ 6 eV [see Fig.~\ref{fig:svo_SE}(f)]. We believe that such oscillations can be attributed to the improved estimator for self-energy function. In general, this estimator can suppress random noises, but sometimes it might induce some sorts of oscillations in the mid-frequency range~\cite{PhysRevB.85.205106,RevModPhys.83.349}. These oscillations are hardly eliminated through increasing the Monte Carlo samplings. We note that another approach to evaluate the self-energy function is through the Dyson equation, instead of Monte Carlo sampling. But this method is numerically unstable. It suffers more significant oscillations at high-frequency region. In addition, we can see that low-frequency part of $\text{Im}\Sigma(i\omega_n)$ for the $t_{2g}$ orbitals exhibits quasi-linear behavior, and the intercept at $i\omega_n \to 0$ is nearly zero, which is consistent with the Fermi liquid theory. 

Next, we turn to the real-frequency self-energy functions. By using the following equation:
\begin{equation}
{m^*}/{m} = {1}/{Z} = 1 - \left. \frac{\partial \text{Re} \Sigma(\omega)}{\partial \omega} \right|_{\omega=0},
\end{equation} 
we can easily calculate the electronic effective masses $m^{*}$ and quasiparticle weights $Z$ for correlated orbitals. We find that ${m^*/m}$ for the $t_{2g}$ orbitals is approximately 1.8, which is consistent with previous DFT+DMFT calculations~\cite{PhysRevB.72.155106,PhysRevB.77.205112}. Furthermore, $\text{Im}\Sigma(\omega = 0)$ for $t_{2g}$ orbitals are close to zero, which indicates that the low-energy electron scattering is trivial. 

Figure.~\ref{fig:svo_dos} shows the spectral functions for V-3$d$'s $t_{2g}$ and $e_{g}$ orbitals. The photoemission spectrum is shown in this figure for comparison~\cite{YOSHIDA201611}. For the $t_{2g}$ orbitals, their spectra exhibit two broad peaks around -1.8~eV and 2.0~eV. They are actually the lower and upper Hubbard bands. Near the Fermi level, besides the quasiparticle peak (at $\omega = 0$), two shoulder peaks appear at $\omega =$ -0.1~eV and 0.4~eV. These features are consistent with the experimental data~\cite{PhysRevLett.93.156402} and previous DFT+DMFT calculations~\cite{PhysRevB.77.205112,PhysRevB.72.155106,PhysRevB.73.155112}. Especially, the lower Hubbard band at $\omega = -1.8$ eV and the left shoulder peak at $\omega = -0.1$ eV are clearly seen in the photoemission spectrum~\cite{YOSHIDA201611}. For the $e_g$ orbitals, their spectra exhibit two broad peaks from 1.0~eV to 8.0~eV. They are totally unoccupied. Overall, the spectra obtained by the two impurity solvers are quite similar. But there are still small differences, because the two solvers work at different temperatures. The peaks in the vicinity of the Fermi level obtained by NORG are sharper and narrower than those obtained by CT-HYB, which implies that the quasiparticles are more coherent at lower temperatures.

\subsection{Unconventional superconductor La$_{3}$Ni$_{2}$O$_{7}$}

Superconductivity has been discovered in the high-pressure phase of La$_{3}$Ni$_{2}$O$_{7}$ with transition temperature up to 80~K~\cite{Sun2023}. Previous experiments and theoretical calculations suggest that the Ni-$3d$ orbitals are correlated~\cite{PhysRevB.108.125105,Yang2024,PhysRevB.109.L081105}. Furthermore, it is the Hund's mechanism that dominates the electronic correlations in La$_{3}$Ni$_{2}$O$_{7}$~\cite{PhysRevB.109.115114,PhysRevB.109.165140}. Around 14 GPa, this material undergoes a structural transition from ${Amam}$ phase to ${Fmmm}$ phase. We note that the five Ni-$3d$ orbitals are also split into three $t_{2g}$ orbitals and two $e_g$ orbitals in the $Fmmm$ phase. The $t_{2g}$ orbitals are fully occupied, while the $e_{g}$ orbitals ($d_{x^2-y^2}$ and $d_{z^2}$) are partially occupied. Here, we just focus on the $Fmmm$ phase of La$_{3}$Ni$_{2}$O$_{7}$, and only the two $e_{g}$ orbitals are explicitly considered in the quantum impurity model. 

Figure~\ref{fig:lno_SE} shows the calculated Green's functions and self-energy functions for Ni's $e_{g}$ orbitals. For $\text{Im}G(i\omega_n)$, both NORG and CT-HYB impurity solvers give consistent results. The functions are convex, implying metallic behaviors. For $\text{Im}\Sigma(i\omega_n)$ of the $d_{x^2-y^2}$ and $d_{z^2}$ orbitals, the results obtained by CT-HYB impurity solver exhibit oscillating behaviors again in the mid-frequency region. Just as discussed above, these oscillations are probably related to the improved estimator for self-energy function~\cite{PhysRevB.85.205106}. For $\text{Re}\Sigma(\omega)$, the quasilinear region is quite small near the Fermi level, the corresponding slope for $d_{z^2}$ is larger than that of $d_{x^2-y^2}$. It is suggested that this material is strongly correlated. And the $d_{z^2}$ orbital is more correlated than the $d_{x^2-y^2}$. For $\text{Im}\Sigma(\omega)$, they deviate from zero at $\omega = 0$, which indicates considerable low-energy electron scattering and violation of the Fermi liquid theory. We note that orbital differentiation and non-Fermi-liquid behavior are two key signatures of Hundness~\cite{PhysRevB.102.125130,annurev-conmatphys-020911-125045}. Thus, the $Fmmm$ phase of La$_{3}$Ni$_{2}$O$_{7}$ is likely a candidate of Hund metal~\cite{PhysRevB.109.L081105,PhysRevB.109.115114,PhysRevB.109.165140}.    

Next, we study the orbital-resolved spectral functions for the Ni's $3d_{x^2-y^2}$ and $3d_{z^2}$ orbitals. As shown in Fig.~\ref{fig:lno_dos}, the spectra obtained by the NORG and CT-HYB impurity solvers are roughly consistent with each other. The spectra from the NORG impurity solver exhibit more structures (i.e. shoulder peaks) near the Fermi level, while the spectra from the CT-HYB impurity solver are broader and smoother. This discrepancy can be ascribed to the finite temperature effect. We find that the quasiparticle peak of the $d_{z^2}$ orbital is narrower than that of the $d_{x^2-y^2}$ orbital. Thus, it is suggested that the $d_{z^2}$ orbital is more renormalized than the $d_{x^2-y^2}$ orbital, which agrees with previous DFT+DMFT calculations using the \texttt{eDMFT} code~\cite{PhysRevB.109.115114}.

\subsection{Mott insulator MnO}

\begin{figure}[ht]
\centering
\includegraphics[width=\columnwidth]{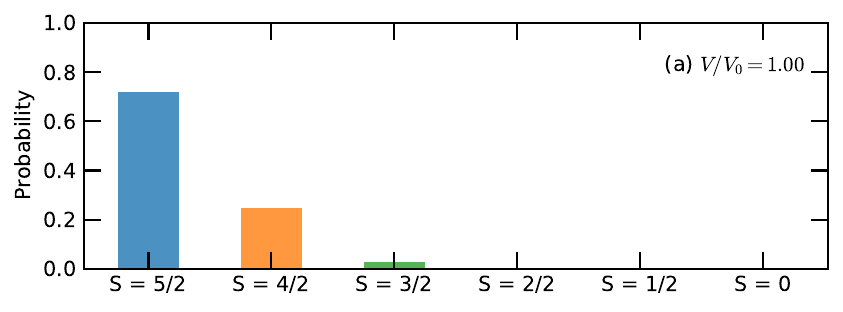}
\includegraphics[width=\columnwidth]{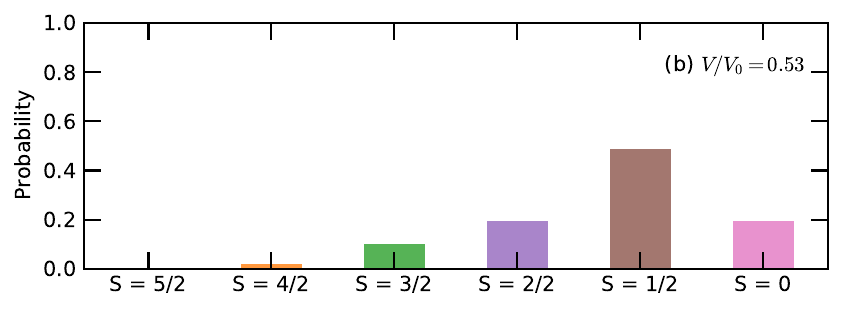}
\caption{Probabilities of spin states in MnO at ambient pressure ($V/V_0 = 1.0$) and high pressure ($V/V_0 = 0.53$). The results are obtained by the CT-HYB impurity solver. \label{fig:mno_prob}}
\end{figure}

The third example is MnO, which is a typical Mott insulator. It crystallizes in the rock-salt structure (the space group is ${Fm\overline{3}m}$). The lattice parameter $a = 4.4457\, \text{\AA}$. At ambient pressure, its band gap is about 2.0~eV. Around 90 $\sim$ 105~GPa, MnO is expected to undergo an insulator-to-metal transition. The corresponding volume collapse, represented as $V/V_0$, is approximately 0.68 $\sim$ 0.63~\cite{noguchi1996shock,mita2001optical,PhysRevB.71.100101}. Previous DFT+DMFT calculations have suggested that the insulator-to-metal transition is orbital-selective, with a simultaneous spin state transition occurring~\cite{Kunes2008}. In the present work, we performed one-shot DFT+DMFT calculations for $V/V_0 = 0.53$ and 1.0 to examine whether the \texttt{Zen} toolkit can successfully reproduce the metallic and insulating phases of MnO.

Figure~\ref{fig:mno_SE} presents the calculated Green's functions and self-energy functions for the Mn's $t_{2g}$ and $e_g$ orbitals. The curves for $\text{Im}G(i\omega_n)$ are concave, indicating insulating characteristics. The CT-HYB impurity solver exhibits significant fluctuations in the mid-frequency region of $\text{Im}\Sigma(i\omega_n)$. However, in the low-frequency range, the results obtained from the NORG and CT-HYB impurity solvers show good agreement. Both $\text{Re}\Sigma(\omega)$ and $\text{Im}\Sigma(\omega)$ display sharp and divergent features between -4.0~eV and 0.0~eV, which are characteristic of a correlated insulator. Notably, the results from the NORG impurity solver closely align with those from the CT-HYB impurity solver in the range of -2.0~eV to 2.0~eV. It appears that the numerical fluctuations in $\text{Im}\Sigma(i\omega_n)$ do not significantly impact the analytically continued $\Sigma(\omega)$.

The spectral functions are illustrated in Fig.~\ref{fig:mno_dos}. At ambient pressure, the calculated spectra exhibit a significant gap (approximately 2.0 eV), which is consistent with the experimental results~\cite{PhysRevB.44.1530}. When the volume is substantially decreased ($V/V_0 = 0.53$), a quasiparticle peak emerges at the Fermi level, indicating that the material transitions into a metallic state. In fact, the Mott insulator-metal transition occurs at larger volumes or lower pressures. Due to the considerable differences in the bandwidths of the $t_{2g}$ and $e_g$ orbitals, the Mott transitions in these orbitals do not occur simultaneously. There exists a significant range of volume (or pressure) in which the $t_{2g}$ orbitals are metallic while the $e_g$ orbitals remain in insulating states. This phenomenon is referred to as the orbital-selective Mott phase~\cite{Kunes2008}. Such behavior is a common characteristic of transition metal monoxides~\cite{PhysRevB.85.245110,PhysRevB.82.195101}. From Figs.~\ref{fig:mno_dos} (b) and (d), we observe that both $t_{2g}$ and $e_g$ orbitals contribute to the quasiparticle peak. Furthermore, by utilizing the NORG impurity solver, additional features can be resolved.

Jan Kune\v{s} \emph{et al.} pointed out that there is collapse of magnetic moment during the pressure-driven Mott transition in MnO~\cite{Kunes2008}. To verify this viewpoint, we studied the spin states and orbital occupancies for Mn's $t_{2g}$ and $e_{g}$ orbitals. Figure~\ref{fig:mno_prob} shows the calculated probabilities of the possible spin states. The magnetic moment $M$ can be calculated as follows:
\begin{equation}
M = \sum_i P_i S_i,
\end{equation}
where $P_i$ is the probability of the $i$-th spin state and $S_i$ is the corresponding total spin. At ambient pressure, the $S=5/2$ state is dominant, and the magnetic moment is about 4.7~$\mu_B$. Now the system is in high-spin state, the major electronic configuration is $t^3_{2g}e^2_{g}$. Under high pressure, the contributions from the $S = 5/2$ state can be ignored. It is the $S = 1/2$ state that makes the predominant contribution. The related electronic configuration becomes $t^{5}_{2g}e^{0}_{g}$. The total magnetic moment $M$ is reduced to 1.3~$\mu_B$. Clearly, there is a pressure-driven high-spin state ($S = 5/2$) to low-spin state ($S= 1/2$) transition in MnO. We confirm previously calculated results again~\cite{Kunes2008}.

\section{concluding remarks\label{sec:summary}}

In this paper, we present a new \emph{ab initio} many-body computational toolkit, \texttt{Zen}, which enables fully self-consistent DFT+DMFT calculations for correlated $d$-electron materials. This toolkit is now interfaced with the \texttt{VASP} code via the PLO scheme and with the \texttt{Quantum ESPRESSO} code via the MLWF scheme. It incorporates two highly efficient impurity solvers: NORG and CT-HYB. What distinguishes \texttt{Zen} from other existing DFT+DMFT packages is the NORG impurity solver, which can handle multi-orbital quantum impurity models with general interactions. It operates at zero temperature and is free from the fermionic sign problem, making it a promising alternative and complement to the CT-HYB impurity solver.

We selected three strongly correlated materials—SrVO$_{3}$, La$_{3}$Ni$_{2}$O$_{7}$, and MnO—to benchmark the \texttt{Zen} toolkit. We conducted systematic DFT+DMFT calculations to investigate the electronic structures of these materials. Our calculated results align closely with experimental data and previous DFT+DMFT calculations, where available. Notably, the results obtained using the NORG impurity solver are consistent with those derived from the CT-HYB impurity solver. Minor discrepancies can be attributed to temperature effects. These benchmarks demonstrate that \texttt{Zen} is a reliable tool for studying strongly correlated materials. In the future, we aim to extend its capabilities to include additional features, such as lattice dynamics calculations~\cite{PhysRevB.102.245104,PhysRevB.102.241108} and the exact double counting term~\cite{PhysRevLett.115.196403}.

\texttt{Zen} is an open-source toolkit. Its source code can be downloaded from GitHub~\cite{github}. We have an open-source package for the NORG impurity solver, which can be downloaded from GitHub~\cite{norg_github}.

\begin{acknowledgments}
This work was supported by National Natural Science Foundation of China (Grants No.~11934020 and No.~12274380) and the Innovation Program for Quantum Science and Technology (Grants No. 2021ZD0302402). Computational resources were provided by Physical Laboratory of High Performance Computing in Renmin University of China.
\end{acknowledgments}

\bibliography{reference}

\begin{thebibliography}{108}%
\makeatletter
\providecommand \@ifxundefined [1]{%
 \@ifx{#1\undefined}
}%
\providecommand \@ifnum [1]{%
 \ifnum #1\expandafter \@firstoftwo
 \else \expandafter \@secondoftwo
 \fi
}%
\providecommand \@ifx [1]{%
 \ifx #1\expandafter \@firstoftwo
 \else \expandafter \@secondoftwo
 \fi
}%
\providecommand \natexlab [1]{#1}%
\providecommand \enquote  [1]{``#1''}%
\providecommand \bibnamefont  [1]{#1}%
\providecommand \bibfnamefont [1]{#1}%
\providecommand \citenamefont [1]{#1}%
\providecommand \href@noop [0]{\@secondoftwo}%
\providecommand \href [0]{\begingroup \@sanitize@url \@href}%
\providecommand \@href[1]{\@@startlink{#1}\@@href}%
\providecommand \@@href[1]{\endgroup#1\@@endlink}%
\providecommand \@sanitize@url [0]{\catcode `\\12\catcode `\$12\catcode
  `\&12\catcode `\#12\catcode `\^12\catcode `\_12\catcode `\%12\relax}%
\providecommand \@@startlink[1]{}%
\providecommand \@@endlink[0]{}%
\providecommand \url  [0]{\begingroup\@sanitize@url \@url }%
\providecommand \@url [1]{\endgroup\@href {#1}{\urlprefix }}%
\providecommand \urlprefix  [0]{URL }%
\providecommand \Eprint [0]{\href }%
\providecommand \doibase [0]{https://doi.org/}%
\providecommand \selectlanguage [0]{\@gobble}%
\providecommand \bibinfo  [0]{\@secondoftwo}%
\providecommand \bibfield  [0]{\@secondoftwo}%
\providecommand \translation [1]{[#1]}%
\providecommand \BibitemOpen [0]{}%
\providecommand \bibitemStop [0]{}%
\providecommand \bibitemNoStop [0]{.\EOS\space}%
\providecommand \EOS [0]{\spacefactor3000\relax}%
\providecommand \BibitemShut  [1]{\csname bibitem#1\endcsname}%
\let\auto@bib@innerbib\@empty
\bibitem [{\citenamefont {Morosan}\ \emph {et~al.}(2012)\citenamefont
  {Morosan}, \citenamefont {Natelson}, \citenamefont {Nevidomskyy},\ and\
  \citenamefont {Si}}]{adma.201202018}%
  \BibitemOpen
  \bibfield  {author} {\bibinfo {author} {\bibfnamefont {E.}~\bibnamefont
  {Morosan}}, \bibinfo {author} {\bibfnamefont {D.}~\bibnamefont {Natelson}},
  \bibinfo {author} {\bibfnamefont {A.~H.}\ \bibnamefont {Nevidomskyy}},\ and\
  \bibinfo {author} {\bibfnamefont {Q.}~\bibnamefont {Si}},\ }\bibfield
  {title} {\bibinfo {title} {{Strongly Correlated Materials}},\ }\href
  {https://doi.org/https://doi.org/10.1002/adma.201202018} {\bibfield
  {journal} {\bibinfo  {journal} {Adv. Mater.}\ }\textbf {\bibinfo {volume}
  {24}},\ \bibinfo {pages} {4896} (\bibinfo {year} {2012})}\BibitemShut
  {NoStop}%
\bibitem [{\citenamefont {Imada}\ \emph {et~al.}(1998)\citenamefont {Imada},
  \citenamefont {Fujimori},\ and\ \citenamefont {Tokura}}]{RevModPhys.70.1039}%
  \BibitemOpen
  \bibfield  {author} {\bibinfo {author} {\bibfnamefont {M.}~\bibnamefont
  {Imada}}, \bibinfo {author} {\bibfnamefont {A.}~\bibnamefont {Fujimori}},\
  and\ \bibinfo {author} {\bibfnamefont {Y.}~\bibnamefont {Tokura}},\
  }\bibfield  {title} {\bibinfo {title} {{Metal-insulator transitions}},\
  }\href {https://doi.org/10.1103/RevModPhys.70.1039} {\bibfield  {journal}
  {\bibinfo  {journal} {Rev. Mod. Phys.}\ }\textbf {\bibinfo {volume} {70}},\
  \bibinfo {pages} {1039} (\bibinfo {year} {1998})}\BibitemShut {NoStop}%
\bibitem [{\citenamefont {Lee}\ \emph {et~al.}(2006)\citenamefont {Lee},
  \citenamefont {Nagaosa},\ and\ \citenamefont {Wen}}]{RevModPhys.78.17}%
  \BibitemOpen
  \bibfield  {author} {\bibinfo {author} {\bibfnamefont {P.~A.}\ \bibnamefont
  {Lee}}, \bibinfo {author} {\bibfnamefont {N.}~\bibnamefont {Nagaosa}},\ and\
  \bibinfo {author} {\bibfnamefont {X.-G.}\ \bibnamefont {Wen}},\ }\bibfield
  {title} {\bibinfo {title} {{Doping a Mott insulator: Physics of
  high-temperature superconductivity}},\ }\href
  {https://doi.org/10.1103/RevModPhys.78.17} {\bibfield  {journal} {\bibinfo
  {journal} {Rev. Mod. Phys.}\ }\textbf {\bibinfo {volume} {78}},\ \bibinfo
  {pages} {17} (\bibinfo {year} {2006})}\BibitemShut {NoStop}%
\bibitem [{\citenamefont {Dai}(2015)}]{RevModPhys.87.855}%
  \BibitemOpen
  \bibfield  {author} {\bibinfo {author} {\bibfnamefont {P.}~\bibnamefont
  {Dai}},\ }\bibfield  {title} {\bibinfo {title} {{Antiferromagnetic order and
  spin dynamics in iron-based superconductors}},\ }\href
  {https://doi.org/10.1103/RevModPhys.87.855} {\bibfield  {journal} {\bibinfo
  {journal} {Rev. Mod. Phys.}\ }\textbf {\bibinfo {volume} {87}},\ \bibinfo
  {pages} {855} (\bibinfo {year} {2015})}\BibitemShut {NoStop}%
\bibitem [{\citenamefont {Weng}\ \emph {et~al.}(2016)\citenamefont {Weng},
  \citenamefont {Smidman}, \citenamefont {Jiao}, \citenamefont {Lu},\ and\
  \citenamefont {Yuan}}]{Weng_2016}%
  \BibitemOpen
  \bibfield  {author} {\bibinfo {author} {\bibfnamefont {Z.~F.}\ \bibnamefont
  {Weng}}, \bibinfo {author} {\bibfnamefont {M.}~\bibnamefont {Smidman}},
  \bibinfo {author} {\bibfnamefont {L.}~\bibnamefont {Jiao}}, \bibinfo {author}
  {\bibfnamefont {X.}~\bibnamefont {Lu}},\ and\ \bibinfo {author}
  {\bibfnamefont {H.~Q.}\ \bibnamefont {Yuan}},\ }\bibfield  {title} {\bibinfo
  {title} {{Multiple quantum phase transitions and superconductivity in
  Ce-based heavy fermions}},\ }\href
  {https://doi.org/10.1088/0034-4885/79/9/094503} {\bibfield  {journal}
  {\bibinfo  {journal} {Rep. Prog. Phys.}\ }\textbf {\bibinfo {volume} {79}},\
  \bibinfo {pages} {094503} (\bibinfo {year} {2016})}\BibitemShut {NoStop}%
\bibitem [{\citenamefont {Bauer}\ and\ \citenamefont
  {Thompson}(2015)}]{annurev-conmatphys-031214-014508}%
  \BibitemOpen
  \bibfield  {author} {\bibinfo {author} {\bibfnamefont {E.}~\bibnamefont
  {Bauer}}\ and\ \bibinfo {author} {\bibfnamefont {J.}~\bibnamefont
  {Thompson}},\ }\bibfield  {title} {\bibinfo {title} {{Plutonium-Based
  Heavy-Fermion Systems}},\ }\href
  {https://doi.org/https://doi.org/10.1146/annurev-conmatphys-031214-014508}
  {\bibfield  {journal} {\bibinfo  {journal} {Annu. Rev. Condens. Matter
  Phys.}\ }\textbf {\bibinfo {volume} {6}},\ \bibinfo {pages} {137} (\bibinfo
  {year} {2015})}\BibitemShut {NoStop}%
\bibitem [{\citenamefont {Anisimov}\ \emph {et~al.}(1991)\citenamefont
  {Anisimov}, \citenamefont {Zaanen},\ and\ \citenamefont
  {Andersen}}]{PhysRevB.44.943}%
  \BibitemOpen
  \bibfield  {author} {\bibinfo {author} {\bibfnamefont {V.~I.}\ \bibnamefont
  {Anisimov}}, \bibinfo {author} {\bibfnamefont {J.}~\bibnamefont {Zaanen}},\
  and\ \bibinfo {author} {\bibfnamefont {O.~K.}\ \bibnamefont {Andersen}},\
  }\bibfield  {title} {\bibinfo {title} {{Band theory and Mott insulators:
  Hubbard U instead of Stoner I}},\ }\href
  {https://doi.org/10.1103/PhysRevB.44.943} {\bibfield  {journal} {\bibinfo
  {journal} {Phys. Rev. B}\ }\textbf {\bibinfo {volume} {44}},\ \bibinfo
  {pages} {943} (\bibinfo {year} {1991})}\BibitemShut {NoStop}%
\bibitem [{\citenamefont {Stewart}(2001)}]{RevModPhys.73.797}%
  \BibitemOpen
  \bibfield  {author} {\bibinfo {author} {\bibfnamefont {G.~R.}\ \bibnamefont
  {Stewart}},\ }\bibfield  {title} {\bibinfo {title} {{Non-Fermi-liquid
  behavior in $d$- and $f$-electron metals}},\ }\href
  {https://doi.org/10.1103/RevModPhys.73.797} {\bibfield  {journal} {\bibinfo
  {journal} {Rev. Mod. Phys.}\ }\textbf {\bibinfo {volume} {73}},\ \bibinfo
  {pages} {797} (\bibinfo {year} {2001})}\BibitemShut {NoStop}%
\bibitem [{\citenamefont {Kent}\ and\ \citenamefont
  {Kotliar}(2018)}]{science.aat5975}%
  \BibitemOpen
  \bibfield  {author} {\bibinfo {author} {\bibfnamefont {P.~R.~C.}\
  \bibnamefont {Kent}}\ and\ \bibinfo {author} {\bibfnamefont {G.}~\bibnamefont
  {Kotliar}},\ }\bibfield  {title} {\bibinfo {title} {{Toward a predictive
  theory of correlated materials}},\ }\href
  {https://doi.org/10.1126/science.aat5975} {\bibfield  {journal} {\bibinfo
  {journal} {Science}\ }\textbf {\bibinfo {volume} {361}},\ \bibinfo {pages}
  {348} (\bibinfo {year} {2018})}\BibitemShut {NoStop}%
\bibitem [{\citenamefont {Vollhardt}(2012)}]{andp.201100250}%
  \BibitemOpen
  \bibfield  {author} {\bibinfo {author} {\bibfnamefont {D.}~\bibnamefont
  {Vollhardt}},\ }\bibfield  {title} {\bibinfo {title} {{Dynamical mean-field
  theory for correlated electrons}},\ }\href
  {https://doi.org/https://doi.org/10.1002/andp.201100250} {\bibfield
  {journal} {\bibinfo  {journal} {Ann. Phys.}\ }\textbf {\bibinfo {volume}
  {524}},\ \bibinfo {pages} {1} (\bibinfo {year} {2012})}\BibitemShut {NoStop}%
\bibitem [{\citenamefont {Georges}\ \emph {et~al.}(1996)\citenamefont
  {Georges}, \citenamefont {Kotliar}, \citenamefont {Krauth},\ and\
  \citenamefont {Rozenberg}}]{RevModPhys.68.13}%
  \BibitemOpen
  \bibfield  {author} {\bibinfo {author} {\bibfnamefont {A.}~\bibnamefont
  {Georges}}, \bibinfo {author} {\bibfnamefont {G.}~\bibnamefont {Kotliar}},
  \bibinfo {author} {\bibfnamefont {W.}~\bibnamefont {Krauth}},\ and\ \bibinfo
  {author} {\bibfnamefont {M.~J.}\ \bibnamefont {Rozenberg}},\ }\bibfield
  {title} {\bibinfo {title} {{Dynamical mean-field theory of strongly
  correlated fermion systems and the limit of infinite dimensions}},\ }\href
  {https://doi.org/10.1103/RevModPhys.68.13} {\bibfield  {journal} {\bibinfo
  {journal} {Rev. Mod. Phys.}\ }\textbf {\bibinfo {volume} {68}},\ \bibinfo
  {pages} {13} (\bibinfo {year} {1996})}\BibitemShut {NoStop}%
\bibitem [{\citenamefont {Song}\ \emph {et~al.}(2008)\citenamefont {Song},
  \citenamefont {Wortis},\ and\ \citenamefont {Atkinson}}]{PhysRevB.77.054202}%
  \BibitemOpen
  \bibfield  {author} {\bibinfo {author} {\bibfnamefont {Y.}~\bibnamefont
  {Song}}, \bibinfo {author} {\bibfnamefont {R.}~\bibnamefont {Wortis}},\ and\
  \bibinfo {author} {\bibfnamefont {W.~A.}\ \bibnamefont {Atkinson}},\
  }\bibfield  {title} {\bibinfo {title} {{Dynamical mean field study of the
  two-dimensional disordered Hubbard model}},\ }\href
  {https://doi.org/10.1103/PhysRevB.77.054202} {\bibfield  {journal} {\bibinfo
  {journal} {Phys. Rev. B}\ }\textbf {\bibinfo {volume} {77}},\ \bibinfo
  {pages} {054202} (\bibinfo {year} {2008})}\BibitemShut {NoStop}%
\bibitem [{\citenamefont {Miranda}\ and\ \citenamefont
  {Dobrosavljević}(2012)}]{acprof:oso}%
  \BibitemOpen
  \bibfield  {author} {\bibinfo {author} {\bibfnamefont {E.}~\bibnamefont
  {Miranda}}\ and\ \bibinfo {author} {\bibfnamefont {V.}~\bibnamefont
  {Dobrosavljević}},\ }\bibfield  {title} {\bibinfo {title} {{Dynamical
  Mean-field Theories of Correlation and Disorder}},\ }in\ \href
  {https://doi.org/10.1093/acprof:oso/9780199592593.003.0006} {\emph {\bibinfo
  {booktitle} {{Conductor-Insulator Quantum Phase Transitions}}}}\ (\bibinfo
  {publisher} {Oxford University Press},\ \bibinfo {year} {2012})\BibitemShut
  {NoStop}%
\bibitem [{\citenamefont {Aoki}\ \emph {et~al.}(2014)\citenamefont {Aoki},
  \citenamefont {Tsuji}, \citenamefont {Eckstein}, \citenamefont {Kollar},
  \citenamefont {Oka},\ and\ \citenamefont {Werner}}]{RevModPhys.86.779}%
  \BibitemOpen
  \bibfield  {author} {\bibinfo {author} {\bibfnamefont {H.}~\bibnamefont
  {Aoki}}, \bibinfo {author} {\bibfnamefont {N.}~\bibnamefont {Tsuji}},
  \bibinfo {author} {\bibfnamefont {M.}~\bibnamefont {Eckstein}}, \bibinfo
  {author} {\bibfnamefont {M.}~\bibnamefont {Kollar}}, \bibinfo {author}
  {\bibfnamefont {T.}~\bibnamefont {Oka}},\ and\ \bibinfo {author}
  {\bibfnamefont {P.}~\bibnamefont {Werner}},\ }\bibfield  {title} {\bibinfo
  {title} {{Nonequilibrium dynamical mean-field theory and its applications}},\
  }\href {https://doi.org/10.1103/RevModPhys.86.779} {\bibfield  {journal}
  {\bibinfo  {journal} {Rev. Mod. Phys.}\ }\textbf {\bibinfo {volume} {86}},\
  \bibinfo {pages} {779} (\bibinfo {year} {2014})}\BibitemShut {NoStop}%
\bibitem [{\citenamefont {Held}(2007)}]{held2007}%
  \BibitemOpen
  \bibfield  {author} {\bibinfo {author} {\bibfnamefont {K.}~\bibnamefont
  {Held}},\ }\bibfield  {title} {\bibinfo {title} {{Electronic structure
  calculations using dynamical mean field theory}},\ }\href
  {https://doi.org/10.1080/00018730701619647} {\bibfield  {journal} {\bibinfo
  {journal} {Adv. Phys.}\ }\textbf {\bibinfo {volume} {56}},\ \bibinfo {pages}
  {829} (\bibinfo {year} {2007})}\BibitemShut {NoStop}%
\bibitem [{\citenamefont {Kotliar}\ \emph {et~al.}(2006)\citenamefont
  {Kotliar}, \citenamefont {Savrasov}, \citenamefont {Haule}, \citenamefont
  {Oudovenko}, \citenamefont {Parcollet},\ and\ \citenamefont
  {Marianetti}}]{RevModPhys.78.865}%
  \BibitemOpen
  \bibfield  {author} {\bibinfo {author} {\bibfnamefont {G.}~\bibnamefont
  {Kotliar}}, \bibinfo {author} {\bibfnamefont {S.~Y.}\ \bibnamefont
  {Savrasov}}, \bibinfo {author} {\bibfnamefont {K.}~\bibnamefont {Haule}},
  \bibinfo {author} {\bibfnamefont {V.~S.}\ \bibnamefont {Oudovenko}}, \bibinfo
  {author} {\bibfnamefont {O.}~\bibnamefont {Parcollet}},\ and\ \bibinfo
  {author} {\bibfnamefont {C.~A.}\ \bibnamefont {Marianetti}},\ }\bibfield
  {title} {\bibinfo {title} {{Electronic structure calculations with dynamical
  mean-field theory}},\ }\href {https://doi.org/10.1103/RevModPhys.78.865}
  {\bibfield  {journal} {\bibinfo  {journal} {Rev. Mod. Phys.}\ }\textbf
  {\bibinfo {volume} {78}},\ \bibinfo {pages} {865} (\bibinfo {year}
  {2006})}\BibitemShut {NoStop}%
\bibitem [{\citenamefont {Anisimov}\ \emph {et~al.}(1997)\citenamefont
  {Anisimov}, \citenamefont {Poteryaev}, \citenamefont {Korotin}, \citenamefont
  {Anokhin},\ and\ \citenamefont {Kotliar}}]{Anisimov_1997}%
  \BibitemOpen
  \bibfield  {author} {\bibinfo {author} {\bibfnamefont {V.~I.}\ \bibnamefont
  {Anisimov}}, \bibinfo {author} {\bibfnamefont {A.~I.}\ \bibnamefont
  {Poteryaev}}, \bibinfo {author} {\bibfnamefont {M.~A.}\ \bibnamefont
  {Korotin}}, \bibinfo {author} {\bibfnamefont {A.~O.}\ \bibnamefont
  {Anokhin}},\ and\ \bibinfo {author} {\bibfnamefont {G.}~\bibnamefont
  {Kotliar}},\ }\bibfield  {title} {\bibinfo {title} {{First-principles
  calculations of the electronic structure and spectra of strongly correlated
  systems: dynamical mean-field theory}},\ }\href
  {https://doi.org/10.1088/0953-8984/9/35/010} {\bibfield  {journal} {\bibinfo
  {journal} {J. Phys.: Condens. Matter}\ }\textbf {\bibinfo {volume} {9}},\
  \bibinfo {pages} {7359} (\bibinfo {year} {1997})}\BibitemShut {NoStop}%
\bibitem [{\citenamefont {Shilenko}\ and\ \citenamefont
  {Leonov}(2023)}]{PhysRevB.108.125105}%
  \BibitemOpen
  \bibfield  {author} {\bibinfo {author} {\bibfnamefont {D.~A.}\ \bibnamefont
  {Shilenko}}\ and\ \bibinfo {author} {\bibfnamefont {I.~V.}\ \bibnamefont
  {Leonov}},\ }\bibfield  {title} {\bibinfo {title} {{Correlated electronic
  structure, orbital-selective behavior, and magnetic correlations in
  double-layer ${\mathrm{La}}_{3}{\mathrm{Ni}}_{2}{\mathrm{O}}_{7}$ under
  pressure}},\ }\href {https://doi.org/10.1103/PhysRevB.108.125105} {\bibfield
  {journal} {\bibinfo  {journal} {Phys. Rev. B}\ }\textbf {\bibinfo {volume}
  {108}},\ \bibinfo {pages} {125105} (\bibinfo {year} {2023})}\BibitemShut
  {NoStop}%
\bibitem [{\citenamefont {Wang}\ \emph {et~al.}(2024)\citenamefont {Wang},
  \citenamefont {Ouyang}, \citenamefont {He},\ and\ \citenamefont
  {Lu}}]{PhysRevB.109.165140}%
  \BibitemOpen
  \bibfield  {author} {\bibinfo {author} {\bibfnamefont {J.-X.}\ \bibnamefont
  {Wang}}, \bibinfo {author} {\bibfnamefont {Z.}~\bibnamefont {Ouyang}},
  \bibinfo {author} {\bibfnamefont {R.-Q.}\ \bibnamefont {He}},\ and\ \bibinfo
  {author} {\bibfnamefont {Z.-Y.}\ \bibnamefont {Lu}},\ }\bibfield  {title}
  {\bibinfo {title} {{Non-Fermi liquid and Hund correlation in
  ${\mathrm{La}}_{4}{\mathrm{Ni}}_{3}{\mathrm{O}}_{10}$ under high pressure}},\
  }\href {https://doi.org/10.1103/PhysRevB.109.165140} {\bibfield  {journal}
  {\bibinfo  {journal} {Phys. Rev. B}\ }\textbf {\bibinfo {volume} {109}},\
  \bibinfo {pages} {165140} (\bibinfo {year} {2024})}\BibitemShut {NoStop}%
\bibitem [{\citenamefont {Aichhorn}\ \emph {et~al.}(2010)\citenamefont
  {Aichhorn}, \citenamefont {Biermann}, \citenamefont {Miyake}, \citenamefont
  {Georges},\ and\ \citenamefont {Imada}}]{PhysRevB.82.064504}%
  \BibitemOpen
  \bibfield  {author} {\bibinfo {author} {\bibfnamefont {M.}~\bibnamefont
  {Aichhorn}}, \bibinfo {author} {\bibfnamefont {S.}~\bibnamefont {Biermann}},
  \bibinfo {author} {\bibfnamefont {T.}~\bibnamefont {Miyake}}, \bibinfo
  {author} {\bibfnamefont {A.}~\bibnamefont {Georges}},\ and\ \bibinfo {author}
  {\bibfnamefont {M.}~\bibnamefont {Imada}},\ }\bibfield  {title} {\bibinfo
  {title} {{Theoretical evidence for strong correlations and incoherent
  metallic state in FeSe}},\ }\href
  {https://doi.org/10.1103/PhysRevB.82.064504} {\bibfield  {journal} {\bibinfo
  {journal} {Phys. Rev. B}\ }\textbf {\bibinfo {volume} {82}},\ \bibinfo
  {pages} {064504} (\bibinfo {year} {2010})}\BibitemShut {NoStop}%
\bibitem [{\citenamefont {Craco}\ and\ \citenamefont
  {Leoni}(2019)}]{PhysRevB.100.121101}%
  \BibitemOpen
  \bibfield  {author} {\bibinfo {author} {\bibfnamefont {L.}~\bibnamefont
  {Craco}}\ and\ \bibinfo {author} {\bibfnamefont {S.}~\bibnamefont {Leoni}},\
  }\bibfield  {title} {\bibinfo {title} {{Theory of two-fluid metallicity in
  superconducting FeSe at high pressure}},\ }\href
  {https://doi.org/10.1103/PhysRevB.100.121101} {\bibfield  {journal} {\bibinfo
   {journal} {Phys. Rev. B}\ }\textbf {\bibinfo {volume} {100}},\ \bibinfo
  {pages} {121101} (\bibinfo {year} {2019})}\BibitemShut {NoStop}%
\bibitem [{\citenamefont {Huang}\ and\ \citenamefont
  {Lu}(2020{\natexlab{a}})}]{PhysRevB.102.125130}%
  \BibitemOpen
  \bibfield  {author} {\bibinfo {author} {\bibfnamefont {L.}~\bibnamefont
  {Huang}}\ and\ \bibinfo {author} {\bibfnamefont {H.}~\bibnamefont {Lu}},\
  }\bibfield  {title} {\bibinfo {title} {{Signatures of Hundness in kagome
  metals}},\ }\href {https://doi.org/10.1103/PhysRevB.102.125130} {\bibfield
  {journal} {\bibinfo  {journal} {Phys. Rev. B}\ }\textbf {\bibinfo {volume}
  {102}},\ \bibinfo {pages} {125130} (\bibinfo {year}
  {2020}{\natexlab{a}})}\BibitemShut {NoStop}%
\bibitem [{\citenamefont {Liu}\ \emph {et~al.}(2020)\citenamefont {Liu},
  \citenamefont {Li}, \citenamefont {Wang}, \citenamefont {Wang}, \citenamefont
  {Wen}, \citenamefont {Jiang}, \citenamefont {Lu}, \citenamefont {Yan},
  \citenamefont {Huang}, \citenamefont {Shen}, \citenamefont {Yin},
  \citenamefont {Wang}, \citenamefont {Yin}, \citenamefont {Lei},\ and\
  \citenamefont {Wang}}]{Liu2020}%
  \BibitemOpen
  \bibfield  {author} {\bibinfo {author} {\bibfnamefont {Z.}~\bibnamefont
  {Liu}}, \bibinfo {author} {\bibfnamefont {M.}~\bibnamefont {Li}}, \bibinfo
  {author} {\bibfnamefont {Q.}~\bibnamefont {Wang}}, \bibinfo {author}
  {\bibfnamefont {G.}~\bibnamefont {Wang}}, \bibinfo {author} {\bibfnamefont
  {C.}~\bibnamefont {Wen}}, \bibinfo {author} {\bibfnamefont {K.}~\bibnamefont
  {Jiang}}, \bibinfo {author} {\bibfnamefont {X.}~\bibnamefont {Lu}}, \bibinfo
  {author} {\bibfnamefont {S.}~\bibnamefont {Yan}}, \bibinfo {author}
  {\bibfnamefont {Y.}~\bibnamefont {Huang}}, \bibinfo {author} {\bibfnamefont
  {D.}~\bibnamefont {Shen}}, \bibinfo {author} {\bibfnamefont {J.-X.}\
  \bibnamefont {Yin}}, \bibinfo {author} {\bibfnamefont {Z.}~\bibnamefont
  {Wang}}, \bibinfo {author} {\bibfnamefont {Z.}~\bibnamefont {Yin}}, \bibinfo
  {author} {\bibfnamefont {H.}~\bibnamefont {Lei}},\ and\ \bibinfo {author}
  {\bibfnamefont {S.}~\bibnamefont {Wang}},\ }\bibfield  {title} {\bibinfo
  {title} {{Orbital-selective Dirac fermions and extremely flat bands in
  frustrated kagome-lattice metal CoSn}},\ }\href
  {https://doi.org/10.1038/s41467-020-17462-4} {\bibfield  {journal} {\bibinfo
  {journal} {Nature Communications}\ }\textbf {\bibinfo {volume} {11}},\
  \bibinfo {pages} {4002} (\bibinfo {year} {2020})}\BibitemShut {NoStop}%
\bibitem [{\citenamefont {Wan}\ \emph {et~al.}(2022)\citenamefont {Wan},
  \citenamefont {Lu},\ and\ \citenamefont {Huang}}]{PhysRevB.105.155131}%
  \BibitemOpen
  \bibfield  {author} {\bibinfo {author} {\bibfnamefont {S.}~\bibnamefont
  {Wan}}, \bibinfo {author} {\bibfnamefont {H.}~\bibnamefont {Lu}},\ and\
  \bibinfo {author} {\bibfnamefont {L.}~\bibnamefont {Huang}},\ }\bibfield
  {title} {\bibinfo {title} {{Temperature dependence of correlated electronic
  states in the archetypal kagome metal CoSn}},\ }\href
  {https://doi.org/10.1103/PhysRevB.105.155131} {\bibfield  {journal} {\bibinfo
   {journal} {Phys. Rev. B}\ }\textbf {\bibinfo {volume} {105}},\ \bibinfo
  {pages} {155131} (\bibinfo {year} {2022})}\BibitemShut {NoStop}%
\bibitem [{\citenamefont {Huang}\ and\ \citenamefont
  {Lu}(2020{\natexlab{b}})}]{PhysRevB.102.155140}%
  \BibitemOpen
  \bibfield  {author} {\bibinfo {author} {\bibfnamefont {L.}~\bibnamefont
  {Huang}}\ and\ \bibinfo {author} {\bibfnamefont {H.}~\bibnamefont {Lu}},\
  }\bibfield  {title} {\bibinfo {title} {{Protracted Kondo screening and kagome
  bands in the heavy-fermion metal ${\mathrm{Ce}}_{3}\mathrm{Al}$}},\ }\href
  {https://doi.org/10.1103/PhysRevB.102.155140} {\bibfield  {journal} {\bibinfo
   {journal} {Phys. Rev. B}\ }\textbf {\bibinfo {volume} {102}},\ \bibinfo
  {pages} {155140} (\bibinfo {year} {2020}{\natexlab{b}})}\BibitemShut
  {NoStop}%
\bibitem [{\citenamefont {Shim}\ \emph {et~al.}(2007)\citenamefont {Shim},
  \citenamefont {Haule},\ and\ \citenamefont {Kotliar}}]{science.1149064}%
  \BibitemOpen
  \bibfield  {author} {\bibinfo {author} {\bibfnamefont {J.~H.}\ \bibnamefont
  {Shim}}, \bibinfo {author} {\bibfnamefont {K.}~\bibnamefont {Haule}},\ and\
  \bibinfo {author} {\bibfnamefont {G.}~\bibnamefont {Kotliar}},\ }\bibfield
  {title} {\bibinfo {title} {{Modeling the Localized-to-Itinerant Electronic
  Transition in the Heavy Fermion System CeIrIn$_5$}},\ }\href
  {https://doi.org/10.1126/science.1149064} {\bibfield  {journal} {\bibinfo
  {journal} {Science}\ }\textbf {\bibinfo {volume} {318}},\ \bibinfo {pages}
  {1615} (\bibinfo {year} {2007})}\BibitemShut {NoStop}%
\bibitem [{\citenamefont {Lu}\ and\ \citenamefont
  {Huang}(2016)}]{PhysRevB.94.075132}%
  \BibitemOpen
  \bibfield  {author} {\bibinfo {author} {\bibfnamefont {H.}~\bibnamefont
  {Lu}}\ and\ \bibinfo {author} {\bibfnamefont {L.}~\bibnamefont {Huang}},\
  }\bibfield  {title} {\bibinfo {title} {{Pressure-driven $4f$
  localized-itinerant crossover in heavy-fermion compound
  ${\mathrm{CeIn}}_{3}$: A first-principles many-body perspective}},\ }\href
  {https://doi.org/10.1103/PhysRevB.94.075132} {\bibfield  {journal} {\bibinfo
  {journal} {Phys. Rev. B}\ }\textbf {\bibinfo {volume} {94}},\ \bibinfo
  {pages} {075132} (\bibinfo {year} {2016})}\BibitemShut {NoStop}%
\bibitem [{\citenamefont {Wallerberger}\ \emph {et~al.}(2019)\citenamefont
  {Wallerberger}, \citenamefont {Hausoel}, \citenamefont {Gunacker},
  \citenamefont {Kowalski}, \citenamefont {Parragh}, \citenamefont {Goth},
  \citenamefont {Held},\ and\ \citenamefont
  {Sangiovanni}}]{WALLERBERGER2019388}%
  \BibitemOpen
  \bibfield  {author} {\bibinfo {author} {\bibfnamefont {M.}~\bibnamefont
  {Wallerberger}}, \bibinfo {author} {\bibfnamefont {A.}~\bibnamefont
  {Hausoel}}, \bibinfo {author} {\bibfnamefont {P.}~\bibnamefont {Gunacker}},
  \bibinfo {author} {\bibfnamefont {A.}~\bibnamefont {Kowalski}}, \bibinfo
  {author} {\bibfnamefont {N.}~\bibnamefont {Parragh}}, \bibinfo {author}
  {\bibfnamefont {F.}~\bibnamefont {Goth}}, \bibinfo {author} {\bibfnamefont
  {K.}~\bibnamefont {Held}},\ and\ \bibinfo {author} {\bibfnamefont
  {G.}~\bibnamefont {Sangiovanni}},\ }\bibfield  {title} {\bibinfo {title}
  {{w2dynamics: Local one- and two-particle quantities from dynamical mean
  field theory}},\ }\href
  {https://doi.org/https://doi.org/10.1016/j.cpc.2018.09.007} {\bibfield
  {journal} {\bibinfo  {journal} {Comput. Phys. Commun.}\ }\textbf {\bibinfo
  {volume} {235}},\ \bibinfo {pages} {388} (\bibinfo {year}
  {2019})}\BibitemShut {NoStop}%
\bibitem [{\citenamefont {Aichhorn}\ \emph {et~al.}(2016)\citenamefont
  {Aichhorn}, \citenamefont {Pourovskii}, \citenamefont {Seth}, \citenamefont
  {Vildosola}, \citenamefont {Zingl}, \citenamefont {Peil}, \citenamefont
  {Deng}, \citenamefont {Mravlje}, \citenamefont {Kraberger}, \citenamefont
  {Martins}, \citenamefont {Ferrero},\ and\ \citenamefont
  {Parcollet}}]{AICHHORN2016200}%
  \BibitemOpen
  \bibfield  {author} {\bibinfo {author} {\bibfnamefont {M.}~\bibnamefont
  {Aichhorn}}, \bibinfo {author} {\bibfnamefont {L.}~\bibnamefont
  {Pourovskii}}, \bibinfo {author} {\bibfnamefont {P.}~\bibnamefont {Seth}},
  \bibinfo {author} {\bibfnamefont {V.}~\bibnamefont {Vildosola}}, \bibinfo
  {author} {\bibfnamefont {M.}~\bibnamefont {Zingl}}, \bibinfo {author}
  {\bibfnamefont {O.~E.}\ \bibnamefont {Peil}}, \bibinfo {author}
  {\bibfnamefont {X.}~\bibnamefont {Deng}}, \bibinfo {author} {\bibfnamefont
  {J.}~\bibnamefont {Mravlje}}, \bibinfo {author} {\bibfnamefont {G.~J.}\
  \bibnamefont {Kraberger}}, \bibinfo {author} {\bibfnamefont {C.}~\bibnamefont
  {Martins}}, \bibinfo {author} {\bibfnamefont {M.}~\bibnamefont {Ferrero}},\
  and\ \bibinfo {author} {\bibfnamefont {O.}~\bibnamefont {Parcollet}},\
  }\bibfield  {title} {\bibinfo {title} {{TRIQS/DFTTools: A TRIQS application
  for ab initio calculations of correlated materials}},\ }\href
  {https://doi.org/https://doi.org/10.1016/j.cpc.2016.03.014} {\bibfield
  {journal} {\bibinfo  {journal} {Comput. Phys. Commun.}\ }\textbf {\bibinfo
  {volume} {204}},\ \bibinfo {pages} {200} (\bibinfo {year}
  {2016})}\BibitemShut {NoStop}%
\bibitem [{\citenamefont {Parcollet}\ \emph {et~al.}(2015)\citenamefont
  {Parcollet}, \citenamefont {Ferrero}, \citenamefont {Ayral}, \citenamefont
  {Hafermann}, \citenamefont {Krivenko}, \citenamefont {Messio},\ and\
  \citenamefont {Seth}}]{PARCOLLET2015398}%
  \BibitemOpen
  \bibfield  {author} {\bibinfo {author} {\bibfnamefont {O.}~\bibnamefont
  {Parcollet}}, \bibinfo {author} {\bibfnamefont {M.}~\bibnamefont {Ferrero}},
  \bibinfo {author} {\bibfnamefont {T.}~\bibnamefont {Ayral}}, \bibinfo
  {author} {\bibfnamefont {H.}~\bibnamefont {Hafermann}}, \bibinfo {author}
  {\bibfnamefont {I.}~\bibnamefont {Krivenko}}, \bibinfo {author}
  {\bibfnamefont {L.}~\bibnamefont {Messio}},\ and\ \bibinfo {author}
  {\bibfnamefont {P.}~\bibnamefont {Seth}},\ }\bibfield  {title} {\bibinfo
  {title} {{TRIQS: A toolbox for research on interacting quantum systems}},\
  }\href {https://doi.org/https://doi.org/10.1016/j.cpc.2015.04.023} {\bibfield
   {journal} {\bibinfo  {journal} {Comput. Phys. Commun.}\ }\textbf {\bibinfo
  {volume} {196}},\ \bibinfo {pages} {398} (\bibinfo {year}
  {2015})}\BibitemShut {NoStop}%
\bibitem [{\citenamefont {Seth}\ \emph {et~al.}(2016)\citenamefont {Seth},
  \citenamefont {Krivenko}, \citenamefont {Ferrero},\ and\ \citenamefont
  {Parcollet}}]{SETH2016274}%
  \BibitemOpen
  \bibfield  {author} {\bibinfo {author} {\bibfnamefont {P.}~\bibnamefont
  {Seth}}, \bibinfo {author} {\bibfnamefont {I.}~\bibnamefont {Krivenko}},
  \bibinfo {author} {\bibfnamefont {M.}~\bibnamefont {Ferrero}},\ and\ \bibinfo
  {author} {\bibfnamefont {O.}~\bibnamefont {Parcollet}},\ }\bibfield  {title}
  {\bibinfo {title} {{TRIQS/CTHYB: A continuous-time quantum Monte Carlo
  hybridisation expansion solver for quantum impurity problems}},\ }\href
  {https://doi.org/https://doi.org/10.1016/j.cpc.2015.10.023} {\bibfield
  {journal} {\bibinfo  {journal} {Comput. Phys. Commun.}\ }\textbf {\bibinfo
  {volume} {200}},\ \bibinfo {pages} {274} (\bibinfo {year}
  {2016})}\BibitemShut {NoStop}%
\bibitem [{\citenamefont {Gaenko}\ \emph {et~al.}(2017)\citenamefont {Gaenko},
  \citenamefont {Antipov}, \citenamefont {Carcassi}, \citenamefont {Chen},
  \citenamefont {Chen}, \citenamefont {Dong}, \citenamefont {Gamper},
  \citenamefont {Gukelberger}, \citenamefont {Igarashi}, \citenamefont
  {Iskakov}, \citenamefont {Könz}, \citenamefont {LeBlanc}, \citenamefont
  {Levy}, \citenamefont {Ma}, \citenamefont {Paki}, \citenamefont {Shinaoka},
  \citenamefont {Todo}, \citenamefont {Troyer},\ and\ \citenamefont
  {Gull}}]{GAENKO2017235}%
  \BibitemOpen
  \bibfield  {author} {\bibinfo {author} {\bibfnamefont {A.}~\bibnamefont
  {Gaenko}}, \bibinfo {author} {\bibfnamefont {A.}~\bibnamefont {Antipov}},
  \bibinfo {author} {\bibfnamefont {G.}~\bibnamefont {Carcassi}}, \bibinfo
  {author} {\bibfnamefont {T.}~\bibnamefont {Chen}}, \bibinfo {author}
  {\bibfnamefont {X.}~\bibnamefont {Chen}}, \bibinfo {author} {\bibfnamefont
  {Q.}~\bibnamefont {Dong}}, \bibinfo {author} {\bibfnamefont {L.}~\bibnamefont
  {Gamper}}, \bibinfo {author} {\bibfnamefont {J.}~\bibnamefont {Gukelberger}},
  \bibinfo {author} {\bibfnamefont {R.}~\bibnamefont {Igarashi}}, \bibinfo
  {author} {\bibfnamefont {S.}~\bibnamefont {Iskakov}}, \bibinfo {author}
  {\bibfnamefont {M.}~\bibnamefont {Könz}}, \bibinfo {author} {\bibfnamefont
  {J.}~\bibnamefont {LeBlanc}}, \bibinfo {author} {\bibfnamefont
  {R.}~\bibnamefont {Levy}}, \bibinfo {author} {\bibfnamefont {P.}~\bibnamefont
  {Ma}}, \bibinfo {author} {\bibfnamefont {J.}~\bibnamefont {Paki}}, \bibinfo
  {author} {\bibfnamefont {H.}~\bibnamefont {Shinaoka}}, \bibinfo {author}
  {\bibfnamefont {S.}~\bibnamefont {Todo}}, \bibinfo {author} {\bibfnamefont
  {M.}~\bibnamefont {Troyer}},\ and\ \bibinfo {author} {\bibfnamefont
  {E.}~\bibnamefont {Gull}},\ }\bibfield  {title} {\bibinfo {title} {{Updated
  core libraries of the ALPS project}},\ }\href
  {https://doi.org/https://doi.org/10.1016/j.cpc.2016.12.009} {\bibfield
  {journal} {\bibinfo  {journal} {Comput. Phys. Commun.}\ }\textbf {\bibinfo
  {volume} {213}},\ \bibinfo {pages} {235} (\bibinfo {year}
  {2017})}\BibitemShut {NoStop}%
\bibitem [{\citenamefont {Shinaoka}\ \emph
  {et~al.}(2017{\natexlab{a}})\citenamefont {Shinaoka}, \citenamefont {Gull},\
  and\ \citenamefont {Werner}}]{SHINAOKA2017128}%
  \BibitemOpen
  \bibfield  {author} {\bibinfo {author} {\bibfnamefont {H.}~\bibnamefont
  {Shinaoka}}, \bibinfo {author} {\bibfnamefont {E.}~\bibnamefont {Gull}},\
  and\ \bibinfo {author} {\bibfnamefont {P.}~\bibnamefont {Werner}},\
  }\bibfield  {title} {\bibinfo {title} {{Continuous-time hybridization
  expansion quantum impurity solver for multi-orbital systems with complex
  hybridizations}},\ }\href
  {https://doi.org/https://doi.org/10.1016/j.cpc.2017.01.003} {\bibfield
  {journal} {\bibinfo  {journal} {Comput. Phys. Commun.}\ }\textbf {\bibinfo
  {volume} {215}},\ \bibinfo {pages} {128} (\bibinfo {year}
  {2017}{\natexlab{a}})}\BibitemShut {NoStop}%
\bibitem [{\citenamefont {Haule}\ \emph {et~al.}(2010)\citenamefont {Haule},
  \citenamefont {Yee},\ and\ \citenamefont {Kim}}]{PhysRevB.81.195107}%
  \BibitemOpen
  \bibfield  {author} {\bibinfo {author} {\bibfnamefont {K.}~\bibnamefont
  {Haule}}, \bibinfo {author} {\bibfnamefont {C.-H.}\ \bibnamefont {Yee}},\
  and\ \bibinfo {author} {\bibfnamefont {K.}~\bibnamefont {Kim}},\ }\bibfield
  {title} {\bibinfo {title} {{Dynamical mean-field theory within the
  full-potential methods: Electronic structure of ${\text{CeIrIn}}_{5}$,
  ${\text{CeCoIn}}_{5}$, and ${\text{CeRhIn}}_{5}$}},\ }\href
  {https://doi.org/10.1103/PhysRevB.81.195107} {\bibfield  {journal} {\bibinfo
  {journal} {Phys. Rev. B}\ }\textbf {\bibinfo {volume} {81}},\ \bibinfo
  {pages} {195107} (\bibinfo {year} {2010})}\BibitemShut {NoStop}%
\bibitem [{\citenamefont {Haule}(2007)}]{PhysRevB.75.155113}%
  \BibitemOpen
  \bibfield  {author} {\bibinfo {author} {\bibfnamefont {K.}~\bibnamefont
  {Haule}},\ }\bibfield  {title} {\bibinfo {title} {{Quantum Monte Carlo
  impurity solver for cluster dynamical mean-field theory and electronic
  structure calculations with adjustable cluster base}},\ }\href
  {https://doi.org/10.1103/PhysRevB.75.155113} {\bibfield  {journal} {\bibinfo
  {journal} {Phys. Rev. B}\ }\textbf {\bibinfo {volume} {75}},\ \bibinfo
  {pages} {155113} (\bibinfo {year} {2007})}\BibitemShut {NoStop}%
\bibitem [{\citenamefont {Haule}(2015)}]{PhysRevLett.115.196403}%
  \BibitemOpen
  \bibfield  {author} {\bibinfo {author} {\bibfnamefont {K.}~\bibnamefont
  {Haule}},\ }\bibfield  {title} {\bibinfo {title} {{Exact Double Counting in
  Combining the Dynamical Mean Field Theory and the Density Functional
  Theory}},\ }\href {https://doi.org/10.1103/PhysRevLett.115.196403} {\bibfield
   {journal} {\bibinfo  {journal} {Phys. Rev. Lett.}\ }\textbf {\bibinfo
  {volume} {115}},\ \bibinfo {pages} {196403} (\bibinfo {year}
  {2015})}\BibitemShut {NoStop}%
\bibitem [{\citenamefont {Shinaoka}\ \emph {et~al.}(2021)\citenamefont
  {Shinaoka}, \citenamefont {Otsuki}, \citenamefont {Kawamura}, \citenamefont
  {Takemori},\ and\ \citenamefont {Yoshimi}}]{10.21468/SciPostPhys.10.5.117}%
  \BibitemOpen
  \bibfield  {author} {\bibinfo {author} {\bibfnamefont {H.}~\bibnamefont
  {Shinaoka}}, \bibinfo {author} {\bibfnamefont {J.}~\bibnamefont {Otsuki}},
  \bibinfo {author} {\bibfnamefont {M.}~\bibnamefont {Kawamura}}, \bibinfo
  {author} {\bibfnamefont {N.}~\bibnamefont {Takemori}},\ and\ \bibinfo
  {author} {\bibfnamefont {K.}~\bibnamefont {Yoshimi}},\ }\bibfield  {title}
  {\bibinfo {title} {{DCore: Integrated DMFT software for correlated
  electrons}},\ }\href {https://doi.org/10.21468/SciPostPhys.10.5.117}
  {\bibfield  {journal} {\bibinfo  {journal} {SciPost Phys.}\ }\textbf
  {\bibinfo {volume} {10}},\ \bibinfo {pages} {117} (\bibinfo {year}
  {2021})}\BibitemShut {NoStop}%
\bibitem [{\citenamefont {Pashov}\ \emph {et~al.}(2020)\citenamefont {Pashov},
  \citenamefont {Acharya}, \citenamefont {Lambrecht}, \citenamefont {Jackson},
  \citenamefont {Belashchenko}, \citenamefont {Chantis}, \citenamefont
  {Jamet},\ and\ \citenamefont {{van Schilfgaarde}}}]{PASHOV2020107065}%
  \BibitemOpen
  \bibfield  {author} {\bibinfo {author} {\bibfnamefont {D.}~\bibnamefont
  {Pashov}}, \bibinfo {author} {\bibfnamefont {S.}~\bibnamefont {Acharya}},
  \bibinfo {author} {\bibfnamefont {W.~R.}\ \bibnamefont {Lambrecht}}, \bibinfo
  {author} {\bibfnamefont {J.}~\bibnamefont {Jackson}}, \bibinfo {author}
  {\bibfnamefont {K.~D.}\ \bibnamefont {Belashchenko}}, \bibinfo {author}
  {\bibfnamefont {A.}~\bibnamefont {Chantis}}, \bibinfo {author} {\bibfnamefont
  {F.}~\bibnamefont {Jamet}},\ and\ \bibinfo {author} {\bibfnamefont
  {M.}~\bibnamefont {{van Schilfgaarde}}},\ }\bibfield  {title} {\bibinfo
  {title} {{Questaal: A package of electronic structure methods based on the
  linear muffin-tin orbital technique}},\ }\href
  {https://doi.org/https://doi.org/10.1016/j.cpc.2019.107065} {\bibfield
  {journal} {\bibinfo  {journal} {Comput. Phys. Commun.}\ }\textbf {\bibinfo
  {volume} {249}},\ \bibinfo {pages} {107065} (\bibinfo {year}
  {2020})}\BibitemShut {NoStop}%
\bibitem [{\citenamefont {Amadon}\ \emph
  {et~al.}(2008{\natexlab{a}})\citenamefont {Amadon}, \citenamefont
  {Lechermann}, \citenamefont {Georges}, \citenamefont {Jollet}, \citenamefont
  {Wehling},\ and\ \citenamefont {Lichtenstein}}]{PhysRevB.77.205112}%
  \BibitemOpen
  \bibfield  {author} {\bibinfo {author} {\bibfnamefont {B.}~\bibnamefont
  {Amadon}}, \bibinfo {author} {\bibfnamefont {F.}~\bibnamefont {Lechermann}},
  \bibinfo {author} {\bibfnamefont {A.}~\bibnamefont {Georges}}, \bibinfo
  {author} {\bibfnamefont {F.}~\bibnamefont {Jollet}}, \bibinfo {author}
  {\bibfnamefont {T.~O.}\ \bibnamefont {Wehling}},\ and\ \bibinfo {author}
  {\bibfnamefont {A.~I.}\ \bibnamefont {Lichtenstein}},\ }\bibfield  {title}
  {\bibinfo {title} {{Plane-wave based electronic structure calculations for
  correlated materials using dynamical mean-field theory and projected local
  orbitals}},\ }\href {https://doi.org/10.1103/PhysRevB.77.205112} {\bibfield
  {journal} {\bibinfo  {journal} {Phys. Rev. B}\ }\textbf {\bibinfo {volume}
  {77}},\ \bibinfo {pages} {205112} (\bibinfo {year}
  {2008}{\natexlab{a}})}\BibitemShut {NoStop}%
\bibitem [{\citenamefont {Amadon}(2012)}]{Amadon_2012}%
  \BibitemOpen
  \bibfield  {author} {\bibinfo {author} {\bibfnamefont {B.}~\bibnamefont
  {Amadon}},\ }\bibfield  {title} {\bibinfo {title} {{A self-consistent
  DFT + DMFT scheme in the projector augmented wave method: applications to
  cerium, Ce$_{2}$O$_{3}$ and Pu$_{2}$O$_{3}$ with the Hubbard I solver and
  comparison to DFT + U}},\ }\href
  {https://doi.org/10.1088/0953-8984/24/7/075604} {\bibfield  {journal}
  {\bibinfo  {journal} {J. Phys.: Condens. Matter}\ }\textbf {\bibinfo {volume}
  {24}},\ \bibinfo {pages} {075604} (\bibinfo {year} {2012})}\BibitemShut
  {NoStop}%
\bibitem [{\citenamefont {Gonze}\ \emph {et~al.}(2020)\citenamefont {Gonze},
  \citenamefont {Amadon}, \citenamefont {Antonius}, \citenamefont {Arnardi},
  \citenamefont {Baguet}, \citenamefont {Beuken}, \citenamefont {Bieder},
  \citenamefont {Bottin}, \citenamefont {Bouchet}, \citenamefont {Bousquet},
  \citenamefont {Brouwer}, \citenamefont {Bruneval}, \citenamefont {Brunin},
  \citenamefont {Cavignac}, \citenamefont {Charraud}, \citenamefont {Chen},
  \citenamefont {Côté}, \citenamefont {Cottenier}, \citenamefont {Denier},
  \citenamefont {Geneste}, \citenamefont {Ghosez}, \citenamefont {Giantomassi},
  \citenamefont {Gillet}, \citenamefont {Gingras}, \citenamefont {Hamann},
  \citenamefont {Hautier}, \citenamefont {He}, \citenamefont {Helbig},
  \citenamefont {Holzwarth}, \citenamefont {Jia}, \citenamefont {Jollet},
  \citenamefont {Lafargue-Dit-Hauret}, \citenamefont {Lejaeghere},
  \citenamefont {Marques}, \citenamefont {Martin}, \citenamefont {Martins},
  \citenamefont {Miranda}, \citenamefont {Naccarato}, \citenamefont {Persson},
  \citenamefont {Petretto}, \citenamefont {Planes}, \citenamefont {Pouillon},
  \citenamefont {Prokhorenko}, \citenamefont {Ricci}, \citenamefont
  {Rignanese}, \citenamefont {Romero}, \citenamefont {Schmitt}, \citenamefont
  {Torrent}, \citenamefont {{van Setten}}, \citenamefont {{Van Troeye}},
  \citenamefont {Verstraete}, \citenamefont {Zérah},\ and\ \citenamefont
  {Zwanziger}}]{GONZE2020107042}%
  \BibitemOpen
  \bibfield  {author} {\bibinfo {author} {\bibfnamefont {X.}~\bibnamefont
  {Gonze}}, \bibinfo {author} {\bibfnamefont {B.}~\bibnamefont {Amadon}},
  \bibinfo {author} {\bibfnamefont {G.}~\bibnamefont {Antonius}}, \bibinfo
  {author} {\bibfnamefont {F.}~\bibnamefont {Arnardi}}, \bibinfo {author}
  {\bibfnamefont {L.}~\bibnamefont {Baguet}}, \bibinfo {author} {\bibfnamefont
  {J.-M.}\ \bibnamefont {Beuken}}, \bibinfo {author} {\bibfnamefont
  {J.}~\bibnamefont {Bieder}}, \bibinfo {author} {\bibfnamefont
  {F.}~\bibnamefont {Bottin}}, \bibinfo {author} {\bibfnamefont
  {J.}~\bibnamefont {Bouchet}}, \bibinfo {author} {\bibfnamefont
  {E.}~\bibnamefont {Bousquet}}, \bibinfo {author} {\bibfnamefont
  {N.}~\bibnamefont {Brouwer}}, \bibinfo {author} {\bibfnamefont
  {F.}~\bibnamefont {Bruneval}}, \bibinfo {author} {\bibfnamefont
  {G.}~\bibnamefont {Brunin}}, \bibinfo {author} {\bibfnamefont
  {T.}~\bibnamefont {Cavignac}}, \bibinfo {author} {\bibfnamefont {J.-B.}\
  \bibnamefont {Charraud}}, \bibinfo {author} {\bibfnamefont {W.}~\bibnamefont
  {Chen}}, \bibinfo {author} {\bibfnamefont {M.}~\bibnamefont {Côté}},
  \bibinfo {author} {\bibfnamefont {S.}~\bibnamefont {Cottenier}}, \bibinfo
  {author} {\bibfnamefont {J.}~\bibnamefont {Denier}}, \bibinfo {author}
  {\bibfnamefont {G.}~\bibnamefont {Geneste}}, \bibinfo {author} {\bibfnamefont
  {P.}~\bibnamefont {Ghosez}}, \bibinfo {author} {\bibfnamefont
  {M.}~\bibnamefont {Giantomassi}}, \bibinfo {author} {\bibfnamefont
  {Y.}~\bibnamefont {Gillet}}, \bibinfo {author} {\bibfnamefont
  {O.}~\bibnamefont {Gingras}}, \bibinfo {author} {\bibfnamefont {D.~R.}\
  \bibnamefont {Hamann}}, \bibinfo {author} {\bibfnamefont {G.}~\bibnamefont
  {Hautier}}, \bibinfo {author} {\bibfnamefont {X.}~\bibnamefont {He}},
  \bibinfo {author} {\bibfnamefont {N.}~\bibnamefont {Helbig}}, \bibinfo
  {author} {\bibfnamefont {N.}~\bibnamefont {Holzwarth}}, \bibinfo {author}
  {\bibfnamefont {Y.}~\bibnamefont {Jia}}, \bibinfo {author} {\bibfnamefont
  {F.}~\bibnamefont {Jollet}}, \bibinfo {author} {\bibfnamefont
  {W.}~\bibnamefont {Lafargue-Dit-Hauret}}, \bibinfo {author} {\bibfnamefont
  {K.}~\bibnamefont {Lejaeghere}}, \bibinfo {author} {\bibfnamefont {M.~A.}\
  \bibnamefont {Marques}}, \bibinfo {author} {\bibfnamefont {A.}~\bibnamefont
  {Martin}}, \bibinfo {author} {\bibfnamefont {C.}~\bibnamefont {Martins}},
  \bibinfo {author} {\bibfnamefont {H.~P.}\ \bibnamefont {Miranda}}, \bibinfo
  {author} {\bibfnamefont {F.}~\bibnamefont {Naccarato}}, \bibinfo {author}
  {\bibfnamefont {K.}~\bibnamefont {Persson}}, \bibinfo {author} {\bibfnamefont
  {G.}~\bibnamefont {Petretto}}, \bibinfo {author} {\bibfnamefont
  {V.}~\bibnamefont {Planes}}, \bibinfo {author} {\bibfnamefont
  {Y.}~\bibnamefont {Pouillon}}, \bibinfo {author} {\bibfnamefont
  {S.}~\bibnamefont {Prokhorenko}}, \bibinfo {author} {\bibfnamefont
  {F.}~\bibnamefont {Ricci}}, \bibinfo {author} {\bibfnamefont {G.-M.}\
  \bibnamefont {Rignanese}}, \bibinfo {author} {\bibfnamefont {A.~H.}\
  \bibnamefont {Romero}}, \bibinfo {author} {\bibfnamefont {M.~M.}\
  \bibnamefont {Schmitt}}, \bibinfo {author} {\bibfnamefont {M.}~\bibnamefont
  {Torrent}}, \bibinfo {author} {\bibfnamefont {M.~J.}\ \bibnamefont {{van
  Setten}}}, \bibinfo {author} {\bibfnamefont {B.}~\bibnamefont {{Van
  Troeye}}}, \bibinfo {author} {\bibfnamefont {M.~J.}\ \bibnamefont
  {Verstraete}}, \bibinfo {author} {\bibfnamefont {G.}~\bibnamefont {Zérah}},\
  and\ \bibinfo {author} {\bibfnamefont {J.~W.}\ \bibnamefont {Zwanziger}},\
  }\bibfield  {title} {\bibinfo {title} {{The Abinit project: Impact,
  environment and recent developments}},\ }\href
  {https://doi.org/https://doi.org/10.1016/j.cpc.2019.107042} {\bibfield
  {journal} {\bibinfo  {journal} {Comput. Phys. Commun.}\ }\textbf {\bibinfo
  {volume} {248}},\ \bibinfo {pages} {107042} (\bibinfo {year}
  {2020})}\BibitemShut {NoStop}%
\bibitem [{\citenamefont {Qu}\ \emph {et~al.}(2022)\citenamefont {Qu},
  \citenamefont {Xu}, \citenamefont {Li}, \citenamefont {Li}, \citenamefont
  {He},\ and\ \citenamefont {Ren}}]{acs.jctc.2c00472}%
  \BibitemOpen
  \bibfield  {author} {\bibinfo {author} {\bibfnamefont {X.}~\bibnamefont
  {Qu}}, \bibinfo {author} {\bibfnamefont {P.}~\bibnamefont {Xu}}, \bibinfo
  {author} {\bibfnamefont {R.}~\bibnamefont {Li}}, \bibinfo {author}
  {\bibfnamefont {G.}~\bibnamefont {Li}}, \bibinfo {author} {\bibfnamefont
  {L.}~\bibnamefont {He}},\ and\ \bibinfo {author} {\bibfnamefont
  {X.}~\bibnamefont {Ren}},\ }\bibfield  {title} {\bibinfo {title} {{Density
  Functional Theory Plus Dynamical Mean Field Theory within the Framework of
  Linear Combination of Numerical Atomic Orbitals: Formulation and
  Benchmarks}},\ }\href {https://doi.org/10.1021/acs.jctc.2c00472} {\bibfield
  {journal} {\bibinfo  {journal} {J. Chem. Theory Comput.}\ }\textbf {\bibinfo
  {volume} {18}},\ \bibinfo {pages} {5589} (\bibinfo {year}
  {2022})}\BibitemShut {NoStop}%
\bibitem [{\citenamefont {Choi}\ \emph {et~al.}(2019)\citenamefont {Choi},
  \citenamefont {Semon}, \citenamefont {Kang}, \citenamefont {Kutepov},\ and\
  \citenamefont {Kotliar}}]{CHOI2019277}%
  \BibitemOpen
  \bibfield  {author} {\bibinfo {author} {\bibfnamefont {S.}~\bibnamefont
  {Choi}}, \bibinfo {author} {\bibfnamefont {P.}~\bibnamefont {Semon}},
  \bibinfo {author} {\bibfnamefont {B.}~\bibnamefont {Kang}}, \bibinfo {author}
  {\bibfnamefont {A.}~\bibnamefont {Kutepov}},\ and\ \bibinfo {author}
  {\bibfnamefont {G.}~\bibnamefont {Kotliar}},\ }\bibfield  {title} {\bibinfo
  {title} {{ComDMFT: A massively parallel computer package for the electronic
  structure of correlated-electron systems}},\ }\href
  {https://doi.org/https://doi.org/10.1016/j.cpc.2019.07.003} {\bibfield
  {journal} {\bibinfo  {journal} {Comput. Phys. Commun.}\ }\textbf {\bibinfo
  {volume} {244}},\ \bibinfo {pages} {277} (\bibinfo {year}
  {2019})}\BibitemShut {NoStop}%
\bibitem [{\citenamefont {Singh}\ \emph {et~al.}(2021)\citenamefont {Singh},
  \citenamefont {Herath}, \citenamefont {Wah}, \citenamefont {Liao},
  \citenamefont {Romero},\ and\ \citenamefont {Park}}]{SINGH2021107778}%
  \BibitemOpen
  \bibfield  {author} {\bibinfo {author} {\bibfnamefont {V.}~\bibnamefont
  {Singh}}, \bibinfo {author} {\bibfnamefont {U.}~\bibnamefont {Herath}},
  \bibinfo {author} {\bibfnamefont {B.}~\bibnamefont {Wah}}, \bibinfo {author}
  {\bibfnamefont {X.}~\bibnamefont {Liao}}, \bibinfo {author} {\bibfnamefont
  {A.~H.}\ \bibnamefont {Romero}},\ and\ \bibinfo {author} {\bibfnamefont
  {H.}~\bibnamefont {Park}},\ }\bibfield  {title} {\bibinfo {title} {{DMFTwDFT:
  An open-source code combining Dynamical Mean Field Theory with various
  density functional theory packages}},\ }\href
  {https://doi.org/https://doi.org/10.1016/j.cpc.2020.107778} {\bibfield
  {journal} {\bibinfo  {journal} {Comput. Phys. Commun.}\ }\textbf {\bibinfo
  {volume} {261}},\ \bibinfo {pages} {107778} (\bibinfo {year}
  {2021})}\BibitemShut {NoStop}%
\bibitem [{\citenamefont {Merkel}\ \emph {et~al.}(2022)\citenamefont {Merkel},
  \citenamefont {Carta}, \citenamefont {Beck},\ and\ \citenamefont
  {Hampel}}]{Merkel2022}%
  \BibitemOpen
  \bibfield  {author} {\bibinfo {author} {\bibfnamefont {M.~E.}\ \bibnamefont
  {Merkel}}, \bibinfo {author} {\bibfnamefont {A.}~\bibnamefont {Carta}},
  \bibinfo {author} {\bibfnamefont {S.}~\bibnamefont {Beck}},\ and\ \bibinfo
  {author} {\bibfnamefont {A.}~\bibnamefont {Hampel}},\ }\bibfield  {title}
  {\bibinfo {title} {{solid\_dmft: gray-boxing DFT+DMFT materials simulations
  with TRIQS}},\ }\href {https://doi.org/10.21105/joss.04623} {\bibfield
  {journal} {\bibinfo  {journal} {J. Open Source Software}\ }\textbf {\bibinfo
  {volume} {7}},\ \bibinfo {pages} {4623} (\bibinfo {year} {2022})}\BibitemShut
  {NoStop}%
\bibitem [{\citenamefont {Gull}\ \emph {et~al.}(2011)\citenamefont {Gull},
  \citenamefont {Millis}, \citenamefont {Lichtenstein}, \citenamefont
  {Rubtsov}, \citenamefont {Troyer},\ and\ \citenamefont
  {Werner}}]{RevModPhys.83.349}%
  \BibitemOpen
  \bibfield  {author} {\bibinfo {author} {\bibfnamefont {E.}~\bibnamefont
  {Gull}}, \bibinfo {author} {\bibfnamefont {A.~J.}\ \bibnamefont {Millis}},
  \bibinfo {author} {\bibfnamefont {A.~I.}\ \bibnamefont {Lichtenstein}},
  \bibinfo {author} {\bibfnamefont {A.~N.}\ \bibnamefont {Rubtsov}}, \bibinfo
  {author} {\bibfnamefont {M.}~\bibnamefont {Troyer}},\ and\ \bibinfo {author}
  {\bibfnamefont {P.}~\bibnamefont {Werner}},\ }\bibfield  {title} {\bibinfo
  {title} {{Continuous-time Monte Carlo methods for quantum impurity models}},\
  }\href {https://doi.org/10.1103/RevModPhys.83.349} {\bibfield  {journal}
  {\bibinfo  {journal} {Rev. Mod. Phys.}\ }\textbf {\bibinfo {volume} {83}},\
  \bibinfo {pages} {349} (\bibinfo {year} {2011})}\BibitemShut {NoStop}%
\bibitem [{\citenamefont {Werner}\ and\ \citenamefont
  {Millis}(2006)}]{PhysRevB.74.155107}%
  \BibitemOpen
  \bibfield  {author} {\bibinfo {author} {\bibfnamefont {P.}~\bibnamefont
  {Werner}}\ and\ \bibinfo {author} {\bibfnamefont {A.~J.}\ \bibnamefont
  {Millis}},\ }\bibfield  {title} {\bibinfo {title} {{Hybridization expansion
  impurity solver: General formulation and application to Kondo lattice and
  two-orbital models}},\ }\href {https://doi.org/10.1103/PhysRevB.74.155107}
  {\bibfield  {journal} {\bibinfo  {journal} {Phys. Rev. B}\ }\textbf {\bibinfo
  {volume} {74}},\ \bibinfo {pages} {155107} (\bibinfo {year}
  {2006})}\BibitemShut {NoStop}%
\bibitem [{\citenamefont {Werner}\ \emph {et~al.}(2006)\citenamefont {Werner},
  \citenamefont {Comanac}, \citenamefont {de' Medici}, \citenamefont {Troyer},\
  and\ \citenamefont {Millis}}]{PhysRevLett.97.076405}%
  \BibitemOpen
  \bibfield  {author} {\bibinfo {author} {\bibfnamefont {P.}~\bibnamefont
  {Werner}}, \bibinfo {author} {\bibfnamefont {A.}~\bibnamefont {Comanac}},
  \bibinfo {author} {\bibfnamefont {L.}~\bibnamefont {de' Medici}}, \bibinfo
  {author} {\bibfnamefont {M.}~\bibnamefont {Troyer}},\ and\ \bibinfo {author}
  {\bibfnamefont {A.~J.}\ \bibnamefont {Millis}},\ }\bibfield  {title}
  {\bibinfo {title} {{Continuous-Time Solver for Quantum Impurity Models}},\
  }\href {https://doi.org/10.1103/PhysRevLett.97.076405} {\bibfield  {journal}
  {\bibinfo  {journal} {Phys. Rev. Lett.}\ }\textbf {\bibinfo {volume} {97}},\
  \bibinfo {pages} {076405} (\bibinfo {year} {2006})}\BibitemShut {NoStop}%
\bibitem [{\citenamefont {Melnick}\ \emph {et~al.}(2021)\citenamefont
  {Melnick}, \citenamefont {Sémon}, \citenamefont {Yu}, \citenamefont
  {D'Imperio}, \citenamefont {Tremblay},\ and\ \citenamefont
  {Kotliar}}]{MELNICK2021108075}%
  \BibitemOpen
  \bibfield  {author} {\bibinfo {author} {\bibfnamefont {C.}~\bibnamefont
  {Melnick}}, \bibinfo {author} {\bibfnamefont {P.}~\bibnamefont {Sémon}},
  \bibinfo {author} {\bibfnamefont {K.}~\bibnamefont {Yu}}, \bibinfo {author}
  {\bibfnamefont {N.}~\bibnamefont {D'Imperio}}, \bibinfo {author}
  {\bibfnamefont {A.-M.}\ \bibnamefont {Tremblay}},\ and\ \bibinfo {author}
  {\bibfnamefont {G.}~\bibnamefont {Kotliar}},\ }\bibfield  {title} {\bibinfo
  {title} {{Accelerated impurity solver for DMFT and its diagrammatic
  extensions}},\ }\href
  {https://doi.org/https://doi.org/10.1016/j.cpc.2021.108075} {\bibfield
  {journal} {\bibinfo  {journal} {Comput. Phys. Commun.}\ }\textbf {\bibinfo
  {volume} {267}},\ \bibinfo {pages} {108075} (\bibinfo {year}
  {2021})}\BibitemShut {NoStop}%
\bibitem [{\citenamefont {Ko\ifmmode~\mbox{\c{c}}\else \c{c}\fi{}er}\ \emph
  {et~al.}(2020)\citenamefont {Ko\ifmmode~\mbox{\c{c}}\else \c{c}\fi{}er},
  \citenamefont {Haule}, \citenamefont {Pascut},\ and\ \citenamefont
  {Monserrat}}]{PhysRevB.102.245104}%
  \BibitemOpen
  \bibfield  {author} {\bibinfo {author} {\bibfnamefont {C.~P.}\ \bibnamefont
  {Ko\ifmmode~\mbox{\c{c}}\else \c{c}\fi{}er}}, \bibinfo {author}
  {\bibfnamefont {K.}~\bibnamefont {Haule}}, \bibinfo {author} {\bibfnamefont
  {G.~L.}\ \bibnamefont {Pascut}},\ and\ \bibinfo {author} {\bibfnamefont
  {B.}~\bibnamefont {Monserrat}},\ }\bibfield  {title} {\bibinfo {title}
  {{Efficient lattice dynamics calculations for correlated materials with
  $\mathrm{DFT}+\mathrm{DMFT}$}},\ }\href
  {https://doi.org/10.1103/PhysRevB.102.245104} {\bibfield  {journal} {\bibinfo
   {journal} {Phys. Rev. B}\ }\textbf {\bibinfo {volume} {102}},\ \bibinfo
  {pages} {245104} (\bibinfo {year} {2020})}\BibitemShut {NoStop}%
\bibitem [{\citenamefont {Abramovitch}\ \emph {et~al.}(2023)\citenamefont
  {Abramovitch}, \citenamefont {Zhou}, \citenamefont {Mravlje}, \citenamefont
  {Georges},\ and\ \citenamefont {Bernardi}}]{PhysRevMaterials.7.093801}%
  \BibitemOpen
  \bibfield  {author} {\bibinfo {author} {\bibfnamefont {D.~J.}\ \bibnamefont
  {Abramovitch}}, \bibinfo {author} {\bibfnamefont {J.-J.}\ \bibnamefont
  {Zhou}}, \bibinfo {author} {\bibfnamefont {J.}~\bibnamefont {Mravlje}},
  \bibinfo {author} {\bibfnamefont {A.}~\bibnamefont {Georges}},\ and\ \bibinfo
  {author} {\bibfnamefont {M.}~\bibnamefont {Bernardi}},\ }\bibfield  {title}
  {\bibinfo {title} {{Combining electron-phonon and dynamical mean-field theory
  calculations of correlated materials: Transport in the correlated metal
  ${\mathrm{Sr}}_{2}{\mathrm{RuO}}_{4}$}},\ }\href
  {https://doi.org/10.1103/PhysRevMaterials.7.093801} {\bibfield  {journal}
  {\bibinfo  {journal} {Phys. Rev. Mater.}\ }\textbf {\bibinfo {volume} {7}},\
  \bibinfo {pages} {093801} (\bibinfo {year} {2023})}\BibitemShut {NoStop}%
\bibitem [{\citenamefont {Khanal}\ and\ \citenamefont
  {Haule}(2020)}]{PhysRevB.102.241108}%
  \BibitemOpen
  \bibfield  {author} {\bibinfo {author} {\bibfnamefont {G.}~\bibnamefont
  {Khanal}}\ and\ \bibinfo {author} {\bibfnamefont {K.}~\bibnamefont {Haule}},\
  }\bibfield  {title} {\bibinfo {title} {{Correlation driven phonon anomalies
  in bulk FeSe}},\ }\href {https://doi.org/10.1103/PhysRevB.102.241108}
  {\bibfield  {journal} {\bibinfo  {journal} {Phys. Rev. B}\ }\textbf {\bibinfo
  {volume} {102}},\ \bibinfo {pages} {241108} (\bibinfo {year}
  {2020})}\BibitemShut {NoStop}%
\bibitem [{\citenamefont {Aryasetiawan}\ and\ \citenamefont
  {Gunnarsson}(1998)}]{faryase1998}%
  \BibitemOpen
  \bibfield  {author} {\bibinfo {author} {\bibfnamefont {F.}~\bibnamefont
  {Aryasetiawan}}\ and\ \bibinfo {author} {\bibfnamefont {O.}~\bibnamefont
  {Gunnarsson}},\ }\bibfield  {title} {\bibinfo {title} {{The GW method}},\
  }\href {https://doi.org/10.1088/0034-4885/61/3/002} {\bibfield  {journal}
  {\bibinfo  {journal} {Rep. Prog. Phys.}\ }\textbf {\bibinfo {volume} {61}},\
  \bibinfo {pages} {237} (\bibinfo {year} {1998})}\BibitemShut {NoStop}%
\bibitem [{\citenamefont {Boehnke}\ \emph {et~al.}(2016)\citenamefont
  {Boehnke}, \citenamefont {Nilsson}, \citenamefont {Aryasetiawan},\ and\
  \citenamefont {Werner}}]{PhysRevB.94.201106}%
  \BibitemOpen
  \bibfield  {author} {\bibinfo {author} {\bibfnamefont {L.}~\bibnamefont
  {Boehnke}}, \bibinfo {author} {\bibfnamefont {F.}~\bibnamefont {Nilsson}},
  \bibinfo {author} {\bibfnamefont {F.}~\bibnamefont {Aryasetiawan}},\ and\
  \bibinfo {author} {\bibfnamefont {P.}~\bibnamefont {Werner}},\ }\bibfield
  {title} {\bibinfo {title} {{When strong correlations become weak: Consistent
  merging of $GW$ and DMFT}},\ }\href
  {https://doi.org/10.1103/PhysRevB.94.201106} {\bibfield  {journal} {\bibinfo
  {journal} {Phys. Rev. B}\ }\textbf {\bibinfo {volume} {94}},\ \bibinfo
  {pages} {201106} (\bibinfo {year} {2016})}\BibitemShut {NoStop}%
\bibitem [{\citenamefont {Zhu}\ and\ \citenamefont
  {Chan}(2021)}]{PhysRevX.11.021006}%
  \BibitemOpen
  \bibfield  {author} {\bibinfo {author} {\bibfnamefont {T.}~\bibnamefont
  {Zhu}}\ and\ \bibinfo {author} {\bibfnamefont {G.~K.-L.}\ \bibnamefont
  {Chan}},\ }\bibfield  {title} {\bibinfo {title} {{Ab Initio Full Cell
  $GW+\mathrm{DMFT}$ for Correlated Materials}},\ }\href
  {https://doi.org/10.1103/PhysRevX.11.021006} {\bibfield  {journal} {\bibinfo
  {journal} {Phys. Rev. X}\ }\textbf {\bibinfo {volume} {11}},\ \bibinfo
  {pages} {021006} (\bibinfo {year} {2021})}\BibitemShut {NoStop}%
\bibitem [{\citenamefont {Galler}\ \emph {et~al.}(2019)\citenamefont {Galler},
  \citenamefont {Thunström}, \citenamefont {Kaufmann}, \citenamefont {Pickem},
  \citenamefont {Tomczak},\ and\ \citenamefont {Held}}]{GALLER2019106847}%
  \BibitemOpen
  \bibfield  {author} {\bibinfo {author} {\bibfnamefont {A.}~\bibnamefont
  {Galler}}, \bibinfo {author} {\bibfnamefont {P.}~\bibnamefont {Thunström}},
  \bibinfo {author} {\bibfnamefont {J.}~\bibnamefont {Kaufmann}}, \bibinfo
  {author} {\bibfnamefont {M.}~\bibnamefont {Pickem}}, \bibinfo {author}
  {\bibfnamefont {J.~M.}\ \bibnamefont {Tomczak}},\ and\ \bibinfo {author}
  {\bibfnamefont {K.}~\bibnamefont {Held}},\ }\bibfield  {title} {\bibinfo
  {title} {{The AbinitioD${\rm \Gamma}$A Project v1.0: Non-local correlations
  beyond and susceptibilities within dynamical mean-field theory}},\ }\href
  {https://doi.org/https://doi.org/10.1016/j.cpc.2019.07.012} {\bibfield
  {journal} {\bibinfo  {journal} {Comput. Phys. Commun.}\ }\textbf {\bibinfo
  {volume} {245}},\ \bibinfo {pages} {106847} (\bibinfo {year}
  {2019})}\BibitemShut {NoStop}%
\bibitem [{\citenamefont {Rohringer}\ \emph {et~al.}(2018)\citenamefont
  {Rohringer}, \citenamefont {Hafermann}, \citenamefont {Toschi}, \citenamefont
  {Katanin}, \citenamefont {Antipov}, \citenamefont {Katsnelson}, \citenamefont
  {Lichtenstein}, \citenamefont {Rubtsov},\ and\ \citenamefont
  {Held}}]{RevModPhys.90.025003}%
  \BibitemOpen
  \bibfield  {author} {\bibinfo {author} {\bibfnamefont {G.}~\bibnamefont
  {Rohringer}}, \bibinfo {author} {\bibfnamefont {H.}~\bibnamefont
  {Hafermann}}, \bibinfo {author} {\bibfnamefont {A.}~\bibnamefont {Toschi}},
  \bibinfo {author} {\bibfnamefont {A.~A.}\ \bibnamefont {Katanin}}, \bibinfo
  {author} {\bibfnamefont {A.~E.}\ \bibnamefont {Antipov}}, \bibinfo {author}
  {\bibfnamefont {M.~I.}\ \bibnamefont {Katsnelson}}, \bibinfo {author}
  {\bibfnamefont {A.~I.}\ \bibnamefont {Lichtenstein}}, \bibinfo {author}
  {\bibfnamefont {A.~N.}\ \bibnamefont {Rubtsov}},\ and\ \bibinfo {author}
  {\bibfnamefont {K.}~\bibnamefont {Held}},\ }\bibfield  {title} {\bibinfo
  {title} {{Diagrammatic routes to nonlocal correlations beyond dynamical mean
  field theory}},\ }\href {https://doi.org/10.1103/RevModPhys.90.025003}
  {\bibfield  {journal} {\bibinfo  {journal} {Rev. Mod. Phys.}\ }\textbf
  {\bibinfo {volume} {90}},\ \bibinfo {pages} {025003} (\bibinfo {year}
  {2018})}\BibitemShut {NoStop}%
\bibitem [{\citenamefont {Toschi}\ \emph {et~al.}(2007)\citenamefont {Toschi},
  \citenamefont {Katanin},\ and\ \citenamefont {Held}}]{PhysRevB.75.045118}%
  \BibitemOpen
  \bibfield  {author} {\bibinfo {author} {\bibfnamefont {A.}~\bibnamefont
  {Toschi}}, \bibinfo {author} {\bibfnamefont {A.~A.}\ \bibnamefont
  {Katanin}},\ and\ \bibinfo {author} {\bibfnamefont {K.}~\bibnamefont
  {Held}},\ }\bibfield  {title} {\bibinfo {title} {{Dynamical vertex
  approximation: A step beyond dynamical mean-field theory}},\ }\href
  {https://doi.org/10.1103/PhysRevB.75.045118} {\bibfield  {journal} {\bibinfo
  {journal} {Phys. Rev. B}\ }\textbf {\bibinfo {volume} {75}},\ \bibinfo
  {pages} {045118} (\bibinfo {year} {2007})}\BibitemShut {NoStop}%
\bibitem [{\citenamefont {Galler}\ \emph {et~al.}(2017)\citenamefont {Galler},
  \citenamefont {Thunstr\"om}, \citenamefont {Gunacker}, \citenamefont
  {Tomczak},\ and\ \citenamefont {Held}}]{PhysRevB.95.115107}%
  \BibitemOpen
  \bibfield  {author} {\bibinfo {author} {\bibfnamefont {A.}~\bibnamefont
  {Galler}}, \bibinfo {author} {\bibfnamefont {P.}~\bibnamefont {Thunstr\"om}},
  \bibinfo {author} {\bibfnamefont {P.}~\bibnamefont {Gunacker}}, \bibinfo
  {author} {\bibfnamefont {J.~M.}\ \bibnamefont {Tomczak}},\ and\ \bibinfo
  {author} {\bibfnamefont {K.}~\bibnamefont {Held}},\ }\bibfield  {title}
  {\bibinfo {title} {{Ab initio dynamical vertex approximation}},\ }\href
  {https://doi.org/10.1103/PhysRevB.95.115107} {\bibfield  {journal} {\bibinfo
  {journal} {Phys. Rev. B}\ }\textbf {\bibinfo {volume} {95}},\ \bibinfo
  {pages} {115107} (\bibinfo {year} {2017})}\BibitemShut {NoStop}%
\bibitem [{\citenamefont {Kresse}\ and\ \citenamefont
  {Joubert}(1999)}]{PhysRevB.59.1758}%
  \BibitemOpen
  \bibfield  {author} {\bibinfo {author} {\bibfnamefont {G.}~\bibnamefont
  {Kresse}}\ and\ \bibinfo {author} {\bibfnamefont {D.}~\bibnamefont
  {Joubert}},\ }\bibfield  {title} {\bibinfo {title} {{From ultrasoft
  pseudopotentials to the projector augmented-wave method}},\ }\href
  {https://doi.org/10.1103/PhysRevB.59.1758} {\bibfield  {journal} {\bibinfo
  {journal} {Phys. Rev. B}\ }\textbf {\bibinfo {volume} {59}},\ \bibinfo
  {pages} {1758} (\bibinfo {year} {1999})}\BibitemShut {NoStop}%
\bibitem [{\citenamefont {Kresse}\ and\ \citenamefont
  {Furthm\"uller}(1996)}]{PhysRevB.54.11169}%
  \BibitemOpen
  \bibfield  {author} {\bibinfo {author} {\bibfnamefont {G.}~\bibnamefont
  {Kresse}}\ and\ \bibinfo {author} {\bibfnamefont {J.}~\bibnamefont
  {Furthm\"uller}},\ }\bibfield  {title} {\bibinfo {title} {{Efficient
  iterative schemes for ab initio total-energy calculations using a plane-wave
  basis set}},\ }\href {https://doi.org/10.1103/PhysRevB.54.11169} {\bibfield
  {journal} {\bibinfo  {journal} {Phys. Rev. B}\ }\textbf {\bibinfo {volume}
  {54}},\ \bibinfo {pages} {11169} (\bibinfo {year} {1996})}\BibitemShut
  {NoStop}%
\bibitem [{\citenamefont {Giannozzi}\ \emph {et~al.}(2009)\citenamefont
  {Giannozzi}, \citenamefont {Baroni}, \citenamefont {Bonini}, \citenamefont
  {Calandra}, \citenamefont {Car}, \citenamefont {Cavazzoni}, \citenamefont
  {Ceresoli}, \citenamefont {Chiarotti}, \citenamefont {Cococcioni},
  \citenamefont {Dabo}, \citenamefont {Corso}, \citenamefont {de~Gironcoli},
  \citenamefont {Fabris}, \citenamefont {Fratesi}, \citenamefont {Gebauer},
  \citenamefont {Gerstmann}, \citenamefont {Gougoussis}, \citenamefont
  {Kokalj}, \citenamefont {Lazzeri}, \citenamefont {Martin-Samos},
  \citenamefont {Marzari}, \citenamefont {Mauri}, \citenamefont {Mazzarello},
  \citenamefont {Paolini}, \citenamefont {Pasquarello}, \citenamefont
  {Paulatto}, \citenamefont {Sbraccia}, \citenamefont {Scandolo}, \citenamefont
  {Sclauzero}, \citenamefont {Seitsonen}, \citenamefont {Smogunov},
  \citenamefont {Umari},\ and\ \citenamefont {Wentzcovitch}}]{Giannozzi_2009}%
  \BibitemOpen
  \bibfield  {author} {\bibinfo {author} {\bibfnamefont {P.}~\bibnamefont
  {Giannozzi}}, \bibinfo {author} {\bibfnamefont {S.}~\bibnamefont {Baroni}},
  \bibinfo {author} {\bibfnamefont {N.}~\bibnamefont {Bonini}}, \bibinfo
  {author} {\bibfnamefont {M.}~\bibnamefont {Calandra}}, \bibinfo {author}
  {\bibfnamefont {R.}~\bibnamefont {Car}}, \bibinfo {author} {\bibfnamefont
  {C.}~\bibnamefont {Cavazzoni}}, \bibinfo {author} {\bibfnamefont
  {D.}~\bibnamefont {Ceresoli}}, \bibinfo {author} {\bibfnamefont {G.~L.}\
  \bibnamefont {Chiarotti}}, \bibinfo {author} {\bibfnamefont {M.}~\bibnamefont
  {Cococcioni}}, \bibinfo {author} {\bibfnamefont {I.}~\bibnamefont {Dabo}},
  \bibinfo {author} {\bibfnamefont {A.~D.}\ \bibnamefont {Corso}}, \bibinfo
  {author} {\bibfnamefont {S.}~\bibnamefont {de~Gironcoli}}, \bibinfo {author}
  {\bibfnamefont {S.}~\bibnamefont {Fabris}}, \bibinfo {author} {\bibfnamefont
  {G.}~\bibnamefont {Fratesi}}, \bibinfo {author} {\bibfnamefont
  {R.}~\bibnamefont {Gebauer}}, \bibinfo {author} {\bibfnamefont
  {U.}~\bibnamefont {Gerstmann}}, \bibinfo {author} {\bibfnamefont
  {C.}~\bibnamefont {Gougoussis}}, \bibinfo {author} {\bibfnamefont
  {A.}~\bibnamefont {Kokalj}}, \bibinfo {author} {\bibfnamefont
  {M.}~\bibnamefont {Lazzeri}}, \bibinfo {author} {\bibfnamefont
  {L.}~\bibnamefont {Martin-Samos}}, \bibinfo {author} {\bibfnamefont
  {N.}~\bibnamefont {Marzari}}, \bibinfo {author} {\bibfnamefont
  {F.}~\bibnamefont {Mauri}}, \bibinfo {author} {\bibfnamefont
  {R.}~\bibnamefont {Mazzarello}}, \bibinfo {author} {\bibfnamefont
  {S.}~\bibnamefont {Paolini}}, \bibinfo {author} {\bibfnamefont
  {A.}~\bibnamefont {Pasquarello}}, \bibinfo {author} {\bibfnamefont
  {L.}~\bibnamefont {Paulatto}}, \bibinfo {author} {\bibfnamefont
  {C.}~\bibnamefont {Sbraccia}}, \bibinfo {author} {\bibfnamefont
  {S.}~\bibnamefont {Scandolo}}, \bibinfo {author} {\bibfnamefont
  {G.}~\bibnamefont {Sclauzero}}, \bibinfo {author} {\bibfnamefont {A.~P.}\
  \bibnamefont {Seitsonen}}, \bibinfo {author} {\bibfnamefont {A.}~\bibnamefont
  {Smogunov}}, \bibinfo {author} {\bibfnamefont {P.}~\bibnamefont {Umari}},\
  and\ \bibinfo {author} {\bibfnamefont {R.~M.}\ \bibnamefont {Wentzcovitch}},\
  }\bibfield  {title} {\bibinfo {title} {{QUANTUM ESPRESSO: a modular and
  open-source software project for quantum simulations of materials}},\ }\href
  {https://doi.org/10.1088/0953-8984/21/39/395502} {\bibfield  {journal}
  {\bibinfo  {journal} {J. Phys.: Condens. Matter}\ }\textbf {\bibinfo {volume}
  {21}},\ \bibinfo {pages} {395502} (\bibinfo {year} {2009})}\BibitemShut
  {NoStop}%
\bibitem [{\citenamefont {Giannozzi}\ \emph {et~al.}(2017)\citenamefont
  {Giannozzi}, \citenamefont {Andreussi}, \citenamefont {Brumme}, \citenamefont
  {Bunau}, \citenamefont {Nardelli}, \citenamefont {Calandra}, \citenamefont
  {Car}, \citenamefont {Cavazzoni}, \citenamefont {Ceresoli}, \citenamefont
  {Cococcioni}, \citenamefont {Colonna}, \citenamefont {Carnimeo},
  \citenamefont {Corso}, \citenamefont {de~Gironcoli}, \citenamefont {Delugas},
  \citenamefont {DiStasio}, \citenamefont {Ferretti}, \citenamefont {Floris},
  \citenamefont {Fratesi}, \citenamefont {Fugallo}, \citenamefont {Gebauer},
  \citenamefont {Gerstmann}, \citenamefont {Giustino}, \citenamefont {Gorni},
  \citenamefont {Jia}, \citenamefont {Kawamura}, \citenamefont {Ko},
  \citenamefont {Kokalj}, \citenamefont {Küçükbenli}, \citenamefont
  {Lazzeri}, \citenamefont {Marsili}, \citenamefont {Marzari}, \citenamefont
  {Mauri}, \citenamefont {Nguyen}, \citenamefont {Nguyen}, \citenamefont {de-la
  Roza}, \citenamefont {Paulatto}, \citenamefont {Poncé}, \citenamefont
  {Rocca}, \citenamefont {Sabatini}, \citenamefont {Santra}, \citenamefont
  {Schlipf}, \citenamefont {Seitsonen}, \citenamefont {Smogunov}, \citenamefont
  {Timrov}, \citenamefont {Thonhauser}, \citenamefont {Umari}, \citenamefont
  {Vast}, \citenamefont {Wu},\ and\ \citenamefont {Baroni}}]{Giannozzi_2017}%
  \BibitemOpen
  \bibfield  {author} {\bibinfo {author} {\bibfnamefont {P.}~\bibnamefont
  {Giannozzi}}, \bibinfo {author} {\bibfnamefont {O.}~\bibnamefont
  {Andreussi}}, \bibinfo {author} {\bibfnamefont {T.}~\bibnamefont {Brumme}},
  \bibinfo {author} {\bibfnamefont {O.}~\bibnamefont {Bunau}}, \bibinfo
  {author} {\bibfnamefont {M.~B.}\ \bibnamefont {Nardelli}}, \bibinfo {author}
  {\bibfnamefont {M.}~\bibnamefont {Calandra}}, \bibinfo {author}
  {\bibfnamefont {R.}~\bibnamefont {Car}}, \bibinfo {author} {\bibfnamefont
  {C.}~\bibnamefont {Cavazzoni}}, \bibinfo {author} {\bibfnamefont
  {D.}~\bibnamefont {Ceresoli}}, \bibinfo {author} {\bibfnamefont
  {M.}~\bibnamefont {Cococcioni}}, \bibinfo {author} {\bibfnamefont
  {N.}~\bibnamefont {Colonna}}, \bibinfo {author} {\bibfnamefont
  {I.}~\bibnamefont {Carnimeo}}, \bibinfo {author} {\bibfnamefont {A.~D.}\
  \bibnamefont {Corso}}, \bibinfo {author} {\bibfnamefont {S.}~\bibnamefont
  {de~Gironcoli}}, \bibinfo {author} {\bibfnamefont {P.}~\bibnamefont
  {Delugas}}, \bibinfo {author} {\bibfnamefont {R.~A.}\ \bibnamefont
  {DiStasio}}, \bibinfo {author} {\bibfnamefont {A.}~\bibnamefont {Ferretti}},
  \bibinfo {author} {\bibfnamefont {A.}~\bibnamefont {Floris}}, \bibinfo
  {author} {\bibfnamefont {G.}~\bibnamefont {Fratesi}}, \bibinfo {author}
  {\bibfnamefont {G.}~\bibnamefont {Fugallo}}, \bibinfo {author} {\bibfnamefont
  {R.}~\bibnamefont {Gebauer}}, \bibinfo {author} {\bibfnamefont
  {U.}~\bibnamefont {Gerstmann}}, \bibinfo {author} {\bibfnamefont
  {F.}~\bibnamefont {Giustino}}, \bibinfo {author} {\bibfnamefont
  {T.}~\bibnamefont {Gorni}}, \bibinfo {author} {\bibfnamefont
  {J.}~\bibnamefont {Jia}}, \bibinfo {author} {\bibfnamefont {M.}~\bibnamefont
  {Kawamura}}, \bibinfo {author} {\bibfnamefont {H.-Y.}\ \bibnamefont {Ko}},
  \bibinfo {author} {\bibfnamefont {A.}~\bibnamefont {Kokalj}}, \bibinfo
  {author} {\bibfnamefont {E.}~\bibnamefont {Küçükbenli}}, \bibinfo {author}
  {\bibfnamefont {M.}~\bibnamefont {Lazzeri}}, \bibinfo {author} {\bibfnamefont
  {M.}~\bibnamefont {Marsili}}, \bibinfo {author} {\bibfnamefont
  {N.}~\bibnamefont {Marzari}}, \bibinfo {author} {\bibfnamefont
  {F.}~\bibnamefont {Mauri}}, \bibinfo {author} {\bibfnamefont {N.~L.}\
  \bibnamefont {Nguyen}}, \bibinfo {author} {\bibfnamefont {H.-V.}\
  \bibnamefont {Nguyen}}, \bibinfo {author} {\bibfnamefont {A.~O.}\
  \bibnamefont {de-la Roza}}, \bibinfo {author} {\bibfnamefont
  {L.}~\bibnamefont {Paulatto}}, \bibinfo {author} {\bibfnamefont
  {S.}~\bibnamefont {Poncé}}, \bibinfo {author} {\bibfnamefont
  {D.}~\bibnamefont {Rocca}}, \bibinfo {author} {\bibfnamefont
  {R.}~\bibnamefont {Sabatini}}, \bibinfo {author} {\bibfnamefont
  {B.}~\bibnamefont {Santra}}, \bibinfo {author} {\bibfnamefont
  {M.}~\bibnamefont {Schlipf}}, \bibinfo {author} {\bibfnamefont {A.~P.}\
  \bibnamefont {Seitsonen}}, \bibinfo {author} {\bibfnamefont {A.}~\bibnamefont
  {Smogunov}}, \bibinfo {author} {\bibfnamefont {I.}~\bibnamefont {Timrov}},
  \bibinfo {author} {\bibfnamefont {T.}~\bibnamefont {Thonhauser}}, \bibinfo
  {author} {\bibfnamefont {P.}~\bibnamefont {Umari}}, \bibinfo {author}
  {\bibfnamefont {N.}~\bibnamefont {Vast}}, \bibinfo {author} {\bibfnamefont
  {X.}~\bibnamefont {Wu}},\ and\ \bibinfo {author} {\bibfnamefont
  {S.}~\bibnamefont {Baroni}},\ }\bibfield  {title} {\bibinfo {title}
  {{Advanced capabilities for materials modelling with Quantum ESPRESSO}},\
  }\href {https://doi.org/10.1088/1361-648X/aa8f79} {\bibfield  {journal}
  {\bibinfo  {journal} {J. Phys.: Condens. Matter}\ }\textbf {\bibinfo {volume}
  {29}},\ \bibinfo {pages} {465901} (\bibinfo {year} {2017})}\BibitemShut
  {NoStop}%
\bibitem [{\citenamefont {Pizzi}\ \emph {et~al.}(2020)\citenamefont {Pizzi},
  \citenamefont {Vitale}, \citenamefont {Arita}, \citenamefont {Blügel},
  \citenamefont {Freimuth}, \citenamefont {G{\'{e}}ranton}, \citenamefont
  {Gibertini}, \citenamefont {Gresch}, \citenamefont {Johnson}, \citenamefont
  {Koretsune}, \citenamefont {Iba{\~{n}}ez-Azpiroz}, \citenamefont {Lee},
  \citenamefont {Lihm}, \citenamefont {Marchand}, \citenamefont {Marrazzo},
  \citenamefont {Mokrousov}, \citenamefont {Mustafa}, \citenamefont {Nohara},
  \citenamefont {Nomura}, \citenamefont {Paulatto}, \citenamefont
  {Ponc{\'{e}}}, \citenamefont {Ponweiser}, \citenamefont {Qiao}, \citenamefont
  {Thöle}, \citenamefont {Tsirkin}, \citenamefont {Wierzbowska}, \citenamefont
  {Marzari}, \citenamefont {Vanderbilt}, \citenamefont {Souza}, \citenamefont
  {Mostofi},\ and\ \citenamefont {Yates}}]{Pizzi2020}%
  \BibitemOpen
  \bibfield  {author} {\bibinfo {author} {\bibfnamefont {G.}~\bibnamefont
  {Pizzi}}, \bibinfo {author} {\bibfnamefont {V.}~\bibnamefont {Vitale}},
  \bibinfo {author} {\bibfnamefont {R.}~\bibnamefont {Arita}}, \bibinfo
  {author} {\bibfnamefont {S.}~\bibnamefont {Blügel}}, \bibinfo {author}
  {\bibfnamefont {F.}~\bibnamefont {Freimuth}}, \bibinfo {author}
  {\bibfnamefont {G.}~\bibnamefont {G{\'{e}}ranton}}, \bibinfo {author}
  {\bibfnamefont {M.}~\bibnamefont {Gibertini}}, \bibinfo {author}
  {\bibfnamefont {D.}~\bibnamefont {Gresch}}, \bibinfo {author} {\bibfnamefont
  {C.}~\bibnamefont {Johnson}}, \bibinfo {author} {\bibfnamefont
  {T.}~\bibnamefont {Koretsune}}, \bibinfo {author} {\bibfnamefont
  {J.}~\bibnamefont {Iba{\~{n}}ez-Azpiroz}}, \bibinfo {author} {\bibfnamefont
  {H.}~\bibnamefont {Lee}}, \bibinfo {author} {\bibfnamefont {J.-M.}\
  \bibnamefont {Lihm}}, \bibinfo {author} {\bibfnamefont {D.}~\bibnamefont
  {Marchand}}, \bibinfo {author} {\bibfnamefont {A.}~\bibnamefont {Marrazzo}},
  \bibinfo {author} {\bibfnamefont {Y.}~\bibnamefont {Mokrousov}}, \bibinfo
  {author} {\bibfnamefont {J.~I.}\ \bibnamefont {Mustafa}}, \bibinfo {author}
  {\bibfnamefont {Y.}~\bibnamefont {Nohara}}, \bibinfo {author} {\bibfnamefont
  {Y.}~\bibnamefont {Nomura}}, \bibinfo {author} {\bibfnamefont
  {L.}~\bibnamefont {Paulatto}}, \bibinfo {author} {\bibfnamefont
  {S.}~\bibnamefont {Ponc{\'{e}}}}, \bibinfo {author} {\bibfnamefont
  {T.}~\bibnamefont {Ponweiser}}, \bibinfo {author} {\bibfnamefont
  {J.}~\bibnamefont {Qiao}}, \bibinfo {author} {\bibfnamefont {F.}~\bibnamefont
  {Thöle}}, \bibinfo {author} {\bibfnamefont {S.~S.}\ \bibnamefont {Tsirkin}},
  \bibinfo {author} {\bibfnamefont {M.}~\bibnamefont {Wierzbowska}}, \bibinfo
  {author} {\bibfnamefont {N.}~\bibnamefont {Marzari}}, \bibinfo {author}
  {\bibfnamefont {D.}~\bibnamefont {Vanderbilt}}, \bibinfo {author}
  {\bibfnamefont {I.}~\bibnamefont {Souza}}, \bibinfo {author} {\bibfnamefont
  {A.~A.}\ \bibnamefont {Mostofi}},\ and\ \bibinfo {author} {\bibfnamefont
  {J.~R.}\ \bibnamefont {Yates}},\ }\bibfield  {title} {\bibinfo {title}
  {{Wannier90 as a community code: new features and applications}},\ }\href
  {https://doi.org/10.1088/1361-648x/ab51ff} {\bibfield  {journal} {\bibinfo
  {journal} {J. Phys.: Condens. Matter}\ }\textbf {\bibinfo {volume} {32}},\
  \bibinfo {pages} {165902} (\bibinfo {year} {2020})}\BibitemShut {NoStop}%
\bibitem [{\citenamefont {He}\ and\ \citenamefont
  {Lu}(2014)}]{PhysRevB.89.085108}%
  \BibitemOpen
  \bibfield  {author} {\bibinfo {author} {\bibfnamefont {R.-Q.}\ \bibnamefont
  {He}}\ and\ \bibinfo {author} {\bibfnamefont {Z.-Y.}\ \bibnamefont {Lu}},\
  }\bibfield  {title} {\bibinfo {title} {{Quantum renormalization groups based
  on natural orbitals}},\ }\href {https://doi.org/10.1103/PhysRevB.89.085108}
  {\bibfield  {journal} {\bibinfo  {journal} {Phys. Rev. B}\ }\textbf {\bibinfo
  {volume} {89}},\ \bibinfo {pages} {085108} (\bibinfo {year}
  {2014})}\BibitemShut {NoStop}%
\bibitem [{\citenamefont {He}\ \emph {et~al.}(2015)\citenamefont {He},
  \citenamefont {Dai},\ and\ \citenamefont {Lu}}]{PhysRevB.91.155140}%
  \BibitemOpen
  \bibfield  {author} {\bibinfo {author} {\bibfnamefont {R.-Q.}\ \bibnamefont
  {He}}, \bibinfo {author} {\bibfnamefont {J.}~\bibnamefont {Dai}},\ and\
  \bibinfo {author} {\bibfnamefont {Z.-Y.}\ \bibnamefont {Lu}},\ }\bibfield
  {title} {\bibinfo {title} {{Natural orbitals renormalization group approach
  to the two-impurity Kondo critical point}},\ }\href
  {https://doi.org/10.1103/PhysRevB.91.155140} {\bibfield  {journal} {\bibinfo
  {journal} {Phys. Rev. B}\ }\textbf {\bibinfo {volume} {91}},\ \bibinfo
  {pages} {155140} (\bibinfo {year} {2015})}\BibitemShut {NoStop}%
\bibitem [{\citenamefont {Huang}\ \emph {et~al.}(2015)\citenamefont {Huang},
  \citenamefont {Wang}, \citenamefont {Meng}, \citenamefont {Du}, \citenamefont
  {Werner},\ and\ \citenamefont {Dai}}]{HUANG2015140}%
  \BibitemOpen
  \bibfield  {author} {\bibinfo {author} {\bibfnamefont {L.}~\bibnamefont
  {Huang}}, \bibinfo {author} {\bibfnamefont {Y.}~\bibnamefont {Wang}},
  \bibinfo {author} {\bibfnamefont {Z.~Y.}\ \bibnamefont {Meng}}, \bibinfo
  {author} {\bibfnamefont {L.}~\bibnamefont {Du}}, \bibinfo {author}
  {\bibfnamefont {P.}~\bibnamefont {Werner}},\ and\ \bibinfo {author}
  {\bibfnamefont {X.}~\bibnamefont {Dai}},\ }\bibfield  {title} {\bibinfo
  {title} {{iQIST: An open source continuous-time quantum Monte Carlo impurity
  solver toolkit}},\ }\href
  {https://doi.org/https://doi.org/10.1016/j.cpc.2015.04.020} {\bibfield
  {journal} {\bibinfo  {journal} {Comput. Phys. Commun.}\ }\textbf {\bibinfo
  {volume} {195}},\ \bibinfo {pages} {140} (\bibinfo {year}
  {2015})}\BibitemShut {NoStop}%
\bibitem [{\citenamefont {Huang}(2017)}]{HUANG2017423}%
  \BibitemOpen
  \bibfield  {author} {\bibinfo {author} {\bibfnamefont {L.}~\bibnamefont
  {Huang}},\ }\bibfield  {title} {\bibinfo {title} {{iQIST v0.7: An open source
  continuous-time quantum Monte Carlo impurity solver toolkit}},\ }\href
  {https://doi.org/https://doi.org/10.1016/j.cpc.2017.08.026} {\bibfield
  {journal} {\bibinfo  {journal} {Comput. Phys. Commun.}\ }\textbf {\bibinfo
  {volume} {221}},\ \bibinfo {pages} {423} (\bibinfo {year}
  {2017})}\BibitemShut {NoStop}%
\bibitem [{\citenamefont {Huang}(2023)}]{HUANG2023108863}%
  \BibitemOpen
  \bibfield  {author} {\bibinfo {author} {\bibfnamefont {L.}~\bibnamefont
  {Huang}},\ }\bibfield  {title} {\bibinfo {title} {{ACFlow: An open source
  toolkit for analytic continuation of quantum Monte Carlo data}},\ }\href
  {https://doi.org/https://doi.org/10.1016/j.cpc.2023.108863} {\bibfield
  {journal} {\bibinfo  {journal} {Comput. Phys. Commun.}\ }\textbf {\bibinfo
  {volume} {292}},\ \bibinfo {pages} {108863} (\bibinfo {year}
  {2023})}\BibitemShut {NoStop}%
\bibitem [{\citenamefont {Bl\"ochl}(1994)}]{PhysRevB.50.17953}%
  \BibitemOpen
  \bibfield  {author} {\bibinfo {author} {\bibfnamefont {P.~E.}\ \bibnamefont
  {Bl\"ochl}},\ }\bibfield  {title} {\bibinfo {title} {{Projector
  augmented-wave method}},\ }\href {https://doi.org/10.1103/PhysRevB.50.17953}
  {\bibfield  {journal} {\bibinfo  {journal} {Phys. Rev. B}\ }\textbf {\bibinfo
  {volume} {50}},\ \bibinfo {pages} {17953} (\bibinfo {year}
  {1994})}\BibitemShut {NoStop}%
\bibitem [{\citenamefont {Marzari}\ and\ \citenamefont
  {Vanderbilt}(1997)}]{PhysRevB.56.12847}%
  \BibitemOpen
  \bibfield  {author} {\bibinfo {author} {\bibfnamefont {N.}~\bibnamefont
  {Marzari}}\ and\ \bibinfo {author} {\bibfnamefont {D.}~\bibnamefont
  {Vanderbilt}},\ }\bibfield  {title} {\bibinfo {title} {{Maximally localized
  generalized Wannier functions for composite energy bands}},\ }\href
  {https://doi.org/10.1103/PhysRevB.56.12847} {\bibfield  {journal} {\bibinfo
  {journal} {Phys. Rev. B}\ }\textbf {\bibinfo {volume} {56}},\ \bibinfo
  {pages} {12847} (\bibinfo {year} {1997})}\BibitemShut {NoStop}%
\bibitem [{\citenamefont {Sakuma}(2013)}]{PhysRevB.87.235109}%
  \BibitemOpen
  \bibfield  {author} {\bibinfo {author} {\bibfnamefont {R.}~\bibnamefont
  {Sakuma}},\ }\bibfield  {title} {\bibinfo {title} {{Symmetry-adapted Wannier
  functions in the maximal localization procedure}},\ }\href
  {https://doi.org/10.1103/PhysRevB.87.235109} {\bibfield  {journal} {\bibinfo
  {journal} {Phys. Rev. B}\ }\textbf {\bibinfo {volume} {87}},\ \bibinfo
  {pages} {235109} (\bibinfo {year} {2013})}\BibitemShut {NoStop}%
\bibitem [{\citenamefont {Lambin}\ and\ \citenamefont
  {Vigneron}(1984)}]{PhysRevB.29.3430}%
  \BibitemOpen
  \bibfield  {author} {\bibinfo {author} {\bibfnamefont {P.}~\bibnamefont
  {Lambin}}\ and\ \bibinfo {author} {\bibfnamefont {J.~P.}\ \bibnamefont
  {Vigneron}},\ }\bibfield  {title} {\bibinfo {title} {{Computation of crystal
  Green's functions in the complex-energy plane with the use of the analytical
  tetrahedron method}},\ }\href {https://doi.org/10.1103/PhysRevB.29.3430}
  {\bibfield  {journal} {\bibinfo  {journal} {Phys. Rev. B}\ }\textbf {\bibinfo
  {volume} {29}},\ \bibinfo {pages} {3430} (\bibinfo {year}
  {1984})}\BibitemShut {NoStop}%
\bibitem [{\citenamefont {Amadon}\ \emph
  {et~al.}(2008{\natexlab{b}})\citenamefont {Amadon}, \citenamefont {Jollet},\
  and\ \citenamefont {Torrent}}]{PhysRevB.77.155104}%
  \BibitemOpen
  \bibfield  {author} {\bibinfo {author} {\bibfnamefont {B.}~\bibnamefont
  {Amadon}}, \bibinfo {author} {\bibfnamefont {F.}~\bibnamefont {Jollet}},\
  and\ \bibinfo {author} {\bibfnamefont {M.}~\bibnamefont {Torrent}},\
  }\bibfield  {title} {\bibinfo {title} {{$\ensuremath{\gamma}$ and
  $\ensuremath{\beta}$ cerium: $\text{LDA}+\text{U}$ calculations of
  ground-state parameters}},\ }\href
  {https://doi.org/10.1103/PhysRevB.77.155104} {\bibfield  {journal} {\bibinfo
  {journal} {Phys. Rev. B}\ }\textbf {\bibinfo {volume} {77}},\ \bibinfo
  {pages} {155104} (\bibinfo {year} {2008}{\natexlab{b}})}\BibitemShut
  {NoStop}%
\bibitem [{\citenamefont {Ouyang}\ \emph
  {et~al.}(2024{\natexlab{a}})\citenamefont {Ouyang}, \citenamefont {He},\ and\
  \citenamefont {Lu}}]{ouyangzhenfeng2024}%
  \BibitemOpen
  \bibfield  {author} {\bibinfo {author} {\bibfnamefont {Z.}~\bibnamefont
  {Ouyang}}, \bibinfo {author} {\bibfnamefont {R.-Q.}\ \bibnamefont {He}},\
  and\ \bibinfo {author} {\bibfnamefont {Z.-Y.}\ \bibnamefont {Lu}},\ }\href
  {https://arxiv.org/abs/2407.08601} {\bibinfo {title} {{DFT+DMFT study of
  correlated electronic structure in the monolayer-trilayer phase of
  La$_3$Ni$_2$O$_7$}}} (\bibinfo {year} {2024}{\natexlab{a}}),\ \Eprint
  {https://arxiv.org/abs/2407.08601} {arXiv:2407.08601 [cond-mat.str-el]}
  \BibitemShut {NoStop}%
\bibitem [{\citenamefont {S\'emon}\ \emph {et~al.}(2014)\citenamefont
  {S\'emon}, \citenamefont {Yee}, \citenamefont {Haule},\ and\ \citenamefont
  {Tremblay}}]{PhysRevB.90.075149}%
  \BibitemOpen
  \bibfield  {author} {\bibinfo {author} {\bibfnamefont {P.}~\bibnamefont
  {S\'emon}}, \bibinfo {author} {\bibfnamefont {C.-H.}\ \bibnamefont {Yee}},
  \bibinfo {author} {\bibfnamefont {K.}~\bibnamefont {Haule}},\ and\ \bibinfo
  {author} {\bibfnamefont {A.-M.~S.}\ \bibnamefont {Tremblay}},\ }\bibfield
  {title} {\bibinfo {title} {{Lazy skip-lists: An algorithm for fast
  hybridization-expansion quantum Monte Carlo}},\ }\href
  {https://doi.org/10.1103/PhysRevB.90.075149} {\bibfield  {journal} {\bibinfo
  {journal} {Phys. Rev. B}\ }\textbf {\bibinfo {volume} {90}},\ \bibinfo
  {pages} {075149} (\bibinfo {year} {2014})}\BibitemShut {NoStop}%
\bibitem [{\citenamefont {Boehnke}\ \emph {et~al.}(2011)\citenamefont
  {Boehnke}, \citenamefont {Hafermann}, \citenamefont {Ferrero}, \citenamefont
  {Lechermann},\ and\ \citenamefont {Parcollet}}]{PhysRevB.84.075145}%
  \BibitemOpen
  \bibfield  {author} {\bibinfo {author} {\bibfnamefont {L.}~\bibnamefont
  {Boehnke}}, \bibinfo {author} {\bibfnamefont {H.}~\bibnamefont {Hafermann}},
  \bibinfo {author} {\bibfnamefont {M.}~\bibnamefont {Ferrero}}, \bibinfo
  {author} {\bibfnamefont {F.}~\bibnamefont {Lechermann}},\ and\ \bibinfo
  {author} {\bibfnamefont {O.}~\bibnamefont {Parcollet}},\ }\bibfield  {title}
  {\bibinfo {title} {{Orthogonal polynomial representation of imaginary-time
  Green's functions}},\ }\href {https://doi.org/10.1103/PhysRevB.84.075145}
  {\bibfield  {journal} {\bibinfo  {journal} {Phys. Rev. B}\ }\textbf {\bibinfo
  {volume} {84}},\ \bibinfo {pages} {075145} (\bibinfo {year}
  {2011})}\BibitemShut {NoStop}%
\bibitem [{\citenamefont {Shinaoka}\ \emph
  {et~al.}(2017{\natexlab{b}})\citenamefont {Shinaoka}, \citenamefont {Otsuki},
  \citenamefont {Ohzeki},\ and\ \citenamefont {Yoshimi}}]{PhysRevB.96.035147}%
  \BibitemOpen
  \bibfield  {author} {\bibinfo {author} {\bibfnamefont {H.}~\bibnamefont
  {Shinaoka}}, \bibinfo {author} {\bibfnamefont {J.}~\bibnamefont {Otsuki}},
  \bibinfo {author} {\bibfnamefont {M.}~\bibnamefont {Ohzeki}},\ and\ \bibinfo
  {author} {\bibfnamefont {K.}~\bibnamefont {Yoshimi}},\ }\bibfield  {title}
  {\bibinfo {title} {{Compressing Green's function using intermediate
  representation between imaginary-time and real-frequency domains}},\ }\href
  {https://doi.org/10.1103/PhysRevB.96.035147} {\bibfield  {journal} {\bibinfo
  {journal} {Phys. Rev. B}\ }\textbf {\bibinfo {volume} {96}},\ \bibinfo
  {pages} {035147} (\bibinfo {year} {2017}{\natexlab{b}})}\BibitemShut
  {NoStop}%
\bibitem [{\citenamefont {Hafermann}\ \emph {et~al.}(2012)\citenamefont
  {Hafermann}, \citenamefont {Patton},\ and\ \citenamefont
  {Werner}}]{PhysRevB.85.205106}%
  \BibitemOpen
  \bibfield  {author} {\bibinfo {author} {\bibfnamefont {H.}~\bibnamefont
  {Hafermann}}, \bibinfo {author} {\bibfnamefont {K.~R.}\ \bibnamefont
  {Patton}},\ and\ \bibinfo {author} {\bibfnamefont {P.}~\bibnamefont
  {Werner}},\ }\bibfield  {title} {\bibinfo {title} {{Improved estimators for
  the self-energy and vertex function in hybridization-expansion
  continuous-time quantum Monte Carlo simulations}},\ }\href
  {https://doi.org/10.1103/PhysRevB.85.205106} {\bibfield  {journal} {\bibinfo
  {journal} {Phys. Rev. B}\ }\textbf {\bibinfo {volume} {85}},\ \bibinfo
  {pages} {205106} (\bibinfo {year} {2012})}\BibitemShut {NoStop}%
\bibitem [{\citenamefont {Jarrell}\ and\ \citenamefont
  {Gubernatis}(1996)}]{JARRELL1996133}%
  \BibitemOpen
  \bibfield  {author} {\bibinfo {author} {\bibfnamefont {M.}~\bibnamefont
  {Jarrell}}\ and\ \bibinfo {author} {\bibfnamefont {J.}~\bibnamefont
  {Gubernatis}},\ }\bibfield  {title} {\bibinfo {title} {{Bayesian inference
  and the analytic continuation of imaginary-time quantum Monte Carlo data}},\
  }\href {https://doi.org/https://doi.org/10.1016/0370-1573(95)00074-7}
  {\bibfield  {journal} {\bibinfo  {journal} {Phys. Rep.}\ }\textbf {\bibinfo
  {volume} {269}},\ \bibinfo {pages} {133} (\bibinfo {year}
  {1996})}\BibitemShut {NoStop}%
\bibitem [{\citenamefont {Fei}\ \emph {et~al.}(2021)\citenamefont {Fei},
  \citenamefont {Yeh},\ and\ \citenamefont {Gull}}]{PhysRevLett.126.056402}%
  \BibitemOpen
  \bibfield  {author} {\bibinfo {author} {\bibfnamefont {J.}~\bibnamefont
  {Fei}}, \bibinfo {author} {\bibfnamefont {C.-N.}\ \bibnamefont {Yeh}},\ and\
  \bibinfo {author} {\bibfnamefont {E.}~\bibnamefont {Gull}},\ }\bibfield
  {title} {\bibinfo {title} {{Nevanlinna Analytical Continuation}},\ }\href
  {https://doi.org/10.1103/PhysRevLett.126.056402} {\bibfield  {journal}
  {\bibinfo  {journal} {Phys. Rev. Lett.}\ }\textbf {\bibinfo {volume} {126}},\
  \bibinfo {pages} {056402} (\bibinfo {year} {2021})}\BibitemShut {NoStop}%
\bibitem [{\citenamefont {Sandvik}(1998)}]{PhysRevB.57.10287}%
  \BibitemOpen
  \bibfield  {author} {\bibinfo {author} {\bibfnamefont {A.~W.}\ \bibnamefont
  {Sandvik}},\ }\bibfield  {title} {\bibinfo {title} {{Stochastic method for
  analytic continuation of quantum Monte Carlo data}},\ }\href
  {https://doi.org/10.1103/PhysRevB.57.10287} {\bibfield  {journal} {\bibinfo
  {journal} {Phys. Rev. B}\ }\textbf {\bibinfo {volume} {57}},\ \bibinfo
  {pages} {10287} (\bibinfo {year} {1998})}\BibitemShut {NoStop}%
\bibitem [{\citenamefont {Shao}\ and\ \citenamefont
  {Sandvik}(2023)}]{SHAO20231}%
  \BibitemOpen
  \bibfield  {author} {\bibinfo {author} {\bibfnamefont {H.}~\bibnamefont
  {Shao}}\ and\ \bibinfo {author} {\bibfnamefont {A.~W.}\ \bibnamefont
  {Sandvik}},\ }\bibfield  {title} {\bibinfo {title} {{Progress on stochastic
  analytic continuation of quantum Monte Carlo data}},\ }\href
  {https://doi.org/https://doi.org/10.1016/j.physrep.2022.11.002} {\bibfield
  {journal} {\bibinfo  {journal} {Phys. Rep.}\ }\textbf {\bibinfo {volume}
  {1003}},\ \bibinfo {pages} {1} (\bibinfo {year} {2023})}\BibitemShut
  {NoStop}%
\bibitem [{\citenamefont {Mishchenko}\ \emph {et~al.}(2000)\citenamefont
  {Mishchenko}, \citenamefont {Prokof'ev}, \citenamefont {Sakamoto},\ and\
  \citenamefont {Svistunov}}]{PhysRevB.62.6317}%
  \BibitemOpen
  \bibfield  {author} {\bibinfo {author} {\bibfnamefont {A.~S.}\ \bibnamefont
  {Mishchenko}}, \bibinfo {author} {\bibfnamefont {N.~V.}\ \bibnamefont
  {Prokof'ev}}, \bibinfo {author} {\bibfnamefont {A.}~\bibnamefont
  {Sakamoto}},\ and\ \bibinfo {author} {\bibfnamefont {B.~V.}\ \bibnamefont
  {Svistunov}},\ }\bibfield  {title} {\bibinfo {title} {{Diagrammatic quantum
  Monte Carlo study of the Fr\"ohlich polaron}},\ }\href
  {https://doi.org/10.1103/PhysRevB.62.6317} {\bibfield  {journal} {\bibinfo
  {journal} {Phys. Rev. B}\ }\textbf {\bibinfo {volume} {62}},\ \bibinfo
  {pages} {6317} (\bibinfo {year} {2000})}\BibitemShut {NoStop}%
\bibitem [{\citenamefont {Huang}\ and\ \citenamefont
  {Liang}(2023)}]{PhysRevB.108.235143}%
  \BibitemOpen
  \bibfield  {author} {\bibinfo {author} {\bibfnamefont {L.}~\bibnamefont
  {Huang}}\ and\ \bibinfo {author} {\bibfnamefont {S.}~\bibnamefont {Liang}},\
  }\bibfield  {title} {\bibinfo {title} {{Stochastic pole expansion method for
  analytic continuation of the Green's function}},\ }\href
  {https://doi.org/10.1103/PhysRevB.108.235143} {\bibfield  {journal} {\bibinfo
   {journal} {Phys. Rev. B}\ }\textbf {\bibinfo {volume} {108}},\ \bibinfo
  {pages} {235143} (\bibinfo {year} {2023})}\BibitemShut {NoStop}%
\bibitem [{\citenamefont {Huang}\ and\ \citenamefont
  {Liang}(2024)}]{PhysRevD.109.054508}%
  \BibitemOpen
  \bibfield  {author} {\bibinfo {author} {\bibfnamefont {L.}~\bibnamefont
  {Huang}}\ and\ \bibinfo {author} {\bibfnamefont {S.}~\bibnamefont {Liang}},\
  }\bibfield  {title} {\bibinfo {title} {{Reconstructing lattice QCD spectral
  functions with stochastic pole expansion and Nevanlinna analytic
  continuation}},\ }\href {https://doi.org/10.1103/PhysRevD.109.054508}
  {\bibfield  {journal} {\bibinfo  {journal} {Phys. Rev. D}\ }\textbf {\bibinfo
  {volume} {109}},\ \bibinfo {pages} {054508} (\bibinfo {year}
  {2024})}\BibitemShut {NoStop}%
\bibitem [{\citenamefont {Yoshida}\ \emph {et~al.}(2016)\citenamefont
  {Yoshida}, \citenamefont {Kobayashi}, \citenamefont {Yoshimatsu},
  \citenamefont {Kumigashira},\ and\ \citenamefont {Fujimori}}]{YOSHIDA201611}%
  \BibitemOpen
  \bibfield  {author} {\bibinfo {author} {\bibfnamefont {T.}~\bibnamefont
  {Yoshida}}, \bibinfo {author} {\bibfnamefont {M.}~\bibnamefont {Kobayashi}},
  \bibinfo {author} {\bibfnamefont {K.}~\bibnamefont {Yoshimatsu}}, \bibinfo
  {author} {\bibfnamefont {H.}~\bibnamefont {Kumigashira}},\ and\ \bibinfo
  {author} {\bibfnamefont {A.}~\bibnamefont {Fujimori}},\ }\bibfield  {title}
  {\bibinfo {title} {{Correlated electronic states of SrVO$_{3}$ revealed by
  angle-resolved photoemission spectroscopy}},\ }\href
  {https://doi.org/https://doi.org/10.1016/j.elspec.2015.11.012} {\bibfield
  {journal} {\bibinfo  {journal} {J. Electron Spectros. Relat. Phenomena}\
  }\textbf {\bibinfo {volume} {208}},\ \bibinfo {pages} {11} (\bibinfo {year}
  {2016})}\BibitemShut {NoStop}%
\bibitem [{\citenamefont {van Elp}\ \emph {et~al.}(1991)\citenamefont {van
  Elp}, \citenamefont {Potze}, \citenamefont {Eskes}, \citenamefont {Berger},\
  and\ \citenamefont {Sawatzky}}]{PhysRevB.44.1530}%
  \BibitemOpen
  \bibfield  {author} {\bibinfo {author} {\bibfnamefont {J.}~\bibnamefont {van
  Elp}}, \bibinfo {author} {\bibfnamefont {R.~H.}\ \bibnamefont {Potze}},
  \bibinfo {author} {\bibfnamefont {H.}~\bibnamefont {Eskes}}, \bibinfo
  {author} {\bibfnamefont {R.}~\bibnamefont {Berger}},\ and\ \bibinfo {author}
  {\bibfnamefont {G.~A.}\ \bibnamefont {Sawatzky}},\ }\bibfield  {title}
  {\bibinfo {title} {{Electronic structure of MnO}},\ }\href
  {https://doi.org/10.1103/PhysRevB.44.1530} {\bibfield  {journal} {\bibinfo
  {journal} {Phys. Rev. B}\ }\textbf {\bibinfo {volume} {44}},\ \bibinfo
  {pages} {1530} (\bibinfo {year} {1991})}\BibitemShut {NoStop}%
\bibitem [{\citenamefont {Perdew}\ \emph {et~al.}(1996)\citenamefont {Perdew},
  \citenamefont {Burke},\ and\ \citenamefont
  {Ernzerhof}}]{PhysRevLett.77.3865}%
  \BibitemOpen
  \bibfield  {author} {\bibinfo {author} {\bibfnamefont {J.~P.}\ \bibnamefont
  {Perdew}}, \bibinfo {author} {\bibfnamefont {K.}~\bibnamefont {Burke}},\ and\
  \bibinfo {author} {\bibfnamefont {M.}~\bibnamefont {Ernzerhof}},\ }\bibfield
  {title} {\bibinfo {title} {{Generalized Gradient Approximation Made
  Simple}},\ }\href {https://doi.org/10.1103/PhysRevLett.77.3865} {\bibfield
  {journal} {\bibinfo  {journal} {Phys. Rev. Lett.}\ }\textbf {\bibinfo
  {volume} {77}},\ \bibinfo {pages} {3865} (\bibinfo {year}
  {1996})}\BibitemShut {NoStop}%
\bibitem [{\citenamefont {Nekrasov}\ \emph {et~al.}(2006)\citenamefont
  {Nekrasov}, \citenamefont {Held}, \citenamefont {Keller}, \citenamefont
  {Kondakov}, \citenamefont {Pruschke}, \citenamefont {Kollar}, \citenamefont
  {Andersen}, \citenamefont {Anisimov},\ and\ \citenamefont
  {Vollhardt}}]{PhysRevB.73.155112}%
  \BibitemOpen
  \bibfield  {author} {\bibinfo {author} {\bibfnamefont {I.~A.}\ \bibnamefont
  {Nekrasov}}, \bibinfo {author} {\bibfnamefont {K.}~\bibnamefont {Held}},
  \bibinfo {author} {\bibfnamefont {G.}~\bibnamefont {Keller}}, \bibinfo
  {author} {\bibfnamefont {D.~E.}\ \bibnamefont {Kondakov}}, \bibinfo {author}
  {\bibfnamefont {T.}~\bibnamefont {Pruschke}}, \bibinfo {author}
  {\bibfnamefont {M.}~\bibnamefont {Kollar}}, \bibinfo {author} {\bibfnamefont
  {O.~K.}\ \bibnamefont {Andersen}}, \bibinfo {author} {\bibfnamefont {V.~I.}\
  \bibnamefont {Anisimov}},\ and\ \bibinfo {author} {\bibfnamefont
  {D.}~\bibnamefont {Vollhardt}},\ }\bibfield  {title} {\bibinfo {title}
  {{Momentum-resolved spectral functions of ${\mathrm{SrVO}}_{3}$ calculated by
  $\mathrm{LDA}+\mathrm{DMFT}$}},\ }\href
  {https://doi.org/10.1103/PhysRevB.73.155112} {\bibfield  {journal} {\bibinfo
  {journal} {Phys. Rev. B}\ }\textbf {\bibinfo {volume} {73}},\ \bibinfo
  {pages} {155112} (\bibinfo {year} {2006})}\BibitemShut {NoStop}%
\bibitem [{\citenamefont {Kune{\v s}}\ \emph {et~al.}(2008)\citenamefont
  {Kune{\v s}}, \citenamefont {Lukoyanov}, \citenamefont {Anisimov},
  \citenamefont {Scalettar},\ and\ \citenamefont {Pickett}}]{Kunes2008}%
  \BibitemOpen
  \bibfield  {author} {\bibinfo {author} {\bibfnamefont {J.}~\bibnamefont
  {Kune{\v s}}}, \bibinfo {author} {\bibfnamefont {A.~V.}\ \bibnamefont
  {Lukoyanov}}, \bibinfo {author} {\bibfnamefont {V.~I.}\ \bibnamefont
  {Anisimov}}, \bibinfo {author} {\bibfnamefont {R.~T.}\ \bibnamefont
  {Scalettar}},\ and\ \bibinfo {author} {\bibfnamefont {W.~E.}\ \bibnamefont
  {Pickett}},\ }\bibfield  {title} {\bibinfo {title} {{Collapse of magnetic
  moment drives the Mott transition in MnO}},\ }\href
  {https://doi.org/10.1038/nmat2115} {\bibfield  {journal} {\bibinfo  {journal}
  {Nat. Mater.}\ }\textbf {\bibinfo {volume} {7}},\ \bibinfo {pages} {198}
  (\bibinfo {year} {2008})}\BibitemShut {NoStop}%
\bibitem [{\citenamefont {Ouyang}\ \emph
  {et~al.}(2024{\natexlab{b}})\citenamefont {Ouyang}, \citenamefont {Wang},
  \citenamefont {Wang}, \citenamefont {He}, \citenamefont {Huang},\ and\
  \citenamefont {Lu}}]{PhysRevB.109.115114}%
  \BibitemOpen
  \bibfield  {author} {\bibinfo {author} {\bibfnamefont {Z.}~\bibnamefont
  {Ouyang}}, \bibinfo {author} {\bibfnamefont {J.-M.}\ \bibnamefont {Wang}},
  \bibinfo {author} {\bibfnamefont {J.-X.}\ \bibnamefont {Wang}}, \bibinfo
  {author} {\bibfnamefont {R.-Q.}\ \bibnamefont {He}}, \bibinfo {author}
  {\bibfnamefont {L.}~\bibnamefont {Huang}},\ and\ \bibinfo {author}
  {\bibfnamefont {Z.-Y.}\ \bibnamefont {Lu}},\ }\bibfield  {title} {\bibinfo
  {title} {{Hund electronic correlation in
  ${\mathrm{La}}_{3}{\mathrm{Ni}}_{2}{\mathrm{O}}_{7}$ under high pressure}},\
  }\href {https://doi.org/10.1103/PhysRevB.109.115114} {\bibfield  {journal}
  {\bibinfo  {journal} {Phys. Rev. B}\ }\textbf {\bibinfo {volume} {109}},\
  \bibinfo {pages} {115114} (\bibinfo {year} {2024}{\natexlab{b}})}\BibitemShut
  {NoStop}%
\bibitem [{\citenamefont {Chen}\ \emph {et~al.}(2024)\citenamefont {Chen},
  \citenamefont {Tian}, \citenamefont {Wang}, \citenamefont {He},\ and\
  \citenamefont {Lu}}]{Chen2407.13737}%
  \BibitemOpen
  \bibfield  {author} {\bibinfo {author} {\bibfnamefont {Y.}~\bibnamefont
  {Chen}}, \bibinfo {author} {\bibfnamefont {Y.-H.}\ \bibnamefont {Tian}},
  \bibinfo {author} {\bibfnamefont {J.-M.}\ \bibnamefont {Wang}}, \bibinfo
  {author} {\bibfnamefont {R.-Q.}\ \bibnamefont {He}},\ and\ \bibinfo {author}
  {\bibfnamefont {Z.-Y.}\ \bibnamefont {Lu}},\ }\href
  {https://arxiv.org/abs/2407.13737} {\bibinfo {title} {{Non-Fermi liquid and
  antiferromagnetic correlations with hole doping in the bilayer two-orbital
  Hubbard model of La$_3$Ni$_2$O$_7$ at zero temperature}}} (\bibinfo {year}
  {2024}),\ \Eprint {https://arxiv.org/abs/2407.13737} {arXiv:2407.13737
  [cond-mat.str-el]} \BibitemShut {NoStop}%
\bibitem [{\citenamefont {Rey}\ \emph {et~al.}(1990)\citenamefont {Rey},
  \citenamefont {Dehaudt}, \citenamefont {Joubert}, \citenamefont
  {Lambert-Andron}, \citenamefont {Cyrot},\ and\ \citenamefont
  {Cyrot-Lackmann}}]{REY1990101}%
  \BibitemOpen
  \bibfield  {author} {\bibinfo {author} {\bibfnamefont {M.}~\bibnamefont
  {Rey}}, \bibinfo {author} {\bibfnamefont {P.}~\bibnamefont {Dehaudt}},
  \bibinfo {author} {\bibfnamefont {J.}~\bibnamefont {Joubert}}, \bibinfo
  {author} {\bibfnamefont {B.}~\bibnamefont {Lambert-Andron}}, \bibinfo
  {author} {\bibfnamefont {M.}~\bibnamefont {Cyrot}},\ and\ \bibinfo {author}
  {\bibfnamefont {F.}~\bibnamefont {Cyrot-Lackmann}},\ }\bibfield  {title}
  {\bibinfo {title} {{Preparation and structure of the compounds SrVO$_{3}$ and
  Sr$_{2}$VO$_{4}$}},\ }\href
  {https://doi.org/https://doi.org/10.1016/0022-4596(90)90119-I} {\bibfield
  {journal} {\bibinfo  {journal} {J. Solid State Chem.}\ }\textbf {\bibinfo
  {volume} {86}},\ \bibinfo {pages} {101} (\bibinfo {year} {1990})}\BibitemShut
  {NoStop}%
\bibitem [{\citenamefont {Nekrasov}\ \emph {et~al.}(2005)\citenamefont
  {Nekrasov}, \citenamefont {Keller}, \citenamefont {Kondakov}, \citenamefont
  {Kozhevnikov}, \citenamefont {Pruschke}, \citenamefont {Held}, \citenamefont
  {Vollhardt},\ and\ \citenamefont {Anisimov}}]{PhysRevB.72.155106}%
  \BibitemOpen
  \bibfield  {author} {\bibinfo {author} {\bibfnamefont {I.~A.}\ \bibnamefont
  {Nekrasov}}, \bibinfo {author} {\bibfnamefont {G.}~\bibnamefont {Keller}},
  \bibinfo {author} {\bibfnamefont {D.~E.}\ \bibnamefont {Kondakov}}, \bibinfo
  {author} {\bibfnamefont {A.~V.}\ \bibnamefont {Kozhevnikov}}, \bibinfo
  {author} {\bibfnamefont {T.}~\bibnamefont {Pruschke}}, \bibinfo {author}
  {\bibfnamefont {K.}~\bibnamefont {Held}}, \bibinfo {author} {\bibfnamefont
  {D.}~\bibnamefont {Vollhardt}},\ and\ \bibinfo {author} {\bibfnamefont
  {V.~I.}\ \bibnamefont {Anisimov}},\ }\bibfield  {title} {\bibinfo {title}
  {{Comparative study of correlation effects in
  $\mathrm{Ca}\mathrm{V}{\mathrm{O}}_{3}$ and
  $\mathrm{Sr}\mathrm{V}{\mathrm{O}}_{3}$}},\ }\href
  {https://doi.org/10.1103/PhysRevB.72.155106} {\bibfield  {journal} {\bibinfo
  {journal} {Phys. Rev. B}\ }\textbf {\bibinfo {volume} {72}},\ \bibinfo
  {pages} {155106} (\bibinfo {year} {2005})}\BibitemShut {NoStop}%
\bibitem [{\citenamefont {Huang}\ and\ \citenamefont
  {Wang}(2012)}]{Huang_2012}%
  \BibitemOpen
  \bibfield  {author} {\bibinfo {author} {\bibfnamefont {L.}~\bibnamefont
  {Huang}}\ and\ \bibinfo {author} {\bibfnamefont {Y.}~\bibnamefont {Wang}},\
  }\bibfield  {title} {\bibinfo {title} {{Dynamical screening in strongly
  correlated metal SrVO$_{3}$}},\ }\href
  {https://doi.org/10.1209/0295-5075/99/67003} {\bibfield  {journal} {\bibinfo
  {journal} {Europhys. Lett.}\ }\textbf {\bibinfo {volume} {99}},\ \bibinfo
  {pages} {67003} (\bibinfo {year} {2012})}\BibitemShut {NoStop}%
\bibitem [{\citenamefont {Sekiyama}\ \emph {et~al.}(2004)\citenamefont
  {Sekiyama}, \citenamefont {Fujiwara}, \citenamefont {Imada}, \citenamefont
  {Suga}, \citenamefont {Eisaki}, \citenamefont {Uchida}, \citenamefont
  {Takegahara}, \citenamefont {Harima}, \citenamefont {Saitoh}, \citenamefont
  {Nekrasov}, \citenamefont {Keller}, \citenamefont {Kondakov}, \citenamefont
  {Kozhevnikov}, \citenamefont {Pruschke}, \citenamefont {Held}, \citenamefont
  {Vollhardt},\ and\ \citenamefont {Anisimov}}]{PhysRevLett.93.156402}%
  \BibitemOpen
  \bibfield  {author} {\bibinfo {author} {\bibfnamefont {A.}~\bibnamefont
  {Sekiyama}}, \bibinfo {author} {\bibfnamefont {H.}~\bibnamefont {Fujiwara}},
  \bibinfo {author} {\bibfnamefont {S.}~\bibnamefont {Imada}}, \bibinfo
  {author} {\bibfnamefont {S.}~\bibnamefont {Suga}}, \bibinfo {author}
  {\bibfnamefont {H.}~\bibnamefont {Eisaki}}, \bibinfo {author} {\bibfnamefont
  {S.~I.}\ \bibnamefont {Uchida}}, \bibinfo {author} {\bibfnamefont
  {K.}~\bibnamefont {Takegahara}}, \bibinfo {author} {\bibfnamefont
  {H.}~\bibnamefont {Harima}}, \bibinfo {author} {\bibfnamefont
  {Y.}~\bibnamefont {Saitoh}}, \bibinfo {author} {\bibfnamefont {I.~A.}\
  \bibnamefont {Nekrasov}}, \bibinfo {author} {\bibfnamefont {G.}~\bibnamefont
  {Keller}}, \bibinfo {author} {\bibfnamefont {D.~E.}\ \bibnamefont
  {Kondakov}}, \bibinfo {author} {\bibfnamefont {A.~V.}\ \bibnamefont
  {Kozhevnikov}}, \bibinfo {author} {\bibfnamefont {T.}~\bibnamefont
  {Pruschke}}, \bibinfo {author} {\bibfnamefont {K.}~\bibnamefont {Held}},
  \bibinfo {author} {\bibfnamefont {D.}~\bibnamefont {Vollhardt}},\ and\
  \bibinfo {author} {\bibfnamefont {V.~I.}\ \bibnamefont {Anisimov}},\
  }\bibfield  {title} {\bibinfo {title} {{Mutual Experimental and Theoretical
  Validation of Bulk Photoemission Spectra of
  ${\mathrm{S}\mathrm{r}}_{1\ensuremath{-}x}{\mathrm{C}\mathrm{a}}_{x}{\mathrm{V}\mathrm{O}}_{3}$}},\
  }\href {https://doi.org/10.1103/PhysRevLett.93.156402} {\bibfield  {journal}
  {\bibinfo  {journal} {Phys. Rev. Lett.}\ }\textbf {\bibinfo {volume} {93}},\
  \bibinfo {pages} {156402} (\bibinfo {year} {2004})}\BibitemShut {NoStop}%
\bibitem [{\citenamefont {Sun}\ \emph {et~al.}(2023)\citenamefont {Sun},
  \citenamefont {Huo}, \citenamefont {Hu}, \citenamefont {Li}, \citenamefont
  {Liu}, \citenamefont {Han}, \citenamefont {Tang}, \citenamefont {Mao},
  \citenamefont {Yang}, \citenamefont {Wang}, \citenamefont {Cheng},
  \citenamefont {Yao}, \citenamefont {Zhang},\ and\ \citenamefont
  {Wang}}]{Sun2023}%
  \BibitemOpen
  \bibfield  {author} {\bibinfo {author} {\bibfnamefont {H.}~\bibnamefont
  {Sun}}, \bibinfo {author} {\bibfnamefont {M.}~\bibnamefont {Huo}}, \bibinfo
  {author} {\bibfnamefont {X.}~\bibnamefont {Hu}}, \bibinfo {author}
  {\bibfnamefont {J.}~\bibnamefont {Li}}, \bibinfo {author} {\bibfnamefont
  {Z.}~\bibnamefont {Liu}}, \bibinfo {author} {\bibfnamefont {Y.}~\bibnamefont
  {Han}}, \bibinfo {author} {\bibfnamefont {L.}~\bibnamefont {Tang}}, \bibinfo
  {author} {\bibfnamefont {Z.}~\bibnamefont {Mao}}, \bibinfo {author}
  {\bibfnamefont {P.}~\bibnamefont {Yang}}, \bibinfo {author} {\bibfnamefont
  {B.}~\bibnamefont {Wang}}, \bibinfo {author} {\bibfnamefont {J.}~\bibnamefont
  {Cheng}}, \bibinfo {author} {\bibfnamefont {D.-X.}\ \bibnamefont {Yao}},
  \bibinfo {author} {\bibfnamefont {G.-M.}\ \bibnamefont {Zhang}},\ and\
  \bibinfo {author} {\bibfnamefont {M.}~\bibnamefont {Wang}},\ }\bibfield
  {title} {\bibinfo {title} {{Signatures of superconductivity near 80 K in a
  nickelate under high pressure}},\ }\href
  {https://doi.org/10.1038/s41586-023-06408-7} {\bibfield  {journal} {\bibinfo
  {journal} {Nature}\ }\textbf {\bibinfo {volume} {621}},\ \bibinfo {pages}
  {493} (\bibinfo {year} {2023})}\BibitemShut {NoStop}%
\bibitem [{\citenamefont {Yang}\ \emph {et~al.}(2024)\citenamefont {Yang},
  \citenamefont {Sun}, \citenamefont {Hu}, \citenamefont {Xie}, \citenamefont
  {Miao}, \citenamefont {Luo}, \citenamefont {Chen}, \citenamefont {Liang},
  \citenamefont {Zhu}, \citenamefont {Qu}, \citenamefont {Chen}, \citenamefont
  {Huo}, \citenamefont {Huang}, \citenamefont {Zhang}, \citenamefont {Zhang},
  \citenamefont {Yang}, \citenamefont {Wang}, \citenamefont {Peng},
  \citenamefont {Mao}, \citenamefont {Liu}, \citenamefont {Xu}, \citenamefont
  {Qian}, \citenamefont {Yao}, \citenamefont {Wang}, \citenamefont {Zhao},\
  and\ \citenamefont {Zhou}}]{Yang2024}%
  \BibitemOpen
  \bibfield  {author} {\bibinfo {author} {\bibfnamefont {J.}~\bibnamefont
  {Yang}}, \bibinfo {author} {\bibfnamefont {H.}~\bibnamefont {Sun}}, \bibinfo
  {author} {\bibfnamefont {X.}~\bibnamefont {Hu}}, \bibinfo {author}
  {\bibfnamefont {Y.}~\bibnamefont {Xie}}, \bibinfo {author} {\bibfnamefont
  {T.}~\bibnamefont {Miao}}, \bibinfo {author} {\bibfnamefont {H.}~\bibnamefont
  {Luo}}, \bibinfo {author} {\bibfnamefont {H.}~\bibnamefont {Chen}}, \bibinfo
  {author} {\bibfnamefont {B.}~\bibnamefont {Liang}}, \bibinfo {author}
  {\bibfnamefont {W.}~\bibnamefont {Zhu}}, \bibinfo {author} {\bibfnamefont
  {G.}~\bibnamefont {Qu}}, \bibinfo {author} {\bibfnamefont {C.-Q.}\
  \bibnamefont {Chen}}, \bibinfo {author} {\bibfnamefont {M.}~\bibnamefont
  {Huo}}, \bibinfo {author} {\bibfnamefont {Y.}~\bibnamefont {Huang}}, \bibinfo
  {author} {\bibfnamefont {S.}~\bibnamefont {Zhang}}, \bibinfo {author}
  {\bibfnamefont {F.}~\bibnamefont {Zhang}}, \bibinfo {author} {\bibfnamefont
  {F.}~\bibnamefont {Yang}}, \bibinfo {author} {\bibfnamefont {Z.}~\bibnamefont
  {Wang}}, \bibinfo {author} {\bibfnamefont {Q.}~\bibnamefont {Peng}}, \bibinfo
  {author} {\bibfnamefont {H.}~\bibnamefont {Mao}}, \bibinfo {author}
  {\bibfnamefont {G.}~\bibnamefont {Liu}}, \bibinfo {author} {\bibfnamefont
  {Z.}~\bibnamefont {Xu}}, \bibinfo {author} {\bibfnamefont {T.}~\bibnamefont
  {Qian}}, \bibinfo {author} {\bibfnamefont {D.-X.}\ \bibnamefont {Yao}},
  \bibinfo {author} {\bibfnamefont {M.}~\bibnamefont {Wang}}, \bibinfo {author}
  {\bibfnamefont {L.}~\bibnamefont {Zhao}},\ and\ \bibinfo {author}
  {\bibfnamefont {X.~J.}\ \bibnamefont {Zhou}},\ }\bibfield  {title} {\bibinfo
  {title} {{Orbital-dependent electron correlation in double-layer nickelate
  La$_{3}$Ni$_{2}$O$_{7}$}},\ }\href
  {https://doi.org/10.1038/s41467-024-48701-7} {\bibfield  {journal} {\bibinfo
  {journal} {Nat. Commun.}\ }\textbf {\bibinfo {volume} {15}},\ \bibinfo
  {pages} {4373} (\bibinfo {year} {2024})}\BibitemShut {NoStop}%
\bibitem [{\citenamefont {Cao}\ and\ \citenamefont
  {Yang}(2024)}]{PhysRevB.109.L081105}%
  \BibitemOpen
  \bibfield  {author} {\bibinfo {author} {\bibfnamefont {Y.}~\bibnamefont
  {Cao}}\ and\ \bibinfo {author} {\bibfnamefont {Y.-f.}\ \bibnamefont {Yang}},\
  }\bibfield  {title} {\bibinfo {title} {{Flat bands promoted by Hund's rule
  coupling in the candidate double-layer high-temperature superconductor
  ${\mathrm{La}}_{3}{\mathrm{Ni}}_{2}{\mathrm{O}}_{7}$ under high pressure}},\
  }\href {https://doi.org/10.1103/PhysRevB.109.L081105} {\bibfield  {journal}
  {\bibinfo  {journal} {Phys. Rev. B}\ }\textbf {\bibinfo {volume} {109}},\
  \bibinfo {pages} {L081105} (\bibinfo {year} {2024})}\BibitemShut {NoStop}%
\bibitem [{\citenamefont {Georges}\ \emph {et~al.}(2013)\citenamefont
  {Georges}, \citenamefont {Medici},\ and\ \citenamefont
  {Mravlje}}]{annurev-conmatphys-020911-125045}%
  \BibitemOpen
  \bibfield  {author} {\bibinfo {author} {\bibfnamefont {A.}~\bibnamefont
  {Georges}}, \bibinfo {author} {\bibfnamefont {L.~d.}\ \bibnamefont
  {Medici}},\ and\ \bibinfo {author} {\bibfnamefont {J.}~\bibnamefont
  {Mravlje}},\ }\bibfield  {title} {\bibinfo {title} {{Strong Correlations from
  Hund’s Coupling}},\ }\href
  {https://doi.org/https://doi.org/10.1146/annurev-conmatphys-020911-125045}
  {\bibfield  {journal} {\bibinfo  {journal} {Annu. Rev. Condens. Matter
  Phys.}\ }\textbf {\bibinfo {volume} {4}},\ \bibinfo {pages} {137} (\bibinfo
  {year} {2013})}\BibitemShut {NoStop}%
\bibitem [{\citenamefont {Noguchi}\ \emph {et~al.}(1996)\citenamefont
  {Noguchi}, \citenamefont {Kusaba}, \citenamefont {Fukuoka},\ and\
  \citenamefont {Syono}}]{noguchi1996shock}%
  \BibitemOpen
  \bibfield  {author} {\bibinfo {author} {\bibfnamefont {Y.}~\bibnamefont
  {Noguchi}}, \bibinfo {author} {\bibfnamefont {K.}~\bibnamefont {Kusaba}},
  \bibinfo {author} {\bibfnamefont {K.}~\bibnamefont {Fukuoka}},\ and\ \bibinfo
  {author} {\bibfnamefont {Y.}~\bibnamefont {Syono}},\ }\bibfield  {title}
  {\bibinfo {title} {{Shock-induced phase transition of MnO around 90GPa}},\
  }\href {https://doi.org/https://doi.org/10.1029/96GL01326} {\bibfield
  {journal} {\bibinfo  {journal} {Geophys. Res. Lett.}\ }\textbf {\bibinfo
  {volume} {23}},\ \bibinfo {pages} {1469} (\bibinfo {year}
  {1996})}\BibitemShut {NoStop}%
\bibitem [{\citenamefont {Mita}\ \emph {et~al.}(2001)\citenamefont {Mita},
  \citenamefont {Sakai}, \citenamefont {Izaki}, \citenamefont {Kobayashi},
  \citenamefont {Endo},\ and\ \citenamefont {Mochizuki}}]{mita2001optical}%
  \BibitemOpen
  \bibfield  {author} {\bibinfo {author} {\bibfnamefont {Y.}~\bibnamefont
  {Mita}}, \bibinfo {author} {\bibfnamefont {Y.}~\bibnamefont {Sakai}},
  \bibinfo {author} {\bibfnamefont {D.}~\bibnamefont {Izaki}}, \bibinfo
  {author} {\bibfnamefont {M.}~\bibnamefont {Kobayashi}}, \bibinfo {author}
  {\bibfnamefont {S.}~\bibnamefont {Endo}},\ and\ \bibinfo {author}
  {\bibfnamefont {S.}~\bibnamefont {Mochizuki}},\ }\bibfield  {title} {\bibinfo
  {title} {{Optical Study of MnO under High Pressure}},\ }\href
  {https://doi.org/https://doi.org/10.1002/1521-3951(200101)223:1<247::AID-PSSB247>3.0.CO;2-R}
  {\bibfield  {journal} {\bibinfo  {journal} {phys. stat. sol. (b)}\ }\textbf
  {\bibinfo {volume} {223}},\ \bibinfo {pages} {247} (\bibinfo {year}
  {2001})}\BibitemShut {NoStop}%
\bibitem [{\citenamefont {Mita}\ \emph {et~al.}(2005)\citenamefont {Mita},
  \citenamefont {Izaki}, \citenamefont {Kobayashi},\ and\ \citenamefont
  {Endo}}]{PhysRevB.71.100101}%
  \BibitemOpen
  \bibfield  {author} {\bibinfo {author} {\bibfnamefont {Y.}~\bibnamefont
  {Mita}}, \bibinfo {author} {\bibfnamefont {D.}~\bibnamefont {Izaki}},
  \bibinfo {author} {\bibfnamefont {M.}~\bibnamefont {Kobayashi}},\ and\
  \bibinfo {author} {\bibfnamefont {S.}~\bibnamefont {Endo}},\ }\bibfield
  {title} {\bibinfo {title} {{Pressure-induced metallization of MnO}},\ }\href
  {https://doi.org/10.1103/PhysRevB.71.100101} {\bibfield  {journal} {\bibinfo
  {journal} {Phys. Rev. B}\ }\textbf {\bibinfo {volume} {71}},\ \bibinfo
  {pages} {100101} (\bibinfo {year} {2005})}\BibitemShut {NoStop}%
\bibitem [{\citenamefont {Huang}\ \emph {et~al.}(2012)\citenamefont {Huang},
  \citenamefont {Wang},\ and\ \citenamefont {Dai}}]{PhysRevB.85.245110}%
  \BibitemOpen
  \bibfield  {author} {\bibinfo {author} {\bibfnamefont {L.}~\bibnamefont
  {Huang}}, \bibinfo {author} {\bibfnamefont {Y.}~\bibnamefont {Wang}},\ and\
  \bibinfo {author} {\bibfnamefont {X.}~\bibnamefont {Dai}},\ }\bibfield
  {title} {\bibinfo {title} {{Pressure-driven orbital selective
  insulator-to-metal transition and spin-state crossover in cubic CoO}},\
  }\href {https://doi.org/10.1103/PhysRevB.85.245110} {\bibfield  {journal}
  {\bibinfo  {journal} {Phys. Rev. B}\ }\textbf {\bibinfo {volume} {85}},\
  \bibinfo {pages} {245110} (\bibinfo {year} {2012})}\BibitemShut {NoStop}%
\bibitem [{\citenamefont {Shorikov}\ \emph {et~al.}(2010)\citenamefont
  {Shorikov}, \citenamefont {Pchelkina}, \citenamefont {Anisimov},
  \citenamefont {Skornyakov},\ and\ \citenamefont
  {Korotin}}]{PhysRevB.82.195101}%
  \BibitemOpen
  \bibfield  {author} {\bibinfo {author} {\bibfnamefont {A.~O.}\ \bibnamefont
  {Shorikov}}, \bibinfo {author} {\bibfnamefont {Z.~V.}\ \bibnamefont
  {Pchelkina}}, \bibinfo {author} {\bibfnamefont {V.~I.}\ \bibnamefont
  {Anisimov}}, \bibinfo {author} {\bibfnamefont {S.~L.}\ \bibnamefont
  {Skornyakov}},\ and\ \bibinfo {author} {\bibfnamefont {M.~A.}\ \bibnamefont
  {Korotin}},\ }\bibfield  {title} {\bibinfo {title} {{Orbital-selective
  pressure-driven metal to insulator transition in FeO from dynamical
  mean-field theory}},\ }\href {https://doi.org/10.1103/PhysRevB.82.195101}
  {\bibfield  {journal} {\bibinfo  {journal} {Phys. Rev. B}\ }\textbf {\bibinfo
  {volume} {82}},\ \bibinfo {pages} {195101} (\bibinfo {year}
  {2010})}\BibitemShut {NoStop}%
\bibitem [{git()}]{github}%
  \BibitemOpen
  \href@noop {} {}\bibinfo {note}
  {Https://github.com/huangli712/Zen}\BibitemShut {NoStop}%
\bibitem [{nor()}]{norg_github}%
  \BibitemOpen
  \href@noop {} {}\bibinfo {note} {Https://github.com/rqHe1/NORG}\BibitemShut
  {NoStop}%
\end{thebibliography}%

\end{document}